\documentclass[twocolumn,tighten]{aastex63_rt42}

\usepackage{graphicx}
\usepackage{amsmath}
\usepackage{multirow}
\usepackage{color}
\usepackage{icomma}
\usepackage{etipopsbl}

\def\cm{\textrm{cm}}

\def\um{\mu\textrm{m}}

\def\Watt{\textrm{W}}

\def\Hz{\textrm{Hz}}

\def\MHz{\textrm{MHz}}
\def\GHz{\textrm{GHz}}

\def\mJy{\textrm{mJy}}
\def\Jy{\textrm{Jy}}

\def\sec{\textrm{s}}
\def\minute{\textrm{min}}
\def\hr{\textrm{hr}}
\def\yr{\textrm{yr}}

\def\Msun{{M}_{\odot}}
\def\Lsun{{L}_{\odot}}

\def\starUnit{\textrm{star}}

\def\meter{\textrm{m}}

\def\kpc{\textrm{kpc}}
\def\Mpc{\textrm{Mpc}}

\def\sec{\textrm{s}}
\def\hr{\textrm{hr}}
\def\yr{\textrm{yr}}

\def\Kelvin{\textrm{K}}

\def\erg{\textrm{erg}}

\def\arcsec{\textrm{arcsec}}
\def\arcmin{\textrm{arcmin}}

\def\ga{\gtrsim}
\def\la{\lesssim}

\def\endash{\text{--}}

\shorttitle{ETI Broadcast Populations II. Single Galaxy Constraints}
\shortauthors{Lacki}

\begin{document}

\title{Artificial Broadcasts as Galactic Populations: II. Comparing Individualist and Collective Bounds on Broadcast Populations in Single Galaxies}

\newcommand{\UCB}{Department of Astronomy,  University of California Berkeley, Berkeley CA 94720}

\correspondingauthor{Brian C. Lacki}
\email{astrobrianlacki@gmail.com}
\author[0000-0003-1515-4857]{Brian C. Lacki}
\affiliation{Breakthrough Listen, \UCB}

\begin{abstract}
The search for extraterrestrial intelligence includes efforts to constrain populations of artificial broadcasts in other galaxies. Previous efforts use individualist methods, searching for single broadcasts with high signal-to-noise ratio. These would be detected as observables with extreme values. This approach is limited to very bright broadcasts and also is subject to confusion, where a large number of broadcasts blend together to form a noise continuum. The mean value of the total emission provides an additional collective bound: the luminosity of the transmitters is no higher than the galaxy's observed luminosity. Using the framework developed in Paper I, I evaluate how confusion affects individualist searches. I then compare individualist and collective approaches for radio broadcasts from the Milky Way, M31, and three Virgo Cluster elliptical galaxies. For current observations, confusion blurs narrowband radio broadcasts together in the Virgo ellipticals when there is one broadcast per gigahertz per 1000 stars. The collective bound implies fewer than $\sim 10^6 (\bLisoBAR/10^{13}\ \Watt)^{-1}$ L-band broadcasts per star gigahertz GHz in the Milky Way and is about $10$ and $400$ times stronger in M31 and M59, respectively. Applying the collective bound to the far-infrared--radio correlation yields constraints on radio broadcast populations in star-forming galaxies throughout the Universe. The collective bound allows us to rule out large regions of broadcast population parameter space even for distant galaxies. It also imposes constraints on gamma-ray, neutrino, and gravitational-wave broadcasts in the nearest galaxies.
\end{abstract}

\keywords{Search for extraterrestrial intelligence -- Technosignatures -- Galaxy luminosities -- Astronomical techniques -- Radio astronomy --- Spatial point processes}

\section{Introduction}

A central debate in the {search for extraterrestrial intelligence} (SETI; \citealt{Tarter01,Worden17}) is whether interstellar travel {boosts} the number of broadcasting societies. Although realistic travel times between the stars are long, they are miniscule compared to the age of the Galaxy. If ETI societies can reliably replicate through interstellar travel and any have the motivation, then the Galaxy could be covered within about a hundred million years (e.g., \citealt{Jones81,Wright14-Paradox}{; \citealt{CarrollNellenback19}}). Essentially, the Galaxy would experience a ``phase transition'' between unpopulated wilderness and a ``metasociety'' of densely packed ETI{s} (Paper I, {Lacki 2024 in press}; {compare} {\citealt{Kuiper77}} {and} \citealt{Cirkovic08}). The apparent lack of evidence for so widespread ETIs in the Milky Way and its implications is the subject of much debate (\citealt{Brin83}; {\citealt{Webb15};} \citealt{Cirkovic18-Book}{; \citealt{Forgan19,Lingam21}}).

Less attention has focused on the technosignature properties of other galaxies if this reasoning is correct, aside from searches for megastructure populations \citep{Annis99,Voros14,Wright14-Paradox,Zackrisson15}. If ETIs are rare but establish a vast number of societies when they occur, then some galaxies may be heavily populated while others are uninhabited. This should carry over to their technosignatures -- some galaxies would be entirely barren of them, {while} others would be brimming with the signs of billions of inhabited worlds. A negative SETI result for one galaxy might mean nothing for another (Paper I).

\begin{figure*}
\centerline{\includegraphics[width=18cm]{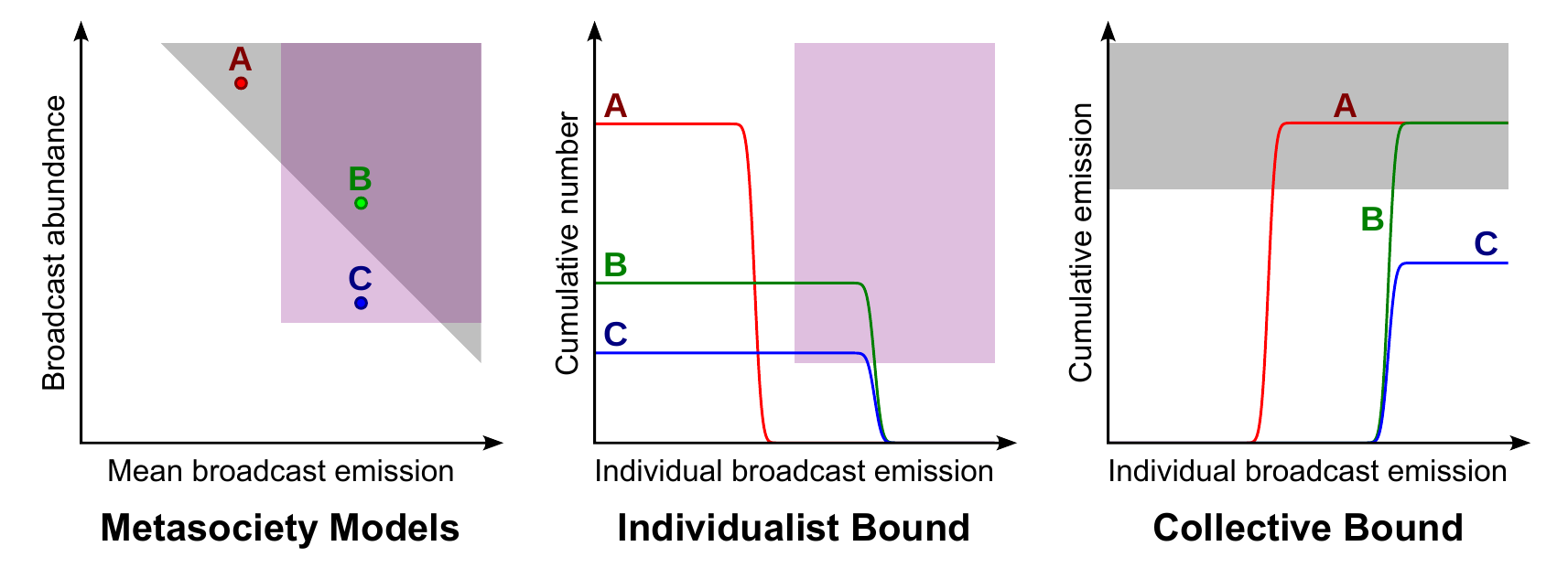}}
\figcaption{{A sketch of how individualist and collective bounds probe the broadcast luminosity distribution in different ways. On left is a standard luminosity {vs.} abundance plot for broadcasts, as seen in many SETI papers. There is a single individualist bound (purple), a result from a typical SETI search, which rules out model metasocieties B and C. A collective bound (grey) rules out model metasocieties A and B. The middle (individualist) and right (collective) panels show the relevant cumulative distributions. Any distribution {that} touches the shaded regions is ruled out. The individualist bound is a very sensitive probe of the cumulative number of broadcasts \emph{brighter} than a luminosity sensitivity, but {it} has no sensitivity below {that,} even if there are very many broadcasts as in model A. Not shown are the effects of confusion, which happens when broadcasts are so numerous as to blend together. The collective bound is weakly sensitive to populations of all luminosities but is unable to constrain model {C,} where there is a moderate number of very bright broadcasts. Note {that} the collective bound's ``power'' is the same for {models} A and B, despite A having far more {broadcasts,} because those broadcasts are much dimmer.}\label{fig:IndividualVsCollective}}
\end{figure*} 

This paper considers the constraints we can set on broadcasts from an individual galactic metasociety. Thus far, radio and optical SETI surveys treat broadcasts as potential rare anomalies that must be sifted out from natural sources, noise, and local interference, employing strategies that look for \emph{individual} candidates that stick out from this background (e.g., {\citealt{Oliver71,Howard04}}; \citealt{Enriquez17}). The individualist approach exploits extreme values in the {broadcast} population statistics. This paper introduces a complementary \emph{collective} approach of using the integrated luminosity of the population: all the broadcasts in the galaxy put together cannot outshine the galaxy's observed total emission (Figure~\ref{fig:IndividualVsCollective}).

Constraints based on aggregate emission are a general strategy for noticing and setting limits on new or unknown phenomena \citep[e.g.,][]{Greggio90,Draine98}, particularly in astroparticle physics \citep[as in][]{Spekkens13,GeringerSameth15}. In SETI, they let us trade abundance with brightness -- a given {aggregate} luminosity can be achieved by a few very bright broadcasts or a great many faint ones -- and allow us to constrain faint transmissions even in distant galaxies. The collective emission of an ETI population can also be relatively insensitive to the detailed properties of individual transmissions thanks to the central limit theorem.  We can use the collective bound to set limits on signals that do not remotely resemble the classical pulses or carrier waves ({\citealt{Kardashev64,Caves94}}; \citealt{Lacki15-ATTs}), even ones that look like white noise, obviating the need for a ``magic frequency'' or ``magic basis.''  Finally, they are not subject to confusion: when an observation covers many broadcasts, they can all blend together into a noisy background with none individually detectable. {It is this quasi-continuum that collective bounds best constrain.}

Indeed, the collective approach follows in the footsteps of searches for \citet{Kardashev64} Type III societies in other galaxies. This {branch of SETI} seeks the collective effects of millions of inhabited solar systems on the galaxy as a whole{, like the waste heat or stellar obscuration from populations of megastructures.} Obviously no such ubiquitous population exists in the Milky Way (\citealt{Jugaku04,Carrigan09}{; \citealt{Suazo22}}), but this does not preclude Type III societies in other galaxies \citep{Wright14-Paradox}.

I present a comparison of the relative strengths and weaknesses of the individualist and collective bounds for radio broadcasts {in individual galaxies}, including a discussion on confusion. Although the focus here is on galaxy-wide populations, collective bounds could be applied to individual stars or planets, even in the absence of interstellar travel.  For example, Earth has a multitude of radio transmitters \citep{Sullivan78}, which could become so numerous and wideband as to be hopelessly confused -- yet Earth could still appear anomalous if it is far too radio-bright for a terrestrial planet.

{
\subsection{Outline of Paper II}
The formalism of Paper I is reviewed in Section~\ref{sec:ReviewI}, and then applied to the measured aggregate emission of metasocieties in Section~\ref{sec:AggEmissionMeasurements}. The next three sections deal with the individualist, {signal-to-noise-based} constraints on ETI broadcasts and how {they are weakened} by confusion: general considerations in Section~\ref{sec:SN}, then the signal-to-noise {ratio} for radio broadcasts in Section~\ref{sec:SNRadio} and optical broadcasts in Section~\ref{sec:SNOptical}. The collective bound is explained in Section~\ref{sec:Collective}. I consider the relative merits of individualist surveys and the collective bounds for some nearby galaxies in Section~\ref{sec:RadioSingleConstraints}. Section~\ref{sec:GeneralCollective} expands the discussion of the collective bound to distant galaxies across the spectrum. The conclusions, in Section~\ref{sec:Conclusion}, are followed by several appendices with detailed derivations.}

\section{Review of concepts}
\label{sec:ReviewI}
Paper I developed a treatment of ETI populations, but this section reviews basic results used in this paper. {At the heart of the formalism is the idea of describing broadcasts and their host societies with random variables. Random variables are a very flexible concept -- even a deterministic quantity can be viewed as a random variable with a degenerate distribution. It is likely that at least some properties of technosignatures are unpredictable to us. The time and circumstances {in which} an ETI evolves on a planet, the exact times it chooses to broadcast and for how long, and the random locations of stars in a galaxy are {all} contingent, and are among the motivations for a statistical treatment.}

{Table~\ref{table:Notation} is a key to the most commonly used variables and notation in the paper.}


\begin{deluxetable*}{cp{13cm}}
\tabletypesize{\footnotesize}
\tablecolumns{2}
\tablewidth{0pt}
\tablecaption{Summary of notation used\label{table:Notation}}
\tablehead{\colhead{Notation} & \colhead{Explanation}}
\startdata
$\IndicatorOf{\fpEvent}$ & Indicator variable for event $\fpEvent$, is $1$ if the event occurs and $0$ otherwise\\
$\CDF{X}(x), \PDF{X}(x)$ & Cumulative distribution function (CDF) and probability density function (PDF) for random variable $X$, evaluated at $x$\\
$\Mean{X}$               & Mean of random variable $X$; by default, is a simple mean\\
$\Var{X}$, {$\SD{X}$} & Variance {and standard deviation} of random variable $X$; by default, {are the} simple {operations}\\
$\Median{X}$             & Median of random variable $X$\\
$\GenLabel, \AltLabel$                            & Generic window\\
$\ModeLabel, \ObsLabel, \PointLabel, \SurvLabel$  & Windows for a mode, observation, pointing, and survey{,} respectively\\
$\TimeLabel (\TimeVar)$                           & Window picking all objects active at a given time $\TimeVar$\\
$\oGenOfAlt, \oNGenAlt$                           & Set of $\GenLabel${-type} windows that make up $\AltLabel$, and the number of windows in that set\\
$\oDurationGen$, $\oTStartGen$, $\oBandwidthGen$, $\oNuMidGen$, $\oDriftRateGen$, $\oPolSetGen$, $\oSkyFieldGen$, $\oVolumeGen$ & Quantities defining a window $\GenLabel$: its duration, starting time, bandwidth, central frequency, effective drift rate (used in dedrifting), set of polarizations covered, sky field covered, physical volume sampled\\
{$\Selection{\GenLabel}{\KMark}$} & \vphantom{S}{Selection that picks objects hosted by $\KMark$ according to window $\GenLabel$}\\
$\jjTuple$               & Parameter tuple describing object $\JMark$; the space of all such tuples is the $\JMark${-}haystack $\jjHaystack$\\
$\jjkDistGen$            & Distribution (intensity) of $\jjTuple$ over ${\jjHaystack}$ for the population that {would} be selected by $\Selection{\GenLabel}{\KMark}$\\
$\jjkSampleGen$, $\jjkNGen$ & Sample of $\JMark$-type objects drawn by $\Selection{\GenLabel}{\KMark}$ and the number of objects in that sample\\
$\jjSingleGen$           & Generic singleton random variable describing an object $\JMark$, with the quantity integrated over the window {$\GenLabel$; when} no window is given, the $\AllLabel$ window is assumed by default\\
$\jjkAggGen$             & Generic aggregate random variable describing the sum of $\jjSingleGen$ for all objects selected by $\Selection{\GenLabel}{\KMark}$\\
$\jjSingleGenREGjkAlt$   & Random variable $\jjSingleGen$ regularized to only include values likely to occur in a sample drawn by {$\Selection{\AltLabel}{\KMark}$; an} aggregate variable can be regularized too.\\
${\jkDOperationAlt{\jjSingleGen}}$ & Selection-relative {operation} of $\jjSingleGen$: describe{s} the distribution when considering the entire population of objects drawn by the selection $\Selection{\AltLabel}{\KMark}$, instead of {for a single object} $\JMark$. {Operations that can stand in for $\DOperationCore$ include the CDF, PDF, minima, maxima, mean, and variance.} An aggregate variable can be substituted for $\jjSingleGen$. When no window is given, it is inherited from the variable ($\GenLabel$ here).\\
${\DOperationAltNULL{\jjSingleGenI}}$         & \vphantom{M}{Multiwindow operation $\DOperationCore$, ranging over all windows $\GenILabel$ in $\AltLabel$}\\
{$\iResponseMeasureGen$}                 & \vphantom{I}{Instrumental response; normalized to $1$ and assumed to depend only on sky position}\\
$\hNGenGal$                                      & Number of stars covered by window $\GenLabel$ in galaxy $\GalIndex$\\
$\azAbund$                                       & Abundance of societies per star in metasociety $\MetaMark$\\
$\bzAbund (\TimeVar, \FreqVar), \bzAbundnu (\TimeVar, \FreqVar)$ & Mean number of broadcasts per star active at time $\TimeVar$ and frequency $\FreqVar$ in metasociety $\MetaMark$, and similarly for the mean number per unit frequency per star\\
$\bzNWMeasureGenIN$                              & \vphantom{N}{Count of broadcasts in $\bzSampleGen$, weighted by instrumental response to $n$th power}\\
$\bDuration$, $\bTStart$, $\bBandwidthTime$, $\bNuMid$, $\bDriftRate$, $\bEiso$, $\bPolQuantity$, $\bPosition$ & The quantities defining a broadcast: its duration, starting time, instantaneous bandwidth, central frequency, drift rate, effective isotropic energy release, polarization properties, and position\\
$\bEmissionisoGen, \bEisoGen, \bPhotonisoGen, \bLisoGen$ & The effective isotropic emission of a broadcast coincident with window $\GenLabel$\\
{$\bFPol(\PolVar)$, $\bFPolGen$}         & \vphantom{F}{Fraction of broadcast emission in polarization $\PolVar$; fraction in polarizations covered by window $\GenLabel$}\\
{$\bFCoherObs$}                          & \vphantom{Q}{Quantity summarizing certain properties related to broadcast coherence (Appendix~\ref{sec:CoherentRadioVar})}\\
{$\bzFDriftGen$}                         & \vphantom{D}{Describes narrowband broadcast drift rate distribution, equals $\zMean{\bDurationGen^2} / (\oDurationGen \zMean{\bDurationGen})$}\\
{$\lTransmittanceRGen$}                  & \vphantom{T}{Transmittance factor: fraction of $\SingleEmissionCore$ emission from broadcast $\BcMark$ that remains after extinction}\\
{$\jjDilutionR$}                         & \vphantom{D}{Dilution factor for $\SingleEmissionCore$ emission from object $\JMark$, equals $1/(4\pi{\yjjDistanceR}^2)$}\\
{$\mzDilutionWMeasureGenIN$}             & \vphantom{W}{Weighted sum of $\lDilutionR \lTransmittanceRGen$ for all broadcasts drawn by $\Selection{\GenLabel}{\MetaMark}$}\\
$\lFluenceGen$, $\lFluenceEGen$, $\lFluenceQGen${, $\lFluxEGen$} & Fluence accumulated over window $\GenLabel$ from broadcast $\BcMark$\\
{$\mzFluenceObs$}                        & \vphantom{G}{Generic type of fluence accumulated over observation $\ObsLabel$ from all broadcasts in metasociety $\MetaMark$}\\
{$\dFluenceObs$}                         & \vphantom{M}{Measurement of total fluence over $\ObsLabel$ from galaxy $\GalIndex$}\\
$\lMeasureGen$, $\lEnergyGen$, $\lPhotonGen$     & Emission accumulated by an instrument over window $\GenLabel$ from broadcast {$\BcMark$}\\
$\mzMeasureGen$, $\mzEnergyGen$, $\mzPhotonGen$  & Emission accumulated by an instrument over window $\GenLabel$ from all broadcasts in metasociety $\MetaMark$\\
$\kMeasureGen$, $\kEnergyGen$, $\kPhotonGen$     & Background collected by instrument over window $\GenLabel$\\
$\qMeasureGen$, $\qEnergyGen$, $\qPhotonGen$     & Total emission accumulated by an instrument over window $\GenLabel$, both signal and noise\\
$\lSNMeasureObsEST${, $\lSNEnergyObsEST$, $\lSNPhotonObsEST$} & Expected mean signal-to-noise {ratio} of a broadcast {in observation $\ObsLabel$}; a scaling variable\\
$\qSNMeasureObsEST${, $\qSNEnergyObsEST$, $\qSNPhotonObsEST$} & Expected maximum signal-to-noise {ratio} expected for measurements\\
\enddata
\tablecomments{{See Paper I for a comprehensive explanation of the notation. Emission variables are listed for generic emission, energy, photons, and{,} when present, power.}}
\end{deluxetable*}


\subsection{\texorpdfstring{Objects: From {the universe} to broadcasts}{Objects: From the universe to broadcasts}}
{The framework interprets populations as nodes on a tree, each level representing a different type of object. Parent objects may host child objects, which form a population of the child level's type. In variables, each type of object is denoted by an uppercase level ($\JMark$, $\KMark$, and $\LMark$ for arbitrary types to describe general relations). The object trees are random but can be characterized statistically.} The tree is rooted in the {model} {universe (type $\UnivMark$), which contains all other objects,} {and} then has four levels of objects below it.  {The universe contains galaxies (object type $\GalMark$){, which in this work are assumed to totally confine ETIs, even with interstellar travel.}\footnote{\vphantom{I}{If there is enough communication or travel between galaxies to allow for intergalactic metasocieties, that could allow for homogenization, in which case the constraints that apply for one galaxy could apply to its neighbors. However, it could also mean that all galaxies near the Milky Way are atypical and {that} their technosignature properties do not reflect those of the Universe at large.}}}

ETIs have the capacity to reproduce themselves, their infrastructure and environments, and their technosignatures. A \emph{metasociety} {(object type $\MetaMark$)} is a collection of ETIs generally sharing a common origin or influence, and can comprise one or many worlds. {All of the societies originating from a single origin through replication share a metasociety.} Depending on {the} scenario, we may treat the galaxy as having one unified metasociety or many small ones {(see Paper I)}. Metasocieties {embody} the ``phase transition'' of a galaxy going from uninhabited to fully populated.

A \emph{society} {(object type $\SocMark$)} is a localized ETI with its infrastructure, capable of producing technosignatures. A natural interpretation is that of a single world or planetary system, although isolated facilities in interstellar space making broadcasts can also count as societies.

A \emph{broadcast} {(object type $\BcMark$)} is an emission of radiation produced as the technosignature of a society. These include deliberate attempts at communication with other (meta)societies, leakage, and even {noncommunicative} releases of energy. Broadcasts are characterized by their time and frequency ranges, among other properties.

{The objects, as nodes on the tree, can be treated in two ways. Sometimes we want to consider objects that are known or postulated to exist, like the Milky Way or the Arecibo message. These are realized objects, and their properties have fixed values. They are labeled either using an obvious designation for the object in question (e.g., $\MWIndex$ for the Milky Way), or by the lowercase letter for the object's type (e.g., $\MetaIndex$ for a generic realized metasociety). Often, however, we will consider objects whose properties are general random variables. These are random objects, labeled by the uppercase letter for the object's type (e.g., $\MetaMark$ for a random metasociety).} {Variables can describe both realized and random objects, and both types can be considered hosts of {descendant} objects, so most relations that work for one kind work for the other (see Paper I for technicalities).}

{The tree structure accounts for the influences between the objects, with influences running from parent to child instead of between ``siblings.'' A key assumption of this work is that the properties of all the children objects of a parent are independent of each other, conditionalized on the properties of the host. For example, the {active times of broadcasts from a single society} are not independent, because if we know {when} one {happens}, we {have an idea when the society was active and a guess for when it made other broadcasts.} But if we already know {when the society was active, knowing the time} of one broadcast provides no {new} information. The basic {motivation} for this is that it lets us treat the effects of shared history or influence between objects while using assumptions of (conditional) independence to simplify analysis. Thus, all the broadcasts in a society may be at a single designated frequency; this is viewed as a property of the society that all the broadcasts independently draw {on}, rather than a dependent shared property of the broadcasts. This conditional independence allows us to {invoke} certain results like the {c}entral {l}imit {t}heorem, which applies to the sum of all the emission from individual broadcasts, and then applies again to the sum of the emission from all the societies.}

{
\subsection{Haystacks and point processes}
{The intrinsic properties of each object $\JMark$ are given by a tuple of parameters $\jjTuple$.} Programs like SETI that seek out rare objects in a vast abstract space of possibilities often refer to seeking these out as ``needles in a cosmic haystack'' (\citealt{Wright18}; see also \citealt{Harwit81,Djorgovski13}. In allusion to this, the parameter space of all tuples for each object type $\JMark$ is its \emph{haystack}, denoted $\jjHaystack$. For example, the broadcast haystack{'s dimensions} include the bandwidth, duration, starting time, energy release, and polarization properties, among others.

Any population of objects is modeled as a point process on the appropriate haystack. A point process is basically a random set of points in a space \citep[for details, see][]{Kingman93,Daley03,Baddeley07,Chiu13,Haenggi13}. The entire population of $\JMark$-objects hosted by a single ancestor $\KMark$ on the tree is the point process denoted $\jjkSample$ (a realized object can be substituted for $\KMark$). $\jjkN$ is a random variable counting the number of objects in $\jjkSample$.

A distribution (or ``intensity'') of objects on the haystack gives the mean number of objects. For the $\JMark$-type objects sharing the ancestor $\KMark$, the distribution is denoted $\jjkDist$. The mean number of objects in a subset $\fpSubset \subset \jjHaystack$ is,
\begin{equation}
\Mean{\jjkN(\fpSubset)} = \int_{\fpSubset} \jjkDist(\jjTuple | \jkTuple) d\jjTuple .
\end{equation}
In general, the distribution depends on the properties of the ancestor $\KMark$, which are specified by the tuple $\jkTuple$. This lets all the child objects share statistical properties. {I}f the ancestor is a realized object $\KIndex$, then $\jkTuple$ is fixed to $\jkTupleK$, and the only remaining dependence of the distribution is on $\jjTuple$ itself.
}

{A Poisson point process is one in which the number of points in any region has a Poisson distribution and non-overlapping regions are independent \citep{Kingman93}. {It is appropriate when objects appear independently of each other according to a single well-defined intensity.} In this paper, the broadcasts of a {realized} society and the societies of a {realized} metasociety are modeled as Poisson point processes: the properties of the immediate ``parent'' of each subpopulation fully specify the intensity. {But when considering the {descendant} objects of a {higher-level} ancestor, the distribution of objects itself is random, because it is a random sum of distributions from each of the parents hosted by that ancestor. A Cox point process is just this sort of random superposition of Poisson point processes \citep{Kingman93}. In this paper, the broadcasts of a metasociety (or galaxy) are modeled generally as a Cox point process.}}

\subsection{\texorpdfstring{Selection{s}}{Selections}}
The two basic steps from the statistical description of a {population} to an observed quantity are selection and measurement. Selection draws a sample. Measurement generates an observable quantity from the gross properties of the sample, introducing noise variance from background noise and microscopic fluctuations in radiation.

Selection{s} are central to the framework {(Paper I)}. They sieve through the vast panoply of objects in all the {universe}'s history, reflecting the limited scope of our programs. A selection $\Selection{\GenLabel}{\KMark}$ consists of a window $\GenLabel$ and a {host} object $\KMark$.

Windows {filter} objects {solely} according to {their tuples' positions in the haystack. They include the probability that an object at a given location will be selected, essentially the completeness of the selection. A common type of window selects {things} according to} fundamental quantities like time ($\TimeLabel(\TimeVar)$) {and} frequency ($\FreqLabel(\FreqVar)$). {Other e}xamples of windows include observations ($\ObsLabel$) and surveys ($\SurvLabel$). Windows also can filter emission and other quantities, defining bounds of integration: for example, the energy fluence of a broadcast intercepted during one observation can be much smaller than the energy fluence over the entire survey, or all of history. If no window is specified for a random variable, the special ALL-window $\AllLabel$ is used by default, picking every object and all emission regardless of parameters.

The {host} specifies a subpopulation of objects to be selected: objects that are not descendants of the {host} on the ``tree'' are not included. The {host} object is specified by its index; when none is given for a random variable, the {universal host $\UnivIndex$ (or $\UnivMark$)} is assumed by default. In order for an object to be selected by $\Selection{\GenLabel}{\KMark}$, its tuple must be a member of $\jjkSample$.

{The random set of $\JMark$-type objects passed by a selection $\Selection{\GenLabel}{\KMark}$ is the random sample $\jjkSampleGen$.} {A $\JMark$-type object passes this selection if its tuple would be included by the window $\GenLabel$, and the object also happens to be a {descendant} of $\KMark$.} {The selection window modifies the distribution of objects within the host to $\jjkDistGen$, which gives the mean number of sampled objects:
\begin{equation}
\Mean{\jjkNGen} = \int_{\jjHaystack} \jjkDistGen(\jjTuple | \jkTuple) d\jjTuple .
\end{equation}
}

\subsection{Random variables and selection-dependent properties}
Singleton variables are random variables describing single objects, like the energy released by an individual broadcast during an observation. The notation for singleton variable {$\jSingleCore$} for object {$\JMark$} has the form {$\jjSingleGen$}, with the variable itself a lowercase character. Generally, the singleton variable is integrated within the bounds of some window $\GenLabel$; a singleton variable {$\jjSingle$} missing its quantity window integrates without external restriction.

Aggregate variables are the sum of singleton variables for a subpopulation of objects drawn by a selection:
\begin{equation}
{\jjkAggGen = \sum_{\jjTuple \in \jjkSampleGen} \jjSingleGen .}
\end{equation}
Number variables {$\jjkNGen$} are special cases of aggregate variables for which {${\jjSingleGen} = 1$.} {The distinction between an aggregate variable and a singleton variable is{,} in the end{,} one of interpretation. The sum of a quantity among all the objects hosted in a parent can be viewed as a variable associated with the parent itself: $\jkSinglePRIMEGen \leftrightarrow \jjkAggGen$. This interconversion lets us transfer definitions between singleton and aggregate variables.}

{The distribution of $\jjSingleGen$ is fully determined by its parameter tuple $\jjTuple$, but the value it actually takes can be otherwise random. So while the effective isotropic luminosity of a broadcast, $\bLiso$, is an innate property, the number of photons we count from inevitably has shot noise that we cannot predict. The simple mean of a random variable is defined:
\begin{equation}
\Mean{\jjSingleGen} \equiv \Mean{\jjSingleGen | \jjTuple} {.}
\end{equation}
The simple variance follows from the definition of variance: $\Var{\jjSingleGen} \equiv \Mean{\jjSingleGen^2 | \jjTuple} - \Mean{\jjSingleGen | \jjTuple}^2$. The definitions apply analogously for aggregate variables. Campbell's formula lets us find the simple mean of an aggregate variable, applying to any point process:
\begin{equation}
\label{eqn:Campbell}
\Mean{\jjkAggGen} = \int_{\jjHaystack} \Mean{\jjSingleGen} \jjkDistGen(\jjTuple | \jkTuple) d\jjTuple 
\end{equation}
\citep[e.g.,][]{Kingman93,Baddeley07}.}

{But we may instead want to know the statistics over all possible objects drawn from a sample -- say, the mean number of photons we collect from a random broadcast sampled by a survey. Given a selection $\Selection{\AltLabel}{\KMark}$, {the} selection-relative probability density function (PDF) of a variable {is}
\begin{equation}
\jkPDFAlt{\jjSingleGen} (x) \equiv \frac{1}{\Mean{\jjkNAlt}} \int_{\jjHaystack} \jjkDistAlt(\jjTuple) \PDF{\jjSingleGen}(x) d\jjTuple ,
\end{equation}
where $\PDF{\jjSingleGen} (x)$ is the unbiased PDF for $\jjSingleGen$ at $x${, implicitly} conditionalized on $\jjTuple$.} The statistical properties of objects and populations can change depending on how they are selected: selection{s} introduce bias. {The selection-relative mean is derived from the selection-relative PDF and is defined {as}
\begin{align}
       \jlMeanAlt{\jjSingleGen}  & = \Mean{\jjSingleGen | \jjTuple \in \jjlSampleAlt}\\
\nonumber \jlMeanAlt{\jjkAggGen} & = \Mean{\jjkAggGen | \jkTuple \in \jklSampleAlt} .
\end{align}
If the values of $\jjSingleGen$ among different objects are independent of each other and the number of objects, then
\begin{equation}
\jlMeanAlt{\jjSingleGen} = \frac{1}{\Mean{\jjlNAlt}} \int_{\jjHaystack} \Mean{\jjSingleGen} \jjlDistAlt(\jjTuple | \jlTuple) d\jjTuple .
\end{equation}
Selection-relative variances are defined using the selection-relative means: $\jlVarAlt{X} = \jlMeanAlt{X^2} - \jlMeanAlt{X}^2$. 

In this work, almost always the selection window is the same as the window in the variable. If we want to know the ``average number of photons collected from a broadcast in an observation'', usually what we mean is the average number from a broadcast sampled by that observation -- we do not care about all the broadcasts that happened a billion years ago that we measure no photons from. Hence, the selection-relative PDF, means, and variances} ``inherit'' the variables they are averaging over: {$\jlPDF{\jjSingleGen} = \jlPDFGen{\jjSingleGen}$, $\jlMean{\jjSingleGen} = \jlMeanGen{\jjSingleGen}$, $\jlVar{\jjSingleGen} = \jlVarGen{\jjSingleGen}$, and similarly for aggregate variables.}

{When $\jjkSample$ is Poissonian, and the individual $\jjSingleGen$ for each object in the population are identically distributed, mutually independent, and independent of $\jjkNGen$, $\jjkAggGen$ has a compound Poisson distribution. This means it has a mean and variance of
\begin{align}
\nonumber \Mean{\jjkAggGen}  & = \Mean{\jjkNGen} \jkMean{\jjSingleGen} \\
          \Var{\jjkAggGen}   & = \Mean{\jjkNGen} \jkMean{\jjSingleGen^2} 
\end{align}
\citep{Adelson66,Barbour01,Bas19}. We can use this to calculate the mean and variance of the aggregate emission of all broadcasts in a society and the aggregate emission of all societies in a galactic metasociety {(see Paper I)}.}

{
\subsection{Working with surveys: regularization and multiwindow operations}
A survey $\SurvLabel$ {typically} consists of {many} observations $\ObsILabel$, perhaps grouped into pointings at different locations in the sky. The set of observations that make up the survey is $\oObsOfSurv$. A lot of the questions we are interested in depend on the variations between observations in the survey: ``{What} is the brightest broadcast fluence in an observation over the {entire} survey?'', for example, or ``{What} is the variance of an observable we measure in a survey?'' 

The first tool {invokes the finite reach of surveys}. This becomes important when we consider probability distributions with very long tails. {Broad fluence distributions naturally arise if there is a wide range in the intrinsic emission of individual broadcasts, their distances, or, for narrowband lines, drift rates, for example.} Certain power laws can have infinite variances and means because of these heavy tails. In practice, however, we are interested in the means and variances of a typical sample. To do this, we can regularize random variables by truncating their distributions at values unlikely to be sampled:
\begin{equation}
\PDF{\jjSingleGenREGjkAlt}(\jSingleCore) \equiv \frac{\PDF{\jjSingleGen}(\jSingleCore)}{\jkPAlt(\jSingleCUTLO \le \jjSingleGen \le \jSingleCUTHI)} \IndicatorOf{\jSingleCUTLO \le \jSingleCore \le \jSingleCUTHI} .
\end{equation}
{The values of $\jSingleCUTHI$ and $\jSingleCUTLO$ are derived from the probability distributions of the maximum and minimum $\jjSingleGen$ in the sample $\jjkSampleAlt$ {using extreme value theory}. The probability that the maximum $\jjSingleGen$, of all the $\jjkNAlt$ values sampled, is above $\jSingleCUTHI$ is a constant ($1/4$ in Paper I); the probability that the minimum $\jjSingleGen$ is below $\jSingleCUTLO$ is also a constant (again, $1/4$ in Paper I).} Thus, $\lFluenceObsREGzSurv$ is the fluence of a broadcast intercepted by the observation $\ObsLabel$, regularized to only include broadcast fluence values likely to be measured in the survey as it covers the metasociety $\MetaMark$.   

The other {issue} is that sometimes we need to find a mean, variance, or maximum of some quantity over {a group of observations}. The {multiwindow} mean and related observations {generalize} selection-relative operations. Selection-relative operations are defined for a single window, while {multiwindow} operations {consider the range of values} over many windows. {The multiwindow mean} is defined
\begin{equation}
\MeanSurv{\jjSingleObs} \equiv \Mean{\Mean{\jjSingleObsI} | \ObsILabel \in \oObsOfSurv} .
\end{equation}
The {multiwindow} mean can be combined with the selection-relative mean for a host,
\begin{equation}
\jlMeanSurvNULL{\jjSingleObs} \equiv \Mean{\jlMean{\jjSingleObsI} | \ObsILabel \in \oObsOfSurv} .
\end{equation}
The {multiwindow} variance follows from the definition of conditional variance:
\begin{align}
\nonumber \VarSurv{\jjSingleObs} & \equiv \MeanSurv{{\jjSingleObs}^2} + \MeanSurv{\jjSingleObs}^2 \\
                              & = \Mean{\Var{\jjSingleObsI} | \ObsILabel \in \oObsOfSurv} + \Var{\Mean{\jjSingleObsI} | \ObsILabel \in \oObsOfSurv}
\end{align}
This expression of the law of total variance says that the {multiwindow} variance results from both the average variation of the variable within each {subwindow} and the variation between {them}. Finally, we have {multiwindow} maxima, which are useful in defining maximum signal-to-noise {ratio}:
\begin{align}
\nonumber     \MaxSurv{\jjSingleObs} & \equiv \Max{\jjSingleObsI | \ObsILabel \in \oObsOfSurv}\\
        \jlMaxSurvNULL{\jjSingleObs} & \equiv \Max{\Max{\jjSingleObsI | \jjTuple \in \jjlSampleObsI} | \ObsILabel \in \oObsOfSurv}
\end{align}
Of course, an aggregate variable can be used in place of the {singleton} variable {in this equation} as well.
}

\subsection{The box and chord models}
The box and chord models give us analytical results for the emission properties of broadcasts. In both models, the window $\GenLabel$ is treated as a contiguous ``box'' covering the time range $\oTStartGen \le \TimeVar \le \oTStartGen + \oDurationGen$, the frequency range $\oNuMidGen - \oBandwidthGen/2 \le \FreqVar \le \oNuMidGen + \oBandwidthGen/2$, and $\oNumPolGen$ independent polarizations ($\PolVar$) from the set $\oPolSetGen$. Here $\oDurationGen$ is the duration of the window, $\oBandwidthGen$ is its bandwidth, $\oTStartGen$ is the {start}{ing} time of the window, and $\oNuMidGen$ is its central frequency. 

{Broadcasts {may} fall into one of four categories according to their duration {$\bDuration$} and (instantaneous) bandwidth {$\bBandwidthTime$}, relative to that of the window.  Lines are broadcasts that are narrowband and long-lived ($\bDuration \gg \oDurationGen$; $\bBandwidthTime \ll \oBandwidthGen$), while pulses are wideband and short-lived ($\bDuration \ll \oDurationGen$; $\bBandwidthTime \gg \oBandwidthGen$). Less commonly considered in SETI are the hisses, wideband and long-lived continuum sources ($\bDuration \gg \oDurationGen$; $\bBandwidthTime \gg \oBandwidthGen$), and the blips, narrowband transients ($\bDuration \ll \oDurationGen$; $\bBandwidthTime \ll \oBandwidthGen$). The box model is used {for} pulses, hisses, and blips, while the chord model allows for the treatment of lines with frequency drift.}

\subsubsection{The box model}
The box model treats broadcasts as uniform ``boxes'' in time-frequency space as well, all with a single duration $\bDurationBox$ and bandwidth $\bBandwidthBox$. {Each broadcast begins at time $\bTStart$ and is centered at frequency $\bNuMid$.} There is no skewness to the box, no drift or dispersion. The {(effective isotropic)} spectral luminosity per polarization of broadcast {$\BcMark$} is unvarying within the box:
\begin{multline}
\bLnupoliso (\TimeVar, \FreqVar, \PolVar) = \bFPol (\PolVar) \frac{\bEiso}{\bDurationBox \bBandwidthBox}\\
{\cdot \IndicatorOf{0 \le \TimeVar - \bTStart \le \bDurationBox} \cdot \IndicatorOf{|\FreqVar - \bNuMid| \le \bBandwidthBox/2)}},
\end{multline}
where $\bEiso$ is the effective isotropic energy released during the broadcast, ${\bFPol}(\PolVar)$ of which is into polarization $\PolVar$. A broadcast is selected by {$\Selection{\GenLabel}{\JMark}$} if its ``box'' overlaps the window's ``box'' and it is part of {$\JMark$'s} broadcast population. The (effective isotropic) energy integrated over the overlap, {$\bEisoGen$}, is proportional to the time-frequency ``area'' of the overlapping region{:
\begin{equation}
{\bEisoGen} = \frac{\bEiso {\bDurationGen \bBandwidthGen}}{\bDurationBox \bBandwidthBox} \sum_{\PolVar \in {\oPolSetGen}} \bFPol (\PolVar) {= \frac{\bEiso \bDurationGen \bBandwidthGen}{\bDurationBox \bBandwidthBox} \bFPolGen},
\end{equation}
where ${\bDurationGen}$ is the length of time and ${\bBandwidthGen}$ is the span of frequencies in which both the broadcast and the window are active.} {In the box model, the} expected number of intercepted broadcasts in {host $\JMark$} (regardless of detectability) is 
\begin{equation}
\label{eqn:MeanNBox}
{\Mean{\bjjNGen}} \approx \bjjRatenu(\oTStartGen, \oNuMidGen) {\cdot} {\Mean{\hjjNGen}} (\oDurationGen + \bDurationBox) (\oBandwidthGen + \bBandwidthBox) ,
\end{equation}
for a broadcast frequency rate per star of {
\begin{equation}
\bjjRatenu (\TimeVar, \FreqVar) = \frac{1}{\Mean{\hjjNTime (\TimeVar)}} \frac{d^2 \Mean{\bjjN}}{d\bTStart d\bNuMid} (\bTStart = \TimeVar, \bNuMid = \FreqVar),
\end{equation}
where $\hjjNTime (\TimeVar)$ is the number of stars existing in the host $\JMark$ at time $\TimeVar$ {and $\hjjNGen$ is the number in $\JMark$ covered by the window $\GenLabel$.}}

\subsubsection{The chord model}
The chord model treats broadcasts as ultranarrowband lines that drift linearly in frequency. Thus, each broadcast acts like a ``chord'' cutting across the window {``box.''} Each broadcast's drift rate {$\bDriftRate$} is constant, but the population of broadcasts has a whole distribution of drift rates. {In this work}, I adopt the uniform drift rate distribution,
\begin{equation}
\label{eqn:UniformDriftDist}
\jjPDF{\bDriftRate} = 1/(2\bjjDriftRateBAR) \cdot {\IndicatorOf{|\bDriftRate - \bjjDriftRateMid| \le \bjjDriftRateBAR}}.
\end{equation}
Note this {is the distribution unbiased by any window selection}; narrowband windows are more likely to pick broadcasts with high drift rates. The chords have no intrinsic luminosity variability {(but see Section~\ref{sec:CoherentRadioNoise})}. Therefore, the total amount of emission from broadcast {$\BcMark$} intercepted {during} the window is directly proportional to the time it takes to cross the box, {$\bDurationGen$.} {Each broadcast has an effective isotropic luminosity $\bLiso$; it releases
\begin{equation}
\Mean{{\bEisoGen}} = \bLiso {\bDurationGen} \sum_{\PolVar \in {\oPolSetGen}} \bFPol (\PolVar) {= \bLiso \bDurationGen \bFPolGen}
\end{equation}
{as energy} during the times, frequencies, and polarizations covered by the observation ${\GenLabel}$.} Selection by $\Selection{\GenLabel}{{\JMark}}$ happens when a broadcast has ${\bDurationGen} > 0$ and is part of ${\JMark}$'s broadcast population.\footnote{As in Paper I, the broadcast's parent society, metasociety, and galaxy must also be selected by the window, although that can be assumed for any broadcast falling within the window for this paper.} From this criterion, the expected number of intercepted broadcasts is
\begin{equation}
\label{eqn:MeanNChord}
\Mean{{\bjjNGen}} \approx {\bjjAbundnu}(\oTStartGen, \oNuMidGen) {\cdot} {\Mean{\hjjNGen}} (\oBandwidthGen + {\jjMean{|\bDriftRate|}} \oDurationGen) ,
\end{equation}
using the instantaneous broadcast frequency abundance per star {
\begin{equation}
\bjjAbundnu (\TimeVar, \FreqVar) = \frac{1}{\Mean{\hjjNTime (\TimeVar)}} \frac{d^2 \Mean{\bjjNTime (\TimeVar)}}{d\bNuMid} (\bNuMid = \FreqVar) \approx \bjjRatenu \jjMean{\bDuration},
\end{equation}
with $\bjjNTime (\TimeVar)$ equal to the number of active broadcasts in $\JMark$ at time $\TimeVar$.}

\subsection{Assumptions Used in This Paper}
\label{sec:StandardAssumptions}
{A number of simplifying assumptions are used in this paper to make calculations tractable, summarized in Table~\ref{table:StandardAssumptions}.}


\begin{deluxetable}{p{2.5cm}p{5cm}}
\tabletypesize{\footnotesize}
\tablecolumns{2}
\tablewidth{0pt}
\tablecaption{{Standard Assumptions Used in This Paper\label{table:StandardAssumptions}}}
\tablehead{\colhead{Assumption} & \colhead{Description}}
\startdata
{Independence}		             & {Objects sharing the same ``parent'' have independent properties, conditionalized on the parent's properties. The number of objects is independent of their properties.}\\
{Interchangeability}           & {The broadcast distribution in one society at one epoch is identical to that in another society or epoch, except translated in space and time.}\\
{Interchangeable observations} & {Within a single metasociety, the statistical properties of broadcasts and societies are the same in every observation considered, regardless of frequency, time, or location.}\\
{Single metasociety}           & {Each galaxy has one metasociety; one distribution characterizes all broadcasts and societies in it. Background and foreground galaxies are ignored.}\\
{Diffuse approximation}        & {The number of sampled broadcasts per society is typically $\ll 1$. Their clumping into societies is ignored, as if the broadcasts themselves are scattered in a diffuse cloud throughout the galaxy according to a Poisson point process.}\\
{Distant galaxy}               & {All objects within a galaxy are at the same distance and the same dilution (fluence-to-emission ratio).}\\
{Negligible extinction}         & {No emission is absorbed or scattered en route to Earth ($\lTransmittanceRGen = 1$).}\\
{Uniform beam}                 & {The instrumental response is uniform across the bandwidth and duration of an observation and across its footprint on the sky ($\bzNWMeasureObsIN = \bzNObs$).}
\enddata
\end{deluxetable}


In this paper, I consider constraints on broadcasts in individual galaxies, specifically their abundance and brightness. All intercepted broadcasts are assumed to arise from the one {galaxy} being targeted. This is important because observable quantities like energy received or photons {counted} cannot tell if the broadcasts are from within the {galaxy} or not.

{There are several different ways of interpreting metasocieties described in Paper I, each reflecting different assumptions about the mutual influence and interstellar replication of societies. In this paper, I adopt the simplest scenario, the single metasociety assumption ($\zNGal = 1$).} We cannot observationally distinguish a {galaxy} with a metasociety with sufficiently few technosignatures from one with no metasociety. Thus, for this paper, uninhabited galaxies have trivial metasocieties with no societies and no broadcasts. The {practical effect is to ignore a} term in the variance of observables related to the ``clumping'' of broadcasts into discrete metasocieties. {The metasociety can be more or less identified with its host galaxy, but I consider astrophysical properties like number of stars to be fixed parameters of the host galaxy. The metasociety $\MetaMark$, on the other hand, contains the parameters describing the ETIs and their broadcasts; these are unknown variables, and the metasociety is considered to be random. SETI programs aimed at galaxies can be viewed as trying to constrain these unknown metasocietal parameters.}

{Societies in the metasociety are described by their position $\aPosition$, and origin time $\aTStart$. Furthermore, it is assumed that societies trace the stellar population, at least in bulk (see discussion in Paper I), with an abundance per star of $\azAbund$. {When the metasocietal properties are fixed,} societies {are described by} a Poisson point process with distribution
\begin{multline}
\label{eqn:SocDist}
\azDist{(\aTuple | \zTuple)} = \azAbund{(\zTuple)} \frac{d{\Mean{\hNTimeGal(\aTStart)}}}{d\hPosition}(\aPosition) \\
\cdot \PDF{\aTStart, \aDuration{, \aParamsOther| \zTuple}} .
\end{multline}
{The $\aParamsOther$ collects societal-level properties of the broadcast distribution, like a shared luminosity.} 

Each society can host broadcasts, which are characterized by their energy release $\bEiso$, quantities relating to polarization of the broadcasts ($\bPolQuantity$), a position ($\bPosition$) identical to that of the transmitting society, and various parameters describing their time/frequency behavior. The distribution of broadcasts for each society is of the form
\begin{multline}
\baDist{(\bTuple | \aTuple)} = \fDirac(\bPosition - \aPosition{(\aTuple)}) \aPDF{\bEiso} \aPDF{\bPolQuantity} \\
\cdot \begin{cases}
		\baRatenuTotal{(\aTuple)} \fDirac(\bDuration - \bDurationBox) \fDirac(\bBandwidthTime - \bBandwidthBox) \fDirac(\bDriftRate) & (\text{box})\\
		\baAbundnuTotal{(\aTuple)} \fDirac(\bTStart) \fDirac(\bDuration - \bDurationInfty) \fDirac(\bBandwidthTime) \aPDF{\bDriftRate} & (\text{chord})
	\end{cases} .
\end{multline}
Each {realized} society's broadcasts also are described by a Poisson point process. I further adopt an interchangeability assumption: the statistical properties of the broadcasts of one society are equivalent to those of another, aside from translation in time and space. {Hence}, every society has the same broadcast energy release distribution $\zPDF{\bEiso}$ {and} the same polarization distribution $\zPDF{\bPolQuantity}$, and all share the same frequency rate or abundance ($\baRatenuTotal = \zMean{\baRatenuTotal}$ in the box model; $\baAbundnuTotal = \zMean{\baAbundnuTotal}$ in the chord model). When I need to pick a $\bEiso$ distribution in this {work}, I assume that all broadcasts have the same effective isotropic energy, and, when relevant, luminosity {(also known as effective isotropic radiated power or EIRP)}:
\begin{equation}
{\aPDF{\bEiso} = \zPDF{\bEiso} = \fDirac(\bEiso - \bEisoBAR) .}
\end{equation}
}

{To simplify matters further, I often employ the} diffuse approximation, {which} ignore{s} the discreteness of societies. This is valid if there are many societies, dividing up a few broadcasts among them ($\Mean{\baNGen} \ll 1$). {In order for it to be false, the broadcasts of an individual society by itself must be confused in a single observation, not just the population in the entire galaxy. Failure requires extraordinary abundances for the fine observations of individualist surveys, but not so much for the coarse observations used in the collective bound.} {Combined with the single metasociety assumption, it means that $\bzSample$ itself is a Poisson point process with distribution
\begin{multline}
\bzDist {(\bTuple | \zTuple)} = \frac{d{\Mean{{\hNTimeGal(\aTStart)}}}}{d\hPosition}(\bPosition) {\zPDF{\bEiso} \zPDF{\bPolQuantity}} \\
\cdot \begin{cases}
		\bzRatenu{(\zTuple)} \fDirac(\bDuration - \bDurationBox) \fDirac(\bBandwidthTime - \bBandwidthBox) \fDirac(\bDriftRate) & (\text{box})\\
    \bzAbundnu{(\zTuple)} \fDirac(\bTStart) \fDirac(\bDuration - \bDurationInfty) \fDirac(\bBandwidthTime) {\zPDF{\bDriftRate}} & (\text{chord})
	\end{cases} .
\end{multline}
}

{Finally, in addition to the other assumptions in Table~\ref{table:StandardAssumptions},} I assume that the {host} {galaxy} {$\GalIndex$} is distant, such that all broadcasts have the same distance ${\yDistanceGal}$ and redshift ${\yRedshiftGal}$. I even apply this to {wide-field} surveys of the Milky Way to get order-of-magnitude constraints, as most stars {in the Galaxy} are of order $\sim 10\ \kpc$ away.

\section{Measurements of aggregate emission}
\label{sec:AggEmissionMeasurements}
{Instruments collect some kind of {emission,} like energy or photons, and report an observable quantity. The most common kind of observable, one at the heart of most SETI analyses, is the integrated amount of collected emission from a target during a {window $\GenLabel$}. This quantity, ${\qMeasureGen}$ for a generic measurable (${\qEnergyGen}$ for collected energy, and ${\qPhotonGen}$ for number of collected photons) has two basic components. First is the aggregate emission from all the broadcasts covered by the observation, denoted ${\mMeasureGen}$ with a superscript $\BcMark$ to indicate its origin in broadcasts. Second is the background ${\kMeasureGen}$, from everything else -- instrumental noise, natural background radiation from Earth, the target system, and things behind it.}

{The mean collected emission of a broadcast is directly proportional to the amount of effective isotropic emission it releases into the universe during the window ${\GenLabel}$, ${\bEmissionisoGen}$. By the time it gets to Earth, propagating an emission distance $\ybDistanceR$, the emission is ``diluted'' by a factor $\lDilutionR \equiv 1/(4 \pi {\ybDistanceR}^2)$ from the inverse square law.} I use the dilution factor in this work because it is more natural when calculating averages, with $\zMean{\lDilutionR} \propto \zMean{{\ybDistanceR}^{-2}}$ occurring frequently. 

{Extinction can also suppress the observed flux. The surviving fraction of emission in $\GenLabel$ after extinction is the transmittance $\lTransmittanceRGen$. Generally, it can be a function of frequency; the transmittance actually depends on the window chosen for wideband broadcasts (e.g., an optical pulse observed with a blue filter will be more subject to dust extinction than one observed with a red filter). Extinction is important for optical light broadcasts more than about a kiloparsec away \citep{Howard04}; it could also {be significant} for more exotic bands at the far ends of the electromagnetic spectrum. Extinction is negligible in radio SETI bands ($\lTransmittanceRGen \approx 1$). It also includes extinction in the Earth's atmosphere, which would set $\lTransmittanceRGen \approx 0$ in much of the infrared spectrum. }

The fluence of broadcast {$\BcMark$} {is the amount of emission received per unit area over the window:
\begin{equation}
{\lFluenceGen} = \frac{{\bEmissionisoGen} {\lTransmittanceRGen}}{4 \pi {\ybDistanceR}^2} = {\bEmissionisoGen} {\lTransmittanceRGen} \lDilutionR .
\end{equation}

The emission is collected by an} instrument with effective area $\iAeff$ and weighted by a {nonconstant} response function of frequency, time, polarization, and location on the sky. I assume that the response function {is $0$ outside the times, frequencies, and polarizations covered by ${\GenLabel}$, and otherwise only {depends on} sky location $\SkyAngVar$:
\begin{multline}
\label{eqn:InstrumentResponse}
{\iResponseMeasureGen} (\TimeVar, \FreqVar, \PolVar, \SkyAngVar) = {\iResponseMeasureGen} (\SkyAngVar) \cdot \IndicatorOf{0 \le \TimeVar - {\oTStartGen} \le {\oDurationGen}} \\
\cdot \IndicatorOf{|\FreqVar - \oNuMidGen| \le {\oBandwidthGen} / 2} \cdot \IndicatorOf{\PolVar \in {\oPolSetGen}}, 
\end{multline}
}where $\max \iResponseMeasureGen = 1$. {Given a fixed broadcast sample, the mean amount of emission measured by the instrument is
\begin{equation}
{\CsxMean{\qMeasureGen}} = \Mean{{\kMeasureGen}} + {\SumBcGen} \Mean{{\lMeasureGen}} = {\SumBcGenADJOIN} \Mean{{\lMeasureGen}} .
\end{equation}
} {On the right,} the noise is regarded as a virtual ``broadcast'' $\BACKIndex${, with a dummy broadcast tuple $\kTuple$}. {Calculations are simplified by regarding the emission} coming from an ``adjoined sample'' with this virtual broadcast, $\bSampleGenADJOIN = \bSampleGen \cup \{\kTuple\}$.

{Of course, we do not know what kinds of broadcasts are present in the sample, so we want to know the statistics of an observable over all possible samples.} Under standard assumptions {(section~\ref{sec:StandardAssumptions})}, the mean intercepted emission {within window $\GenLabel$} is given by
\begin{equation}
\Mean{\qMeasureGen} = \Mean{\kMeasureGen} + \iAeff {\Mean{\bzNWMeasureGenIONE} \zMean{{\lFluenceGen}}} . 
\end{equation}
This also uses a weighted number of broadcasts,
\begin{equation}
{\bzNWMeasureGenIN} \equiv {\zSumBcGen} \iResponseMeasureGen(\lSkyLocation)^n
\end{equation}
with means
\begin{equation}
\Mean{{\bzNWMeasureGenIN}} \equiv {\azAbund \zMean{\baNGen}} \int_{\SkyAngVar \in \oSkyFieldGen} \frac{d{\Mean{\hNGenGal}}}{d\SkyAngVar} [\iResponseMeasureGen(\SkyAngVar)]^n d\SkyAngVar 
\end{equation}
for different exponents $n$ {(Appendix~\ref{sec:AppendixAggregateEmission})}. In turn, ${\azAbund}$ is the mean instantaneous number of societies per star, $\zMean{\baNGen}$ is the mean number of selected broadcasts per society, and $\hNGenGal$ is the number of stars in the window.

According to the law of total variance (Paper I; \citealt{Wasserman04}), the variance in the intercepted emission
\begin{equation}
\label{eqn:TotalVariance}
\Var{\qMeasureGen} = \Var{\CsxMean{\qMeasureGen}} + \Mean{\CsxVar{\qMeasureGen}} 
\end{equation}
splits into a sample variance term and a noise variance term, respectively. Here $\CsxMean{\qMeasureGen}$ and $\CsxVar{\qMeasureGen}$ refer to the mean and {variance, respectively,} for a fixed sample from selection $\Selection{\GenLabel}{{\MetaMark}}$\footnote{{Actually, the selection associated with $\qMeasureGen${, $\bSampleGen$, and $\aSampleGen$} is $\Selection{\GenLabel}{\UnivMark}$ {(see Paper I)}, but all broadcasts are assumed to arise from the single metasociety $\MetaMark$ in this paper.}}, over all possible realizations of the measurement noise. The sample variance for observations of a distant {metasociety} is then
\begin{multline}
\label{eqn:SampleVar}
\Var{\CsxMean{\qMeasureGen}} \approx \iAeff^2 {\Mean{\bzNWMeasureGenITWO} \left(\zMean{\lFluenceGen^2} \right.} \\
                                     {\left. + \zMean{\baNGen} \zMean{\lFluenceGen}^2\right)}
\end{multline}
in single metasociety scenarios.

\section{Individualist constraints: Signal-to-noise ratio and confusion}
\label{sec:SN}

SETI surveys -- and indeed most surveys for astrophysical objects -- take an individualist approach, looking for single events that stand out from the background with high signal-to-noise {ratio}. Individualist constraints use the \emph{extreme} values of an observable $\qMeasureObs$, but they disregard the rest of {its} distribution. They {can rule out} even a single very bright broadcast, but say nothing about even a vast number of very faint broadcasts. In this section, I calculate effective signal-to-noise ratios using the statistics of energy and photon measurements.  This lets me explore the effects of confusion with the framework.

\subsection{The effective number of independent measurements}
Surveys are groups of observations. An archetypal survey consists of $\oNPointSurv$ {nonoverlapping} pointings ($\PointLabel$), each covering a fixed (possibly noncontiguous) field on the sky.  While aimed at one location on the sky, the instrument can make independent observations in $\oNBeamPoint$ resolution elements ({$\BeamILabel$}) on the sky.  The sample of stars covered by these resolution elements is essentially fixed even at different times or frequencies.  The number of stars covered in the survey is simply the sum over all resolution elements and pointings, and
\begin{equation}
\label{eqn:NstarSurvey}
\Mean{\hNSurvGal} \approx \oNPointSurv \oNBeamPoint {\MeanSurv{\hNBeamIGal}} .
\end{equation}
Unless a very bright source spills into multiple resolution elements, measurements made in different {resolution elements} and pointings are independent, taking samples of different societies.

In the box and chord models, a survey has $\oNObsSurv = \oNPointSurv \oNBeamPoint \oDurationBeam \oBandwidthBeam / (\oDurationObs \oBandwidthObs)$ observations, but their broadcast samples are not necessarily independent. If broadcasts are very wideband, the same {ones} will be present across many adjacent frequency channels; very {long lived} broadcasts will be present in many sequential snapshots. These measurements have high sample covariance, missing the sampling fluctuations between them. I estimate the effective number of independent measurements as $\oNObsSurvEFF \equiv \Mean{\bNSurv}/\Mean{\bNObs}$:
\begin{equation}
\label{eqn:NMeasureEff}
\oNObsSurvEFF \approx \begin{cases}
               \displaystyle \oNPointSurv \oNBeamPoint \frac{(\oDurationBeam + \bDurationBox) (\oBandwidthBeam + \bBandwidthBox)}{(\oDurationObs + \bDurationBox) (\oBandwidthObs + \bBandwidthBox)} & \text{(Box)} \\
							 \displaystyle \oNPointSurv \oNBeamPoint \frac{\oBandwidthBeam + \oDurationBeam {\zMean{|\bDriftRate|}}}{\oBandwidthObs + \oDurationObs {\zMean{|\bDriftRate|}}} & \text{(Chord)}.
							 \end{cases}
\end{equation}

\subsection{Signal-to-noise ratio definitions}
\label{sec:SNDefs}
Most SETI surveys {identify candidate ``hits'' with} a differential measurement, looking for abnormally large fluctuations in measurements that are not expected statistically from noise. {The signal-to-noise ratio is thus a key quantity in estimating the sensitivity reach of a SETI program. One can estimate the baseline value of an observable and the usual size of the fluctuations empirically from a collection of related observations. The group of observations analyzed to yield these estimators is not necessarily the survey as a whole, or even all the observations of a particular pointing. {Instead, the idea is generally that the observations within a group are comparable, with statistically interchangeable distributions of background (and presumably broadcasts).} {Hence, we generally want to compare only observations covering the same field at the same time; the noise can change at different times and at different points in the sky.} Breakthrough Listen's \texttt{turboSETI} estimates these quantities in narrowband line searches by comparison with other fine channels in a coarse channel, for instance \citep{Enriquez17}. The estimation in other programs{, particularly those searching for other types of signal, could use different groupings of observations; $\GroupFORObsLabel$} stands in for the actual group of observations used.}

{The signal-to-noise {test statistic} for each measurement is}
\begin{equation}
\label{eqn:SNMeasured}
{\qSNMeasureObs} = {\frac{\text{deviation from mean}}{\text{typical fluctuation}}} = {\frac{\qMeasureObs - {\dMeanMeasureObsFROMGroupFORObs}}{\displaystyle {\dSDMeasureObsFROMGroupFORObs}}},
\end{equation}
where ${\dMeanMeasureObsFROMGroupFORObs}$ and ${\dSDMeasureObsFROMGroupFORObs}$ are estimators for the mean and standard deviation of $\qMeasureObs$ respectively. Although the sample mean and sample standard deviation of $\qMeasureObs$ {within {$\GroupFORObsLabel$}} might be used to evaluate equation~\ref{eqn:SNMeasured}, this is not necessary: in \citet{Enriquez17}, for example, the trimmed sample mean and standard deviation are used to {discard} spectral features induced by the processing. {The most extreme values of the sample are excluded (those outside the {5th--95th} percentile range in \citealt{Enriquez17}) before the sample mean and standard deviation are calculated \citep[see][]{Stigler73,Castillo05}. This makes them more robust to the outliers that are inevitable artifacts of data reduction.}

A candidate detection is found for observation $\ObsLabel$ if $\qSNObs > \qSNThreshSurv$, where the signal-to-noise threshold $\qSNThreshSurv$ is calculated to have a negligible {false-alarm} rate over the entire survey $\SurvLabel$. A null result happens when
\begin{equation}
\label{eqn:IndividualistMethod}
{\MaxSurv{\qSNMeasureObs} = \MaxSurv{\frac{\qMeasureObs - {\dMeanMeasureObsFROMGroupFORObs}}{\displaystyle {\dSDMeasureObsFROMGroupFORObs}}} \le \qSNThreshSurv} .
\end{equation}

{We calculate the expected sensitivity of a survey by estimating the maximum $\qSNMeasureObs$ expected from a population and comparing it to $\qSNThreshSurv$. If the latter is greater, then {no detections are expected}. Now, the actual value of $\qSNMeasureObs$ is a nonlinear combination of several factors that is not easily tractable.} {If different observations have wildly different broadcast populations or noise properties, then the estimated variance itself can vary a lot; the signal-to-noise will be greater in those groups of observations where {it is small}. A full accounting of these effects may best be found through numerical simulation.}

{Our goal here is to understand the basic behavior of signal-to-noise {ratio} and survey sensitivity, how it rises and falls as the number of broadcasts increases. For that reason, I assume that all observations of a metasociety being considered are interchangeable: the $\qMeasureObsI$ for each observation $\ObsILabel$ all have the same statistical properties, with a constant background noise level and similar broadcast populations sampled by each.} {I also posit that} the {background} noise fluctuations in $\qMeasureObs$ are negligible compared to the contribution of a detectable broadcast. In other words, the large fluctuations that result in a detection happen because the observation covers more broadcasts or brighter broadcasts than typical. {The maximum signal-to-noise {ratio} is then {estimated:}}
\begin{equation}
\label{eqn:SNTheoryEst}
\qSNMeasureObsEST {= \frac{\Median{\MaxSurv{\mzMeasureObs}} - \Mean{{\mzMeasureObsREGzSurv}}}{{\SD{\qMeasureObsREGzSurv}}}},
\end{equation}
{with $\ObsLabel$ understood to stand in for any representative observation of the galaxy, because the observations are assumed to be interchangeable.\footnote{\vphantom{T}{The inverted hat is used here to mean a theoretical estimate, in contrast to an estimator derived from actual data.}} This estimate should be adequate as long as the variance between observations is not too great.}

This lends itself to two simple approximations for different {$\lFluenceObs$} distributions, applying if the expected {numbers} of covered broadcasts and societies, $\Mean{{\bzNSurv}}$ and $\Mean{{\azNSurv}}$, are $\ga 1$. If the {fluence} distribution is {narrow,} then fluctuations in the brightness are due mainly to variability in the number of broadcasts intercepted by a broadcast:
\begin{multline}
\label{eqn:SNFluctuationsNarrow}
{\Median{\MaxSurv{\qMeasureObs}} - \Mean{\qMeasureObsREGSurv}} \\
\approx {\left(\Median{\MaxSurv{\bzNWMeasureObsIONE}} - \Mean{\bzNWMeasureObsIONE}\right)} \Mean{\lMeasureObsREGzSurv}.
\end{multline}
{So the biggest signal is the greatest expected excess in the number of broadcasts times the expected emission from a single broadcast.} Appendix~\ref{sec:MaxPoisson} presents approximations for {the median excess number of broadcasts}, but when $\Mean{{\bzNObs}} \ga 1$, it has {a} $\sim \Mean{{\bzNObs}}^{{1/2}}$ dependence up to a logarithmic factor. {However, the variance grows at least as quickly, forcing the signal-to-noise {ratio} to remain below the threshold for detection.}

In a broad {fluence} distribution, however, the fluctuations are determined by a single broadcast that dominates all the others, with ${\MaxSurv{\mzMeasureObs} \approx \zMaxSurvNULL{\lMeasureObs} + \Mean{\mzMeasureObs}}$:
\begin{equation}
\label{eqn:SNFluctuationsBroad}
{\Median{\MaxSurv{\qMeasureObs}} - \Mean{{\qMeasureObsREGzSurv}} \approx \Median{\zMaxSurvNULL{\lMeasureObs}}} .
\end{equation}
{Power-law} distributions (${\zPDF{\lMeasureObs}} \propto \lMeasureObs^{-{\bPowerLawCore}}$) behave like broad distributions when ${\bPowerLawCore} < 3$, as long as $\Median{{\zMaxSurvNULL{\lMeasureObs}}}$ is much smaller than the maximum possible $\lMeasureObs$. Note, however, that any distribution with ${\bPowerLawCore} \le 1$ must have a maximum cutoff {that} is saturated quickly. 

{I define a single variable, an effective number of broadcasts, to encapsulate both approximations:}
\begin{multline}
\label{eqn:SNNeff}
{\bzNObsEFF \equiv \max\left[\left(\Median{\MaxSurv{\bzNWMeasureObsIONE}} - \Mean{\bzNWMeasureObsIONE}\right), \right.}\\
{\left. \frac{\Median{\zMaxSurvNULL{\lMeasureObs}}}{\Mean{\lMeasureObsREGzSurv}}\right]},
\end{multline}
with
\begin{equation}
{\qSNMeasureObsEST \approx \frac{\bzNObsEFF \Mean{\lMeasureObsREGzSurv}}{{\SD{\qMeasureObsREGSurv}}}} .
\end{equation}

Ideally, we would like a very large number of \emph{independent} measurements to both accurately estimate the mean and observe rare maxima in $\qMeasureObs$. As noted, wideband and long-duration broadcasts reduce the number of effective independent measurements to $\oNObsSurvEFF$ (equation~\ref{eqn:NMeasureEff}).

\subsection{The sparse limit}
The common assumption in SETI is that broadcasts are very rare, with $\Mean{{\bzNObs}} \ll 1$. In the sparse limit, if any broadcasts are intercepted at all, {{all} observation{s have} at most one ($\MaxSurv{\bzNObs} = 1$)} and {the maximum signal-to-noise occurs for the observation containing the brightest broadcast ($\MaxSurv{\mzMeasureObs} = \zMaxSurvNULL{\lMeasureObs}$)}.  Then, a null result implies either
\begin{equation}
\label{eqn:IndividualistClassic}
{\frac{\zMaxSurvNULL{\lMeasureObs}}{{\SD{\qMeasureObs}}} \la \qSNThreshSurv} ~\mathrm{or}~{\bzNSurv} = 0 :
\end{equation}
{they are either too faint or too rare to observe.} The two conditions in fact correspond to the luminosity and rate limits that are so commonly quoted in SETI (e.g., \citealt{Enriquez17} and references therein, as shown in that work's Figure 7).

Usually, the broadcasts are assumed to be sparse enough that the background noise is expected to dominate the variance, with $\Var{\kMeasureObs}$ used to estimate total variance. {Sufficiently powerful broadcasts can dominate the variance, however, long before $\Mean{\bzNObs}$ increases past $1$, just because they are contributing so much energy.} Trimmed {means and standard deviations} help address this problem by discarding a certain percentage of the observations with the highest values of $\qMeasureObs$, but once $\Mean{{\bzNObs}}$ rises past that fraction, the effect of broadcasts on the noise must be taken into account.

\subsection{The confusion limit}
As $\Mean{{\bzNObs}}$ increases past $1$, however, $\mzMeasureObs$ {nearly} converges to a stable distribution by the central limit theorem {under certain general conditions \citep{Embrechts13}.\footnote{\vphantom{T}{These conditions do not precisely hold because $\bzNObs$ is itself a random variable; the limiting distribution is a mixture of stable distributions. Nonetheless, in the diffuse approximation for Poissonian broadcasts, $\bzNObs$ itself converges to a narrow Gaussian, so the effects of the spread are weak.}} If $\zMean{\lMeasureObs}$ exists, then once there are enough broadcasts to sample the emission distribution well in each observation ($\zMean{\lMeasureObs} \approx \zMean{\lMeasureObsREGzSurv}$), $\mzMeasureObs$ should approach $\Mean{\bzNObs} \zMean{\lMeasureObs}$ by the law of large numbers. This is a quantity that continues growing with $\Mean{\bzNObs}$. T}he mean {emission from} all the broadcasts intercepted by an observation {eventually becomes much} greater than the brightest single broadcast. In this confusion limit, the broadcasts effectively blend together into another noise background and the individualist approach of equation~\ref{eqn:IndividualistMethod} fails. This leads to a new problem in SETI: we might fail to detect ETIs not because they are too rare but because they are too common!

{To be clear, confusion only sets in when there are too many broadcasts per observation -- generally a separate epoch, channel, and beam in radio. The number of broadcasts covered in a survey, or even a single pointing, can be far greater, while still leaving ``empty'' observations that contrast with the {occupied} observations. The more fine-grained the survey, the less of an issue confusion should be. Modern surveys may cover billions of observations and could detect many millions of broadcasts before confusion sets in.}

The convergence happens even in the absence of noise and is an effect of the underlying distribution of samples. Confusion is implicit in the sample variance $\Var{\CsoMean{\qMeasureObs}}$. Because the variance is necessarily larger than the sample variance alone, equation~\ref{eqn:SampleVar} gives a hard upper bound on the estimated signal-to-noise ratio:
\begin{equation}
\label{eqn:SNSampleVar}
\qSNMeasureObsEST \la \bzNObsEFF \left[\Mean{{\bzNWMeasureObsITWO}} \frac{\zMean{{\lFluenceObsREGzSurv}^2}}{\zMean{\lFluenceObsREGzSurv}^2}\right]^{-1/2}
\end{equation}
for distant {galaxies} with one metasociety {under the diffuse approximation}{.} {Confusion resulting from the sample variance alone is dubbed here ``sample confusion'', to contrast with the (rarer) {``noise} confusion{''} from noise variance.} {Sample c}{onfusion can be said to set in when the right-hand side falls below the detection threshold $\qSNThreshSurv$.}

Assuming that the fluence distribution is narrow and the diffuse approximation {applies}, this converges to
\begin{equation}
\label{eqn:SNSampleVarAlt}
\qSNMeasureObsEST \la \left(\Median{{\MaxSurv{{\bzNWMeasureObsIONE}}}} - {\Mean{{\bzNWMeasureObsIONE}}}\right) / \sqrt{{\Mean{\bzNWMeasureObsITWO}}} {.}
\end{equation}
{The sample variance is driven by a kind of shot noise -- not in photons or electrons, but in the broadcasts themselves, akin to the ``graininess'' of good optical images of galaxies arising from variations in the number of discrete bright stars \citep{Tonry88}. But the numerator falls and rises as $\Median{\MaxSurv{\bzNObs}}/\Mean{\bzNObs}$; this ratio is large only if most observations have no broadcasts so we can compare unoccupied and occupied observations.} Thus, {the estimated maximum signal-to-noise {ratio} decreases {at least} as ${\Mean{{\bzNObs}}^{1/2}}$, until $\Mean{{\bzNObs}} \ga {1/\qSNThreshSurv^2}$ by which point it is too small} to pass the stringent cuts used to eliminate false positives.

Broad $\lFluenceObs$ distributions may be essentially immune to confusion even when $\Mean{\bzNObs} \gg 1$, because increasing $\Mean{\bzNObs}$ also results in much brighter broadcasts being intercepted. The maximum fluence broadcast rises above the background of confused broadcasts present in all observations for a broad enough distribution. {Often, however, there is a maximum fluence, limited by the maximum available power, the nearest star, or zero drift rate. {These brightest broadcasts must be submerged in the aggregate background in order for confusion to set in.} Define a window $\VsigLabel$ that samples from $\ObsLabel$ but only passes broadcasts from this end of the distribution:
\begin{equation}
\label{eqn:SNSampleVarBrightest}
\bjjSampleVsig = \{\bTuple | \bTuple \in \bjjSampleObs~{and}~\lFluenceObs \approx \max \lFluenceObs\}.
\end{equation}
This effectively \emph{induces} a narrow fluence distribution in the remaining sample; if there's already a narrow fluence distribution, the $\VsigLabel$ window is equivalent to the observation window. Now, if the bright subset of broadcasts picked by $\VsigLabel$ are confused on their own, then the entire sample is confused -- there are no even brighter broadcasts to stick out of the blended emission, and the faint broadcasts only add to the {``noise.''} Thus, a very conservative constraint for confusion is that it sets in when}
\begin{equation}
\label{eqn:ConfusionLimit}
{\Mean{\bzNVsig}} \ga 1/{\qSNThreshSurv}^2 {,}
\end{equation}
{an approximate result for equation~\ref{eqn:SNSampleVarAlt} applied to a population of only these brightest broadcasts, with the signal-to-noise required to be above $\qSNThreshSurv$.}

\subsection{\texorpdfstring{{The rise and fall of signal-to-noise ratio}}{The rise and fall of signal-to-noise ratio}}
\label{sec:SNBehavior}
Consider the behavior of $\qSNEnergyObsEST$  when $\lFluenceObs$ has a narrow distribution. As ${\Mean{\bzNObs}}$ increases, we {can discern} six regimes:
\begin{itemize}
\item In a null regime, ${\MaxSurv{\bzNObs}} = 0$. Derived signal-to-noise ratios have values of order unity, resulting entirely from fluctuations in the background{.}
\item {{When considering possible hits, m}ost SETI analyses work in a strongly sparse regime, where} the broadcasts are rare enough {to} have no effect on {the estimated background}. Furthermore, ${\MaxSurv{\bzNObs}} = 1$ because there are too few observations for multiple broadcasts to ever ``touch'' the same observation {window}. Thus, $\qSNEnergyObsEST$ remains constant at a value that can be $\gg 1$. 
\item In a moderately sparse regime, the background noise continues to dominate the variance. Although the typical observation still has no broadcasts, some have at least one, and ${\MaxSurv{\bzNObs}} > 1$. $\qSNEnergyObsEST$ is rising in this regime because ${\MaxSurv{\bzNObs}}$ is growing while ${\Mean{\bzNObs}}$ remains {below} one. The onset of this regime occurs at ${\Mean{\bzNObs}} \sim {(\oNObsSurvEFF)^{-1/2}}$.
\item In a {transition} regime, the broadcasts themselves dominate the variance of $\qEnergyObs$. The slow rise of ${\MaxSurv{\bzNObs}}$ is overcome by the rising variance. $\qSNEnergyObsEST$ falls as ${\Mean{\bzNObs}^{-1/2}}$ {to $\Mean{\bzNObs}^{-1}$} in this regime, although it may be still high enough to claim a detection. {Using the trimmed means and standard deviations} effectively delay{s} the onset of the {transition} regime so that it has no dependence on $\oNObsSurvEFF$.
\item In {the} confusion regime, the typical observation has at least one broadcast. $\qSNEnergyObsEST$ continues to decline due to the sample variance, as according to equation~\ref{eqn:SNSampleVar}. The rate of the decline slows down because ${\MaxSurv{\bzNObs} - \Mean{\bzNObs}}$ starts growing as ${{\MeanSurv{\bzNObs}}}^{1/2}$. At this point $\qSNEnergyObsEST$ is a factor of order unity, too small to make any individualist detections.
\item {Finally, radio broadcasts experience a {mutual} interference regime, where} the broadcasts are so numerous that the noise variance from their wave noise dominates over the sample variance. $\qSNEnergyObsEST$ becomes indistinguishable from the null case. Equation~\ref{eqn:SNTheoryEst} no longer describes the numerator of $\qSNEnergyObs$ -- the wave noise fluct{u}ations from the interference between the broadcasts, not the Poisson fluctuations, {dominate} variations in $\qEnergyObs$.
\end{itemize}
{In short, the signal-to-noise ratio at first rises as broadcasts start being sampled and then falls as they overlap with each other.} The borders between these regimes depend on {the number of independent observations, the construction of observations}, and the luminosities of the broadcasts. Not all of these regimes occur for all parameters {or all types of broadcasts}. In broad distributions, $\qSNEnergyObsEST$ is governed by the broadcasts with the greatest fluence. The strong sparse regime can split into two stages: signal-to-noise {ratio} increases at first, as signals with increasing energy per observation are intercepted, followed by a regime of flat signal-to-noise {ratio}, where all observations have 0 or 1 broadcast with near-maximal fluence.

\section{\texorpdfstring{{Individualist constraints: radio broadcasts}}{Individualist limits constraints: radio broadcasts}}
\label{sec:SNRadio}

\subsection{Wave noise, modes, and amplitudes}
\label{sec:Amplitudes}
{Natural electromagnetic radiation} can be regarded as the sum of many microscopic emitters {with} different locations, frequencies, and phases. The mutual interference between all these sources results in the amplitude of the detected radiation fluctuating chaotically instead of maintaining a constant {magnitude}. In quantum terms, chaotic light displays photon bunching \citep[e.g.,][]{Foellmi09,Tan14,Zmuidzinas15}. These wave noise fluctuations greatly dominate over photon shot noise at radio wavelengths because the {photon occupation number in each field mode (cell in phase space)} is much greater than one \citep{Radhakrishnan99}. Thermal noise in the receiver usually overwhelms over the wave noise from sources, with a greater noise temperature, but it too is the chaotic sum of many microscopic fluctuations. Artificial broadcasts may be coherent (\citealt{Hippke21}; see also Appendix B of \citealt{Cordes97}), but wave noise is still present in the background and in the mutual interference of many broadcasts.

Although we generally measure energy in radio SETI, {the} statistics of the {intercepted} energy follow from the underlying wave amplitudes of the electric field, commonly measured as voltages \citep{Wilson09}. {Each amplitude is measured for an} individual mode $\ModeLabel$ of the electromagnetic field. {{A mode is an independent oscillator of the electromagnetic field.} A photon can be localized to a single mode, but no further.} {In terms of temporal properties, a mode includes only one polarization and ha{s} a bandwidth-duration product} $\oDurationMode \oBandwidthMode = 1$ \citep{Yamamoto86,Nityananda94,Caves94,Hippke21}. Amplitudes add linearly: the amplitude measured in mode $\ModeLabel$ is 
\begin{equation}
\qAmplitudeMode = \kAmplitudeMode + \SumBcMode {\lAmplitudeMode} .
\end{equation}
The background amplitude $\kAmplitudeMode$ is an independent random variable with a zero-mean complex Gaussian distribution \citep[e.g.,][]{Wilson09}. The same is true for the amplitudes of broadcasts that are incoherent, although {each has its own amplitude} variance.  

Square-law detectors derive the energy in the mode by taking the square of its complex modulus, $\qEnergyMode = {\qAmplitudeMode} {{\qAmplitudeMode}}^{\ast}$.\footnote{This is the power if voltage is being directly squared{, up to a constant factor}; I adopt a convention that $|\qAmplitudeMode| = \sqrt{\qEnergyMode}$.} As a thermal noise, the background energy $\kEnergyMode = |{\kAmplitudeMode}|^2$ per mode has a mean value $\ikTBack$, which is the background temperature (includ{ing} system noise and natural background radio flux in the field) multiplied by Boltzmann's constant {(\citealt{Radhakrishnan99}).}\footnote{\vphantom{T}{This also follows from the Rayleigh-Jeans law, with one mode covering a solid angle equal to the wavelength squared divided by collecting area \citep{Zmuidzinas03}.}} The energy intercepted from a broadcast, ${\lEnergyMode}$, {is the product of fluence, collecting area, and instrumental response}.\footnote{Note that $\iResponseEnergyMode$ is the \emph{energy} (power) beam pattern of the instrument, not the amplitude (voltage) response $\iResponseAmplitudeMode = \sqrt{\iResponseEnergyMode}$.} Thus,
\begin{equation}
\label{eqn:RadioModeCondMeanEnergy}
\CsuMean{\qEnergyMode} = \ikTBack + \SumBcMode \iAeff \iResponseEnergyMode ({\lSkyLocation}) {\lFluenceEMode} ,
\end{equation}
and, for distant {metasocieties},
\begin{equation}
\label{eqn:RadioModeMeanEnergy}
\Mean{\qEnergyMode} \approx \ikTBack + \Mean{\bzNWEnergyModeIONE} \iAeff {\zMean{\lFluenceEMode}} .
\end{equation}
The variance depends on the number and type of broadcasts.

\subsection{Energy measured in an observation}
Where complications arise is that usually {the analysis does not work directly} with these mode energies either. The energy in $\oNModeObs$ modes per observation -- covering different times, frequencies, or polarizations -- is summed together to yield an observed energy $\qEnergyObs$:
\begin{equation}
\qEnergyObs = \SumModeInObs \qEnergyMode ,
\end{equation}
where $\oModesOfObs$ is the set of modes that are summed for the observation $\ObsLabel$. {It can be assumed that the modes are interchangeable, with statistically equivalent populations of broadcasts, because observations are so fine-grained.} The mean energy in the observation follows simply enough from equation~\ref{eqn:RadioModeMeanEnergy} and the linearity of expectation:
\begin{equation}
\label{eqn:MeanEoRadio}
\Mean{\qEnergyObs} \approx \oNModeObs \left[\ikTBack + \Mean{{\bzNWEnergyModeIONE}} \iAeff {\zMean{\lFluenceEMode}}\right] ,
\end{equation}
assuming {that} the target {galaxy} is distant.

The sample variance for a distant {galaxy} with a single metasociety is
\begin{multline}
\label{eqn:RadioSampleVariance}
\Var{\CsoMean{\qEnergyObs}} \approx \iAeff^2 \Mean{{\bzNWEnergyObsITWO}} \left[ {\zMean{{\lFluenceEObs}^2}} \right.\\
                                    \left. { + \zMean{\baNObs} \zMean{\lFluenceEObs}^2} \right] .
\end{multline}
$\zMean{\baNObs}$ is the mean number of broadcasts per society, representing a clumping effect.

In Appendix~\ref{sec:RadioNoiseVar}, I show that the noise variance for distant {metasocieties} is 
\begin{multline}
\Mean{\CsoVar{\qEnergyObs}} \approx \SumBcObs \DoubleSumModeInObs {\Cov{\lEnergyModeONE, \lEnergyModeTWO}}\\
+ \oNModeObs \left[\left(\ikTBack + \iAeff \Mean{{\bzNWEnergyModeIONE}} {\zMean{\lFluenceEMode}}\right)^2 \right.\\
\left. + \iAeff^2 \Mean{{\bzNWEnergyModeITWO}} {\zMean{\lFluenceEMode^2}}\right] .
\end{multline}
by working with the amplitudes. The {covariance $\Cov{\lEnergyModeONE, \lEnergyModeTWO}$ for broadcast $\BcMark$ equals $\Mean{\lEnergyModeONE \lEnergyModeTWO} - \Mean{\lEnergyModeONE} \Mean{\lEnergyModeTWO}$} and is related to the coherence properties of the broadcast.

{The noise variance includes the mutual {interference} of the background and numerous broadcasts, resulting in a quasi-thermal background even if the broadcasts are coherent. This interference puts an additional upper limit to the signal-of-noise {ratio} of
\begin{equation}
\label{eqn:NoiseConfusion}
\qSNEnergyObsEST \la \frac{\bzNObsEFF}{\sqrt{\oNModeObs} \Mean{\bzNWEnergyModeIONE}} .
\end{equation}
This can lead to a ``noise confusion'', which in some cases can set in before the previously derived sample confusion (equation~\ref{eqn:SNSampleVarAlt}).}

\subsection{Incoherent radio broadcasts}
\label{sec:IncoherentRadioNoise}
The incoherent case applies to all known natural phenomena, including astrophysical masers where radio amplitude statistics have been measured \citep{Evans72}. It may also apply to artificial broadcasts if they are sufficiently broadband, as expected if they are rich in information \citep[see][]{Caves94,Messerschmitt15}. We can model a broadcast as a burst of white noise with a (possibly frequency drifting) bandpass filter applied to it, as in the box model; if the instantaneous bandwidth {$\bBandwidthTime$} is wider than $\oDurationMode^{-1}$, the broadcast is incoherent. Furthermore, when many coherent broadcasts are {confused}, the {aggregate emission} also behaves like an incoherent source.

The noise variance calculation is presented in Appendix~\ref{sec:RadioNoiseVar}, but essentially it may be calculated from the fact that the noise is independent between modes. We find for distant {galaxies} under the diffuse approximation 
\begin{multline}
\label{eqn:VarERadioIncoherent}
\Var{\qEnergyObs} \approx \oNModeObs \left(\ikTBack + \iAeff \Mean{{\bzNWEnergyModeIONE}} {\zMean{\lFluenceEMode}}\right)^2 \\
+ \iAeff^2 \left[\oNModeObs \Mean{{\bzNWEnergyModeITWO}} {\zMean{\lFluenceEMode^2}} + \Mean{{\bzNWEnergyObsITWO}} {\zMean{{\lFluenceEObs}^2}}\right]
\end{multline}
(see Appendix~\ref{sec:IncoherentRadioVar} for the full expression).

\subsubsection{\texorpdfstring{{Confusion and} searching for continuum sources in the box model}{Confusion and searching for continuum sources in the box model}}
\begin{figure}
\centerline{\includegraphics[width=8.5cm]{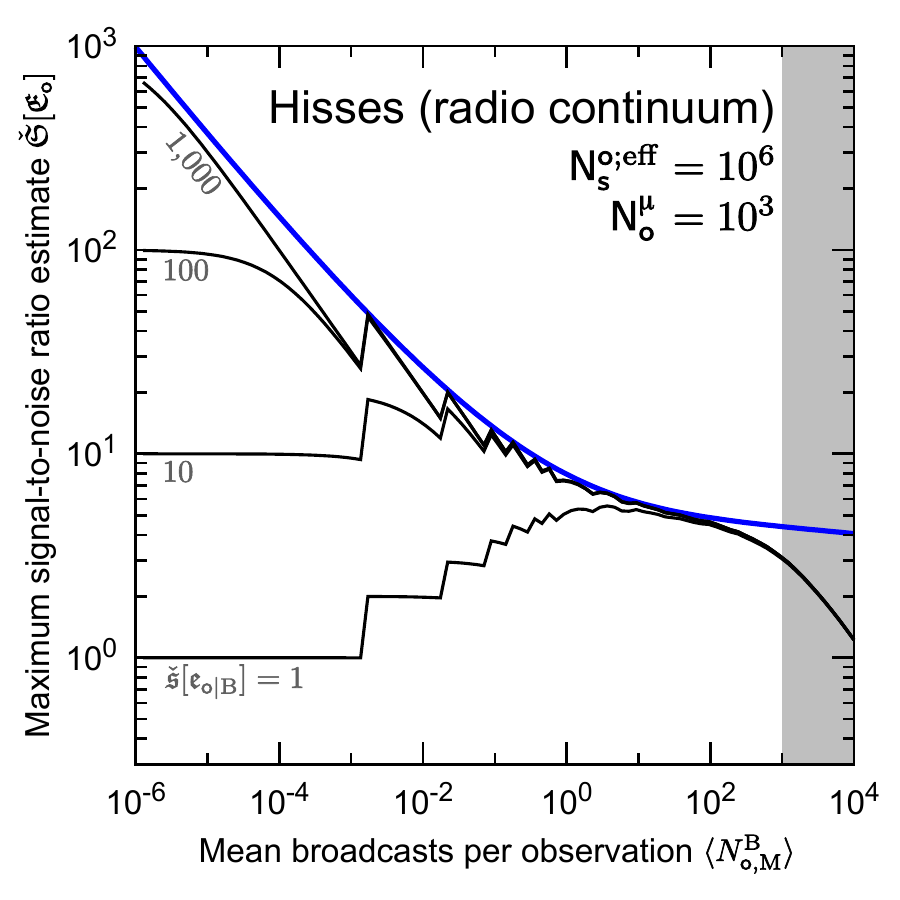}}
\figcaption{Growth and fall of {total signal-to-noise {ratio} $\qSNEnergyObsEST$} as {the number of broadcasts per observation} increases and confusion sets in{, as depicted for} {for radio continuum sources, in the hiss regime of the box model}. I {adopt the standard assumptions of Table~\ref{table:StandardAssumptions}. All broadcasts have the same luminosity. Each individual source, in the absence of other sources, would have a signal-to-noise ratio of $\lSNEnergyObsEST$ ranging from $1$ to {$1000$}. The b}lue line indicate{s} the bound from sample variance. The {{mutual} interference} regime, where the approximation for $\qSNEnergyObsEST$ breaks down, is shaded in grey. {The survey contains $10^6$ independent observations, where each observation is summed from $10^3$ electromagnetic modes.}\label{fig:SNHiss}} 
\end{figure}

In the box model, {hisses are {flat-spectrum} continuum sources} ($\bDurationBox \gg \oDurationObs$, $\bBandwidthBox \gg \oBandwidthObs$). When observing a distant {galaxy} in both polarizations,
\begin{multline}
\qSNEnergyObsEST \approx {\bzNObsEFF} \lSNEnergyObsEST \left[\left(1 + \frac{\lSNEnergyObsEST \Mean{{\bzNWEnergyObsIONE}}}{\sqrt{\oNModeObs}}\right)^2 \right.\\
\left. + \lSNEnergyObsEST^2 \Mean{{\bzNWEnergyObsITWO}} \frac{{\zMean{(\bLnuisoREGzSurv)^2}}}{{\zMean{\bLnuisoREGzSurv}^2}} \left(\frac{1}{\oNModeObs} + 1\right)\right]^{-1/2}
\end{multline}
where I have applied the diffuse approximation, and 
\begin{equation}
\label{eqn:lSNDef}
\lSNEnergyObsEST \equiv \frac{\iAeff {\zMean{\lFluenceEObsREGzSurv}}}{\sqrt{\oNModeObs} \ikTBack} \approx \frac{\iAeff {\zMean{\bLnuisoREGzSurv}} {\zMean{\lDilutionE {\lTransmittanceE}}} \sqrt{\oNModeObs}}{{2} \ikTBack}
\end{equation}
is the expected signal-to-noise {ratio} for a single broadcast at $\iResponseEnergyObs = 1$ in the presence of background noise only. {If individual broadcasts in isolation are detectable ($\lSNEnergyObsEST \ga 1$)}, the {threshold of the transition region (section~\ref{sec:SNBehavior}) is passed when} $\Mean{{\bzNObs}} \ga 1/\lSNEnergyObsEST^2$. {The confusion regime occurs around when $\Mean{\bzNObs} \ga 1/\qSNThreshSurv^2$, which translates to
\begin{equation}
\label{eqn:ConfusionThreshHiss}
\bzAbund \ga \left[\qSNThreshSurv^2 \Mean{\hNObsGal}\right]^{-1}
\end{equation}
for hisses. Broadcast  mutual interference becomes the dominant noise source for $\Mean{{\bzNObs}} \ga \oNModeObs$.} Figure~\ref{fig:SNHiss} shows how ${\qSNEnergyObsEST}$ varies with $\Mean{{\bzNObs}}$ under standard assumptions -- in particular, how the {transition} regime is absent for $\lSNEnergyObsEST = 1$, while the moderate sparse regime vanishes for high $\lSNEnergyObsEST$.
	
\subsection{Fully and partially coherent radio broadcasts}
\label{sec:CoherentRadioNoise}

Unlike known natural sources, artificial radio broadcasts can be coherent. Although a perfect coherent {carrier} contains negligible information, it conveys one important fact very well -- the existence of technology at that location -- perhaps ``advertising'' dimmer information-rich broadcasts in the vicinity.

Perfectly coherent broadcasts are characterized as perfect chirps, with a constant {luminosity and a well-defined phase $\lPhase$ at any time}: ${\lAmplitude}(\TimeVar{, \PolVar}) = |{\lAmplitude (\PolVar)}| \exp(i {\lPhase}(\TimeVar{, \PolVar}))$. The measured amplitudes can vary from mode to mode as the broadcast drifts across channels, but there is {zero} noise variance{, ignoring the miniscule Poissonian photon shot noise} -- ${\lEnergy}$ is exactly proportional to ${\lFluenceE}$ -- and thus all covariance terms are zero.

\label{sec:PartiallyCoherentRadioNoise}

The partially coherent case is an important one for actual ultranarrowband broadcasts. Any modulation of a perfect chirp will broaden the instantaneous bandwidth. This leads to fluctuations that are correlated on long {timescales}. In fact, even if the broadcast itself is perfectly coherent, scintillation in the interstellar medium introduces observed variability (\citealt{Cordes97}{; \citealt{Brzycki23}}).  In the partially coherent case, the coherence timescale ${\bTCoher} \approx {\bBandwidthTime^{-1}}$ is greater than $\oDurationMode$. Modes of the same polarization and separated by less than ${\bTCoher}$ in time then have ${\Cov{\lEnergyModeONE, \lEnergyModeTWO} \sim \Mean{\lEnergyModeONE} {\Mean{\lEnergyModeTWO}}}$. The fall-off of the covariance can be described by the well-known ${\bGTwoPol}(\fDelta \TimeVar)$ function (e.g., \citealt{Foellmi09}{; see Appendix~\ref{sec:AppendixChord}}). 

Both the fully coherent and partially coherent cases can be covered in a single formula with the use of a factor describing the self-interference:
\begin{multline}
\label{eqn:VarERadioPartCoherent}
\Var{\qEnergyObs} \approx \oNModeObs \left(\ikTBack + \iAeff \Mean{{\bzNWEnergyModeIONE}} {\zMean{\lFluenceEMode}}\right)^2 \\
+ \iAeff^2 \Mean{{\bzNWEnergyObsITWO}} {\zMean{{\lFluenceEObs}^2}} (1 + {\zMean{\bFCoherObs}})
\end{multline} 
in the diffuse approximation for {a} distant {galaxy}. The coherence term ${\bFCoherObs}$ ranges from $0$ for fully coherent broadcasts to $1$ for {polarized,} partially coherent broadcasts with slow (${\bTCoher} \gg \oDurationObs$), {high-amplitude} fluctuations (see Appendix~\ref{sec:RadioNoiseVar}).

Although the variance is missing a term related to self-interference, {it} is greatly increased by strong, {long-timescale} fluctuations, with ${\zMean{\bFCoherObs}} \approx 1$ essentially doubling the sample variance. It may seem odd that a perfectly coherent broadcast with ${\bBandwidthTime} = 0$ has less noise variance than expected from Gaussian statistics, while a partially coherent broadcast with ${\bTCoher \gg \oDurationObs}$ can have more, despite being indistinguishable -- especially since this is the regime of actual coherent broadcasts from space modulated by strong scattering. The discrepancy basically amounts to whether we treat the fluctuations as noise variance or sample variance.

Suppose we took a snapshot of a large population of partially coherent broadcasts, all with infinite {life spans} {and fully polarized}, all with the same ${\bTCoher} \gg \oDurationObs$, and all with the same \emph{time-averaged} luminosity. Because they are only partially coherent, the \emph{measured} brightness of the different broadcasts will vary -- in fact, we might generally expect them to have an exponential distribution just as if they are incoherent \citep[compare with][]{Cordes97}.\footnote{{The wave noise of an unpolarized broadcast does not have an exponential distribution -- the power received in \emph{each} polarization has an exponential distribution{, but the polarizations can vary independently of each other.} As a result, the total power in both polarizations has a $\fchi^2_4$ distribution. An exponential distribution still applies if the modulation in both polarizations is identical, as might apply for interstellar scintillation.}} The partial coherence approach treats the variations as the result of microscopic fluctuations on an underlying constant luminosity. {Thus,} the variations are included in the noise variance. 

But since the fluctuations are much too slow to observe, from an empirical point of view, we could conclude that the broadcasts are perfectly coherent but that \emph{the luminosities themselves differ}. That is, the exponential distribution in fluence reflects an exponential distribution in luminosity, and thus we regard the fluctuations as sample variance, resulting from a ${\zMean{\bLiso^2}} = 2 {\zMean{\bLiso}}^2$ term. Either approach is consistent as long as we choose one convention and stick with it, to avoid {double-counting} the fluctuations. 

If we take many snapshots of the population separated by $\gg {\bTCoher}$, we will observe the partially coherent broadcasts varying in luminosity. The sample variance approach would interpret this as intrinsic variability in the broadcast luminosities themselves; the instantaneous distribution of $\bLiso$ in the population would reflect the temporal distribution of $\bLiso$ for each individual broadcast. Since the box and chord models assume a {nonvarying} luminosity, I proceed with the {noise} interpretation {for} partially coherent broadcasts {(degenerate $\bLiso$, $\zMean{\bFCoherObs} = 1$). Despite this, I group the corresponding variance term with the sample variance because it behaves in exactly the same way, considering this noise term to contribute to sample confusion instead of noise confusion.}

\subsubsection{\texorpdfstring{{Confusion and} lines in the chord model}{Confusion and lines in the chord model}}
Ultranarrowband line searches look for lines with different drift rates, essentially {seeking} concentrations of energy when summing along skewed lines in spectrograms ({for example,} by applying a frequency shift to each time step, as in \citealt{Siemion13}). The dedrifting is performed on data that sample time and frequency ($\DatumLabel${-type windows}). With coherent dedispersion, each data point corresponds to one mode. More often, dedrifting is performed on a spectrogram where each data point is the sum of the energy in several modes, summed sequentially in time {\citep[as in][]{Lebofsky19}}. After dedrifting, a detected line appears on a spectrogram like a broadened line with no drift. {An observation is constructed by summing several sequential data points together into an observation. Each observation has an associated (de)drift rate $\oDriftRateObs$ applied to it. A} dedrifted line {$\BcMark$} has
\begin{equation}
\label{eqn:DedriftedChordEnergy}
{\Mean{\lEnergyObs (\text{dedrifted})}} \approx \frac{\iResponseEnergyObs({\lSkyLocation}) \iAeff {\bLiso \lDilutionE {\lTransmittanceE} \bFPolObs} \oDurationObs}{\max(1, {|\bDriftRate|} \oDurationDatum / \oBandwidthObs)} .
\end{equation}
The second term {in the denominator} accounts for smearing resulting from the line crossing in and out of each datum point within $\oDurationDatum$ \citep{Sheikh19,Margot21}. Any remaining lines continue to have {a range of} drift rates.

{In order to understand how a wider spread in drift rate affects detectability, I express the signal-to-noise of an observation} in terms of {the counterfactual} expected number of broadcasts and individual broadcast signal-to-noise {ratio} if the drift rate for all broadcasts were forced to zero. The former quantity is 
\begin{equation}
{\Mean{\bzNObsZD} =  \Mean{\hNObsGal} \bzAbundnu \oBandwidthObs}
\end{equation}
(equation~\ref{eqn:MeanNChord}), and {is the expected number of broadcasts per observation channel at any one instant. It is a direct proxy for the abundance of transmitters. T}he latter is
\begin{align}
\nonumber \lSNEnergyObsESTZD & \equiv \BUILDlSNEST{{\lEnergyObs | \bDriftRate = {\oDriftRateObs =} 0}}\\
\label{eqn:lSNZeroDef}
                             & = \frac{\iAeff {\zMean{\bLisoREGzSurv} \zMean{\lDilutionE {\lTransmittanceE}}} \oNumPolObs \oDurationObs}{{2} \sqrt{\oNModeObs} \ikTBack} .
\end{align}
There are fewer broadcasts in the absence of drift, but they individually would have higher {signal-to-noise ratios}{, and the product is invariant ($\lSNEnergyObsESTZD \Mean{\bzNObsZD} = \lSNEnergyObsEST \Mean{\bzNObs}$; see Appendix~\ref{sec:ChordConservation})}.

\begin{figure*}
\centerline{\includegraphics[width=14cm]{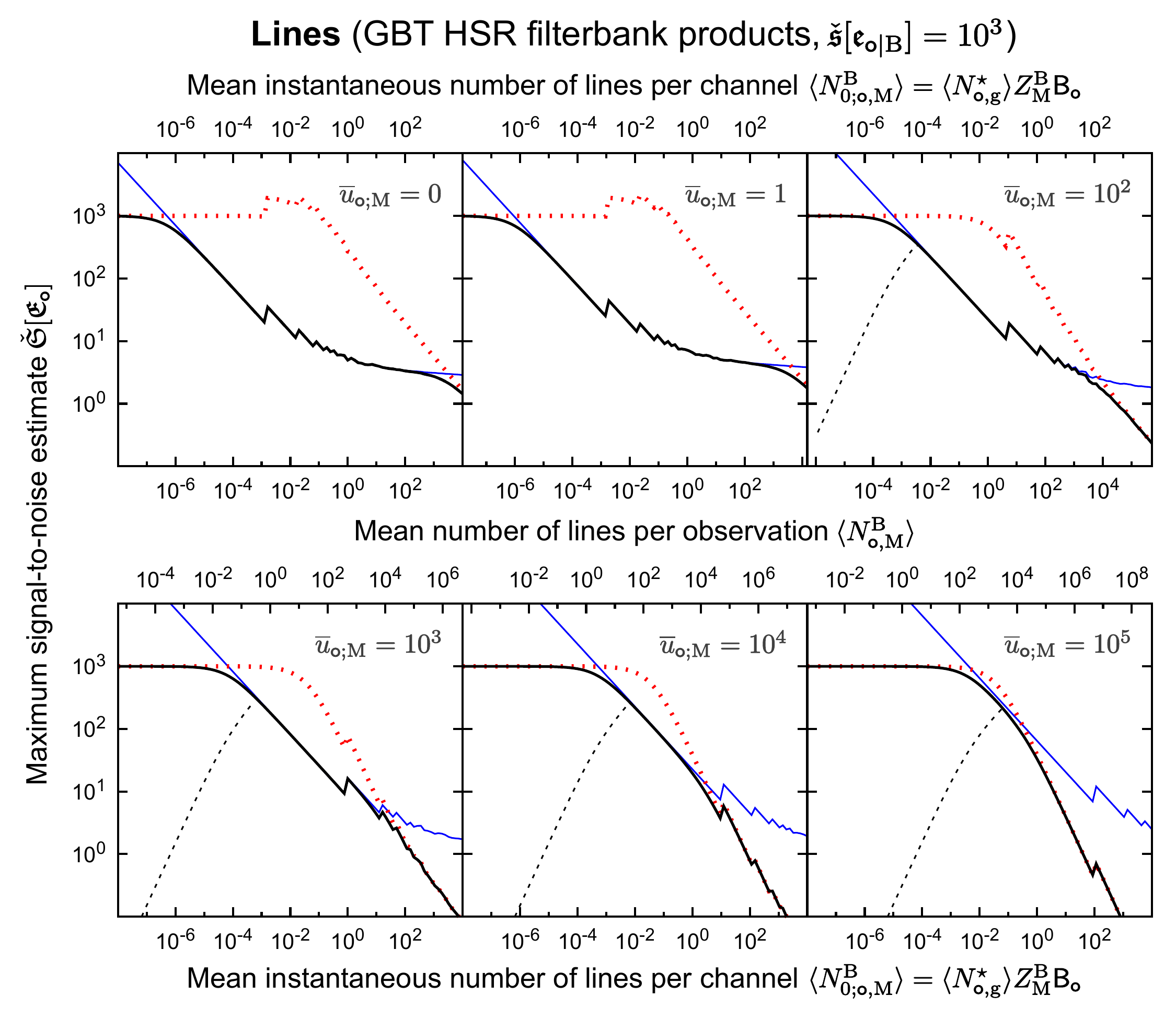}}
\figcaption{{The impact of confusion on $\qSNEnergyObsEST${, the estimated maximum signal-to-noise {ratio} found in the survey,} when observing a population of narrowband lines, as compared for different drift rate distributions.} I {adopt the standard assumptions of Table~\ref{table:StandardAssumptions}}. {The lines all have the same luminosity}{; individually they would have signal-to-noise {1000} if they were not drifting ($\lSNEnergyObsESTZD = {1000}$).} Their drift rate distribution has a spread of $\bzDriftRateBAR = \bzDriftNatBARObs \oBandwidthObs / \oDurationObs$, centered on $\bDriftRate = 0$. {No dedrifting has been applied ($\oDriftRateObs = 0$).} The blue lines show the limits from sample variance alone{, and determines when sample confusion occurs. The red dotted lines show limits from the quasi-thermal noise variance caused by the mutual interference {and self-noise} of all the broadcasts and background, and determines when noise confusion occurs.} Wider spreads in drift rates delay the onset of {sample} confusion{, but not noise confusion}. {Different line styles indicate whether or not we conditionalize on having found a line: solid for ${\bzIndicatorObs} = 1$ {(a line has definitely been found)}, and dashed for ${\bzIndicatorObs} = 0$ {(only serendipitous lines are included)}.} {The observations here are modeled according to the high spectral resolution filterbank products used in Breakthrough Listen for the Green Bank Telescope, for a single coarse channel (two polarizations, $\oDurationDatum = 51 \oDurationMode$, $\oDurationObs = 16 \oDurationDatum$, $\oBandwidthSurv = 2^{20} \oBandwidthObs$).} \label{fig:SNLines}}
\end{figure*}

{For} a randomly chosen trial drift rate $\oDriftRateObs$, {there may be lines that serendipitously are dedrifted, but a SETI analysis seeks for drift rates that yield the highest signal-to-noise {ratio}. {Hence,} sensitivity is generally better evaluated by supposing that the analysis has found one of these lines and chosen $\oDriftRateObs = \bDriftRate$ for that line.} {We specify whether or not we have found a line with $\bzIndicatorObs$: it is $1$ when we consider only an observation that definitely has a dedrifted line in it, and $0$ if we have a typical observation with a typical sample of serendipitous lines.} For observations in both polarizations of a distant {galaxy with transmitters of identical luminosity}, the diffuse approximation gives us
\begin{multline}
\label{eqn:SNEstDedrift}
\qSNEnergyObsEST \approx \lSNEnergyObsESTZD \left[\frac{{\bzIndicatorObs}}{\max(1, {|}\oDriftRateObs{|} \oDurationDatum/\oBandwidthObs)} + \fDelta {\bzNObsEFF}\right]\\
{\cdot} \left[\left(1 + \frac{\lSNEnergyObsESTZD \Mean{{\bzNWEnergyObsIONEZD}}}{\sqrt{\oNModeObs}} \right)^2 \right.\\
\left. + \lSNEnergyObsESTZD^2 \Mean{{\bzNWEnergyObsITWOZD}} {\bzFDriftObs} (1 + {\zMean{\bFCoherObs}})\right]^{-1/2} ,
\end{multline}
{where} ${\bzFDriftObs \equiv \zMean{\bDurationObs^2} / (\oDurationObs \zMean{\bDurationObs})}$ {and $\fDelta {\bzNObsEFF}$ is a quantity describing the contribution of serendipitous lines} (see Appendix~\ref{sec:ChordSNFull} for details{)}. {The value of $\bzFDriftObs$ is difficult to calculate when a dedrifting algorithm is applied, because of the way it ``slices'' {lines,} complicating the analysis (Appendix~\ref{sec:ChordConservation}). I present results for the case when no dedrifting is applied in Figure~\ref{fig:SNLines}, in which case there is no smearing. Basically, however, the signal-to-noise {ratio} for high intrinsic drift rate lines is expected to be suppressed, but {it} may be more resistant to confusion.}

Dedrifting {serves to accumulate the energy in} $\max[1, {|}\oDriftRateObs{|} \oDurationDatum/{\oBandwidthObs}]$ channels when {we do find a line (}${\bzIndicatorObs} = 1${)}. {Sample variance is generally suppressed; when $\oDriftRateObs = 0$,} ${\bzFDriftObs} \propto {\bzDriftRateBAR}^{-1} \ln {\bzDriftRateBAR}$ -- the mean collected fluence is the same, but it is divided among a larger number of quickly transiting broadcasts. In the sparse regime, the smearing within each time point decreases {the} signal-to-noise {ratio} of individual broadcasts \citep[as noted in][]{Sheikh19,Margot21}. {The reduction of sample variance can be expected to delay sample confusion, however, compared to the low drift rate case.}

We can evaluate when {sample} confusion sets in by again using the sample variance as a minimum variance and finding when $\qSNEnergyObsEST \la {\qSNThreshSurv}$. {The sample variance is increased by a factor $(1 + {\zMean{\bFCoherObs}})$ because the slow wave noise fluctuations effectively spread out the intrinsic luminosity distribution (section~\ref{sec:CoherentRadioNoise}).} As discussed in Appendix~\ref{sec:ChordSensitivityDetails}, confusion is evaluated for $\oDriftRateObs = 0$, because some lines will have nearly zero drift rate by chance, and these{, with minimal leakage into other channels, will be the last to be confused as $\Mean{\bzNObs}$ increases}. {Using the numerator from equation~\ref{eqn:SNEstDedrift} to define the expected maximum signal, {detection fails when}
\begin{equation}
\qSNThreshSurv \ga \frac{\bzIndicatorObs + \fDelta {\bzNObsEFF}}{\sqrt{\Mean{\bzNWEnergyObsITWOZD}\bzFDriftObs (1 + {\zMean{\bFCoherObs}})}} ,
\end{equation}
with the standard assumptions and when all broadcasts have the same luminosity. Let us focus on the signal from a single detected {nondrifting} line ($\bzIndicatorObs = 1$, $\fDelta {\bzNObsEFF} = 0$). By solving for $\bzNWEnergyObsITWOZD \approx \Mean{\bzNObsZD} = \bzAbundnu {\Mean{\hNObsGal}} \oBandwidthObs$, I find that} {sample} confusion results in a null detection when
\begin{equation}
\label{eqn:ConfusionThreshChord}
{\bzAbundnu} \ga \left[{\qSNThreshSurv}^2 {\Mean{\hNObsGal}} \oBandwidthObs {\bzFDriftObs} (1 + {\zMean{\bFCoherObs}})\right]^{-1}
\end{equation}

{When the drift rate spread becomes big enough, most lines are spread over many observations, and sample variance is so low that noise confusion becomes the impediment to detection when there are many broadcasts. This is seen in Figure~\ref{fig:SNLines} for the highest $\bzDriftNatBARObs (\bzDriftRateBAR \oDurationObs / \oBandwidthObs)$ cases. Using the quasi-thermal noise term in the variance, and again evaluating when no dedrifting is applied with standard assumptions, detection fails when
\begin{equation}
\qSNThreshSurv \la \sqrt{\oNModeObs} \frac{\bzIndicatorObs + \fDelta \bzNObsEFF}{\Mean{\bzNWEnergyObsIONEZD}} .
\end{equation}
I again solve for $\bzNWEnergyObsIONEZD \approx \Mean{\bzNObsZD}$, finding that noise confusion prevents a detection for
\begin{equation}
\label{eqn:NoiseConfusionThreshChord}
\bzAbundnu \ga \sqrt{\oNModeObs} \left[\qSNThreshSurv \Mean{\hNObsGal} \oBandwidthObs\right]^{-1} .
\end{equation}
Since most radio SETI observations use filterbank products that integrate many modes together, noise confusion is only a concern when the drift rate spread is quite high. For a zero-centered drift rate distribution, noise confusion happens first if $\bzDriftRateBAR/\ln(\bzDriftRateBAR \oDurationObs/\oBandwidthObs) \ga \sqrt{\oNModeObs} \qSNThreshSurv (1 + \zMean{\bFCoherObs}) \oBandwidthObs/\oDurationObs$, which I estimate to be $\bzDriftRateBAR \ga 14\ \Hz\,\sec^{-1}$ for line searches with the Green Bank Telescope.}

\subsection{Special considerations for radio interferometers}
\label{sec:RadioInterferometers}
	
Interferometric arrays are playing an increasing role in radio SETI \citep[e.g.,][]{Rampadarath12,Harp16,Tremblay20}.  Nominally, they provide increased angular resolution, which should decrease ${\Mean{\hNObsGal}}${, {$\Mean{\bzNObs}$},} and thus confusion.  But the situation is complicated by the presence of strong sidelobes in the synthesized ``dirty'' beam that span the primary (antenna) beam. 

Signals from the $\iNAntenna$ antennas in an array are {combined either coherently or incoherently. I}ncoherent summing simply adds together the voltages from all the antennas. The summed beam pattern is the same as the primary beam pattern, an advantage when searching for bright signals over a wide field.  Sensitivity increases as $\iNAntenna^{1/2}$ if detector noise dominates the variance in $\qEnergyMode$, but celestial radio noise hampers sensitivity if it dominates the variance \citep{Kudale17}, analogous to the self-noise limit for single-dish observations \citep{Radhakrishnan99}. {Sample c}onfusion occurs when there is one broadcast {on average} in the time-frequency window of the observation anywhere within the primary beam.

{Coherent beamforming processes the voltages to synthesize narrow beams with} complex sidelobes. The ``thinned array curse'' implies that, given a fixed collection of antennas, the maximum gain is invariant with respect to how those antennas are arranged \citep{Forward84}.\footnote{Originally derived for energy transmitters, it also applies to receivers by reciprocity.}  Increasing the spacing of the antennas merely shrinks the size of the main lobe. It can be shown then that the mean response over the primary beam is $1/\iNAntenna$, and for a sparse array with filling factor $\ll 1$, almost all of this response comes from the sidelobes.

{There are two basic strategies for beamforming that are used. Coherent summing synthesizes tied-array beams by applying} phase correction{s} to each antenna before {adding} the voltages. {It} requires no time-averaging and is used for observations of transients and pulsars \citep{Stappers11}. It also is the basis of Breakthrough Listen's {million-star} survey with MeerKAT \citep{Czech21}.  However, computational limits currently allow only a few dozen beams to be formed in the primary beam, sampling only a small part of it. {Aperture synthesis, the more well-known technique,} perform{s} a Fourier transform on the visibilities formed by multiplying the voltages of two antennas together and time-averaging. This approach generates a map of the entire primary beam but is not suited for detection of rapid transients because of the averaging and computational cost. In addition, only baselines between distinct antennas are included, so there is no sensitivity to diffuse backgrounds that cover the entire primary beam: these observations only detect spatial \emph{fluctuations} in broadcasts and could fail to detect a heavily populated galaxy {covering a large enough sky area}.

{In both approaches, the dirty beam sidelobes have an amplitude $1/\iNAntenna$.}\footnote{\vphantom{I}{In aperture synthesis,} the lack of zero-spacing data shifts the beam pattern to have zero mean overall with negative responses over much of the sidelobes \citep{Kogan99}.} The relatively strong sidelobes lead to much greater confusion than would normally be expected, hampering deconvolution of the dirty map.  Under most circumstances relevant to radio astronomy, the sky can be regarded as basically empty, with only a few sources covering a small fraction of the primary beam, so this is not an issue (\citealt{Hogbom74}). If there are more sources in the primary beam {than independent measurements}, however, it becomes impossible to disentangle them. {Coherent summing yields only one independent measurement per antenna per observation, $\iNAntenna$ in total; aperture synthesis provides only one per baseline, $\iNBaseline$ in total \citep{Hogbom74,Schwarz78}. Thus,} confusion necessarily sets in when
\begin{equation}
\label{eqn:ConfusionInterferom}
\Mean{{\bzNObsPRIM}} \ga \begin{cases}
\iNAntenna  & (\text{coherent summing})\\
\iNBaseline & (\text{aperture synthesis})
\end{cases} .
\end{equation}
The minimum number of baselines is $\iNAntenna (\iNAntenna - 1)/2$ from a single snapshot. Rotational synthesis using multiple snapshots to increase $\iNBaseline$ {is} limited to long-lasting broadcasts.

\section{\texorpdfstring{{Individualist constraints: optical broadcasts}}{Individualist constraints: optical broadcasts}}
\label{sec:SNOptical}

\subsection{Noise variance in optical photon counting}
\label{sec:OpticalNoise}
At frequencies beyond radio, the mean number of photons arriving {in an electromagnetic field mode (the photon occupation number)} is much less than {$1$}. {When that happens}, the photon shot noise completely overwhelms the photon bunching effect of wave noise \citep{Radhakrishnan99}.\footnote{Under certain circumstances, {photon counts} can have \emph{sub-Poissonian} statistics, with the Fock states having no number fluctuations at all \citep[e.g.,][]{Foellmi09}.  This is a purely quantum phenomenon, requiring photon detectors {to measure}, and is not expected from natural astrophysical sources, but it could be a technosignature \citep{Hippke21}.}

Instruments at these higher frequencies nowadays count photons, with the measured quantity being the number of photons collected during an observation, $\qPhotonObs$. The number of photons {from the background, $\kPhotonObs$,} can be regarded as {a} pure Poisson random variable. {In the absence of modulation, the number of photons from each broadcast $\BcMark$, $\lPhotonObs$, is also Poissonian.} This greatly simplifies analysis compared to the radio case, because the sum of a fixed number of {independent} Poisson random variables is another Poisson random variable {\citep{Kingman93}}.  I assume the broadcasts have constant intrinsic luminosity, as in the box model, with no significant modulation on timescales $\ga \oDurationObs$. Let ${\lFluenceQObs}$ be the photon fluence from each broadcast ${\BcMark}$ within the observation $\ObsLabel$. For {a} distant {metasociety}, the mean is
\begin{equation}
\label{eqn:MeanOptical}
\Mean{\qPhotonObs} \approx \Mean{\kPhotonObs} + \iAeff {\Mean{\bzNWPhotonObsIONE} \zMean{\bPhotonisoObs} \zMean{\lDilutionQ {\lTransmittanceQObs}} .}
\end{equation}
{{In the absence of attenuation, i}t can also be shown (Appendix~\ref{sec:PhotonDerivation}) that the} variance under the diffuse approximation {is} 
\begin{multline}
\label{eqn:VarOptical}
\Var{\qPhotonObs} \approx \Mean{\kPhotonObs} + \iAeff {\zMean{\bPhotonisoObs} \zMean{\lDilutionQ {\lTransmittanceQ}} \Mean{\bzNWPhotonObsIONE}} \\
{+ \iAeff^2 \zMean{(\bPhotonisoObs)^2} \zMean{\lDilutionQ^2 {\lTransmittanceQ^2}} \Mean{\bzNWPhotonObsITWO}} ,
\end{multline}
which includes the mean number $\Mean{\kPhotonObs}$ of noise photons from the sky background, dark current, and readout noise. Slow modulation increases the variance, which can be modeled either {as} an intrinsic spread in the photon fluence of the broadcasts or by an additional self-noise term {(much like as in Section~\ref{sec:PartiallyCoherentRadioNoise})}.

\subsection{\texorpdfstring{{The behavior of signal-to-noise ratio for Poissonian photon counts}}{The behavior of signal-to-noise ratio for Poissonian photon counts}}

It is straightforward to estimate {the} signal-to-noise {ratio} with photon counting instruments at high frequencies using equation~\ref{eqn:VarOptical}. The expected signal-to-noise {ratio} of an isolated broadcast at $\iResponsePhotonObs = 1$,
\begin{equation}
\lSNPhotonObsEST {\equiv} \frac{\iAeff {\zMean{\lFluenceQObsREGzSurv}}}{\sqrt{\Mean{\kPhotonObs}}} {=
\frac{\iAeff \zMean{\bPhotonisoObsREGzSurv} \zMean{\lDilutionQ {\lTransmittanceQObs}}}{\sqrt{\Mean{\kPhotonObs}}}},
\end{equation}
provides another convenient scaling variable. Then, for distant {galaxies} and using the diffuse approximation,
\begin{multline}
\qSNPhotonObsEST = {\bzNObsEFF} \lSNPhotonObsEST {\cdot} \left(1 + \lSNPhotonObsEST \frac{\Mean{{\bzNWPhotonObsIONE}}}{\sqrt{\Mean{\kPhotonObs}}} \right. \\
\left. + \lSNPhotonObsEST^2 \Mean{{\bzNWPhotonObsITWO}} \frac{{\zMean{(\bPhotonisoObsREGzSurv)^2}}}{{\zMean{\bPhotonisoObsREGzSurv}^2}}\right)^{-1/2} .
\end{multline}

{Sample variance exceeds background noise when
\begin{equation}
\Mean{\bzNWPhotonObsITWO} > \left(\lSNPhotonObsEST^2 \frac{\zMean{(\bPhotonisoObsREGzSurv)^2}}{\zMean{\bPhotonisoObsREGzSurv}^2}\right)^{-1} ,
\end{equation}
{similar to the radio case.} Moreover, sample variance dominates the photon shot noise from the broadcasts when
\begin{equation}
\zMean{\lPhotonObs} > \left[\frac{\Mean{\bzNWPhotonObsIONE}}{\Mean{\bzNWPhotonObsITWO}} \left(1 + \frac{\zVar{\bPhotonisoObsREGzSurv}}{\zMean{\bPhotonisoObsREGzSurv}^2}\right)\right]^{-1}.
\end{equation}
Under normal circumstances, this second inequality holds as} long as at least ${\sim 1}$ photon is expected to be detected per broadcast.

{This means that} there is no {{mutual} interference} regime {for optical broadcasts} -- $\qSNPhotonObsEST$ remains at a factor of a few from sample variance alone. {Noise confusion by itself never prevents detection.} {Because each broadcast is contributing many photons, the photons can be thought of as coming in groups, one for each broadcast. This introduces an intrinsic ``graininess'' to the photon count statistics, reminiscent of the surface brightness fluctuations observed in distant galaxies from discrete bright stars \citep{Tonry88,Raimondo05}. Now, there still {is} a {(sample)} confusion regime, so this cannot be exploited to make an individual detection and does not intrinsically favor optical over radio. {Perhaps t}he graininess might be discerned in the photon count statistics when in the confusion regime, and the lack of interference {can extend} the range of {that kind of} technique in optical. In any case, the confusion regime already implies extremely high abundances of broadcasts, so these considerations only apply to an arguably contrived region of parameter space.}

\section{The collective bound: The total emission of the galaxy}
\label{sec:Collective}

The collective bound, by contrast, is based on the total emission received from all the broadcasts in the target {galaxy}.  The concept is very simple: the broadcast from ETIs cannot outshine the galaxy as a whole {(including the broadcasts)} when we're looking at it. {To be clear, this is an entirely empirical constraint, applying the observed emission rather than how much emission we expect from natural processes. It is possible that ETIs might vastly increase the apparent luminosity of a galaxy by harnessing its central black hole or invoking unknown physics -- {or,} more practically, by beaming emission in our direction -- but the observed emission necessarily includes this increased luminosity, along with any additional natural emission. We already know that there are no trillion $\Lsun$ radio beacons in M31, for instance, because the radio luminosity of M31 is a lot lower than a trillion Suns, even though such beacons could hypothetically be constructed.} 

It is impossible to make a detection using the collective approach alone, since any emission we detect could be natural, but {an estimated fluence $\dFluenceObs$ derived from measurement} does let us conclude that most likely ${\mzFluenceObs} \la \dFluenceObs$.  Statistical analysis lets us constrain a combination of the abundance and brightness distribution. 

Observations in the literature effectively report fluences, generally cast as a flux.  Now, the variance in ${\mzFluenceObs}$ includes sample variance and noise variance.  Noise variance is included in the reported errors $\fDelta\dFluenceObs$ on the reported fluence. {Given a cumulative fluence distribution $\CDF{\mzFluenceObs}$, a confidence interval only allows those models for which}
\begin{equation}
{\fpPBAR \le}~{\CDF{\mzFluenceObs}(\dFluenceObs + C_{\ObsLabel} \fDelta \dFluenceObs)},
\end{equation}
where ${C_{\ObsLabel}}$ is some {predefined} constant {and $\fpPBAR \approx 0$ is a conservative probability threshold}. {That is, we want to include all models where there is any significant chance that the broadcast population is fainter than the observed luminosity.} {When only an upper limit on the emission is known, it can be {{substituted} for} $\dFluenceObs + C_{\ObsLabel} \fDelta \dFluenceObs$. Taking the inverse CDF of both sides, we find
\begin{equation}
\label{eqn:InvCDFCollBound}
\InvCDF{\mzFluenceObs}(\fpPBAR) \le \dFluenceObs + C_{\ObsLabel} \fDelta \dFluenceObs .
\end{equation}
If the broadcast fluence has a {narrow distribution with a} well-defined variance, we can approximate the left-hand side as
\begin{equation}
\label{eqn:InvCDFFluenceEstimate}
\InvCDF{\mzFluenceObs}(\fpPBAR) \approx \Mean{{\mzFluenceObs}} - C_{\MetaMark} {\SD{\mzFluenceObs}}
\end{equation}
for some suitable constant $C_{\MetaMark}$ {that absorbs the dependence on $\fpPBAR$}.}\footnote{{If the aggregate fluence has a power-law tail, the distribution is likely to be highly asymmetrical and this approximation will fail{, even using the regularized mean and variance}.}}

{Now, say that \emph{all} of the observed fluence {from} from broadcasts, with no natural background. We could estimate the number of broadcasts that are contributing to the observed emission simply as
\begin{equation}
\label{eqn:NEstimatedCollective}
\dbzNObs = \frac{\dFluenceObs + C_{\ObsLabel} \fDelta \dFluenceObs}{\zMean{{\lFluenceObs}}} = \frac{\oBandwidthObs(\dLnu + C_{\ObsLabel} \fDelta \dLnu)}{\zMean{{\bLiso}}},
\end{equation}
the number of typical broadcasts that can ``fit'' into the emission (in terms of fluence and spectral luminosity, respectively). This naive estimate is in fact central to the collective bound. Under our usual assumptions of a single galactic metasociety, with interchangeable broadcasts and societies and Poissonian $\baNObs{(\aTuple)}$ and $\azNObs{(\zTuple)}$

\begin{equation}
\Mean{{\mzFluenceObs}} {\approx} \Mean{\bzNObs} \zMean{{\lFluenceObs}}
\end{equation}
{and}
\begin{multline}
\Var{{\mzFluenceObs}} {\approx} \Mean{\bzNObs} \left(\zMean{{\lFluenceObs^2}} + \zMean{\baNObs} \zMean{{\lFluenceObs}}^2\right)
\end{multline}
Plugging these into {the {left-hand} side of} equation~\ref{eqn:InvCDFFluenceEstimate} {by use of equation~\ref{eqn:InvCDFCollBound}}, we find the expression for the collective bound:
\begin{equation}
\label{eqn:Collective}
\Mean{\bzNObs} {\la} \dbzNObs + \frac{C^{\prime 2}}{2} \left(1 + \sqrt{1 + \frac{4}{C^{\prime 2}} \dbzNObs}\right),
\end{equation}
where 
\begin{equation}
C^{\prime} \equiv C_{\MetaMark} \sqrt{\frac{\zMean{{\lFluenceObs^2}}}{\zMean{{\lFluenceObs}}^2} + \zMean{\baNObs}} .
\end{equation}
}

{The meaning of equation~\ref{eqn:Collective} depends on the average brightness of a broadcast. When $\zMean{{\lFluenceObs}}$ is very small, many broadcasts can ``fit'' into the observed emission, and equation~\ref{eqn:Collective} reduces to $\Mean{\bzNObs} \le \dbzNObs + C^{\prime} \sqrt{\dbzNObs}$, the maximum number that are allowed before the aggregate population outshines the actual emission (with allowances for Poissonian fluctuations).}

When $\zMean{\lFluenceObs}$ is very bright, even one broadcast would produce more emission than is observed, and is thus ruled out by how faint the galaxy actually is. {In this limit, when the broadcast fluence distribution is narrow, $\dbzNObs \to 0$ and equation~\ref{eqn:Collective} reduces to $\Mean{\bzNObs} \la C^{\prime 2}$. When $\zMean{\baNObs} \la 1$, $C^{\prime} \approx C_{\MetaMark}$, so the bound says that broadcasts are just too rare to have been caught in the observation: $\Mean{\bzNObs} \la C_{\MetaMark}^2$. If $\zMean{\baNObs} \ga 1$, broadcasts outnumber societies and $C^{\prime} \approx C_{\MetaMark} \zMean{\baNObs}$; equation~\ref{eqn:Collective} simplifies to $\Mean{\azNObs} \la C_{\MetaMark}^2$. Equation~\ref{eqn:Collective} thus contains the discreteness bound (Paper I) -- it cannot rule out a model that predicts a good chance of there being no societies or broadcasts covered by an observation, because no artificial emission is expected then.}

In the box model, {equations}~\ref{eqn:Collective} {and~\ref{eqn:MeanNBox}} {give} us the constraint that either {$\Mean{\azNObs} \approx 0$} or
\begin{equation}
\label{eqn:FundamentalCollectiveBox}
{\bzRatenu} \la \frac{{\dbzNObs + (C^{\prime 2}/2) \left[1 + \sqrt{1 + 4 \dbzNObs/C^{\prime 2}}\right]}}{\Mean{\hNObsGal} (\oDurationObs + {\bDurationBox}) (\oBandwidthObs + \bBandwidthBox)} {.}
\end{equation}
{Applying the collective bound instead to the chord model using equation~\ref{eqn:MeanNChord}, the bound is}
\begin{equation}
\label{eqn:FundamentalCollectiveChord}
{\bzAbundnu} \la \frac{{\dbzNObs + (C^{\prime 2}/2) \left[1 + \sqrt{1 + 4 \dbzNObs/C^{\prime 2}}\right]}}{\Mean{\hNObsGal} \oBandwidthObs (1 + {\zMean{|\bDriftRate|}}\oDurationObs/\oBandwidthObs{)}}
\end{equation}
unless ${\Mean{\azNObs} \approx 0}$. {Observations of continuum galactic luminosity use wideband observations, so $(1 + \zMean{|\bDriftRate|}\oDurationObs/\oBandwidthObs) \approx 1$ in practice.}

The collective bound is sensitive to the {full brightness} distribution instead of only the most extreme values, and thus explore{s} a different range of parameter space than individualist searches. {Observations used to derive galactic luminosities} are usually relatively wideband {by SETI standards}{; they also have long integration times compared to the time resolution of pulse searches}. This increases the fluence collected from noise and background emission, burying the brightest individual signals, but the mean total fluence in ETI broadcasts also increase{s} proportionally. Collective bounds apply to arbitrarily faint broadcasts as long as they are numerous enough, even when they fail to limit rare but very bright single broadcasts.  The collective approach therefore has the distinct capability of setting constraints on {pervasive Kardashev} Type I--II transmitters in very distant galaxies, far below the detection threshold of typical SETI surveys.

The other advantage of this seemingly trivial limit is that it is unaffected by confusion. In fact, it demands we are near or in the confusion limit for the observation we are applying -- the longer, wideband observations used to evaluate $\dLnu$ reach this limit much sooner than typical SETI measurements. This mostly closes the gap in parameter space opened by the confusion limit in individualist approaches (section~\ref{sec:RadioSingleConstraints}).

\section{Constraints on artificial radio transmissions in nearby galaxies}
\label{sec:RadioSingleConstraints}

A comparison of the collective bound with individualist constraints using extant or planned SETI observations illustrates their relative merits, with the latter becoming weaker at greater distance. {I present these comparisons for narrowband line searches in L-band, near 1.4 GHz. This region of the spectrum has historically been favored in radio SETI: background noise from synchrotron emission is quiet in this regime, and it contains the ``water hole'', a band between radio lines presumed to be well-known to alien astrophysicists wanting to make contact \citep{Cocconi59,Oliver71}. The need to know which frequency to ``meet at'' was important when {back ends} covered at best only a few {megahertz} at a time, but modern instrumentation largely sidesteps the problem. Although there have been narrowband radio SETI surveys at much lower and higher frequencies, L-band is used here because it has the most results to compare with \citep[see][]{Tarter85,Enriquez17}, including extragalactic results \citep{Horowitz93,Shostak96,Gray17}. But it is also commonly used to study galaxies' radio continuum emission (e.g., \citealt{Yun01} among many others). {Thus,} it is a natural choice for collective bounds derived from the literature. The basic ideas should apply for other types of broadcasts, however.}

The broadcasts are assumed to {all} have {the same} luminosity $\bLisoBAR$ and an instantaneous abundance per unit frequency ${\bzAbundnu}$. The power of individualist searches to detect lines is estimated using the chord model, assuming a single line (${\bzIndicatorObs = 1}${;} equation~\ref{eqn:SNEstDedrift}) {and no dedrifting ($\oDriftRateObs = 0$)}. The diffuse approximation is also applied. {Sample c}onfusion is evaluated according to {equations}~\ref{eqn:ConfusionThreshChord} and~\ref{eqn:ConfusionInterferom}{; noise confusion is also considered (equation~\ref{eqn:NoiseConfusionThreshChord}) as a robust limit against background broadcasts}.

{The range of mean broadcast luminosities considers spans many orders of magnitude. As a rough guide, I compare the results for different Kardashev scale levels. As in \citet{Kardashev64}, the  levels measure the \emph{broadcast} effective isotropic power, with Type I representing a {``planetary''-scale} beacon ($\sim 10^{-10}\ \Lsun$), Type II a {``stellar''-scale} beacon ($\sim 1\ \Lsun$), and Type III a {``galactic''-scale} beacon ($\sim 10^{10}\ \Lsun$). {When considered quantitatively, I adopt a Sagan-like normalization of $1\ \Lsun$ for Type II, with a ratio of $10^{10}$ between each class \citep[see][]{Cirkovic15}.}}

\subsection{Milky Way}
\label{sec:MilkyWay}
The great majority of targeted SETI programs have observed {things within our own Galaxy}, focusing on nearby stars.  At distances of parsecs instead of megaparsecs, we are able to detect far weaker transmissions -- comparable to our planetary radars, although still well above any permanent broadcast we maintain \citep{Enriquez17}. There is, however, a great diversity of surveys, ranging from observations of one or two very nearby stars to large-scale surveys of millions of stars in the Galactic Center (Table~\ref{table:MWObs}).

\begin{deluxetable*}{lccccccccccccc}
\tabletypesize{\footnotesize}
\tablecolumns{13}
\tablewidth{0pt}
\tablecaption{Summary of individualist Milky Way searches for lines covering 1.4 GHz\label{table:MWObs}}
\tablehead{\colhead{Survey} & \colhead{Instrument} & \colhead{$\oBandwidthObs$} & \colhead{$\oBandwidthPoint$} & \colhead{$\oDurationDatum$} & \colhead{$\oDurationObs$} & \colhead{$\oDurationPoint$} & \colhead{${\qSNThreshSurv}$} & \colhead{$\oNPointSurv$} & \colhead{${\Mean{\hNObsGal}}$} & \colhead{${\Mean{\hNSurvGal}}$} & \colhead{$\bLiso$} & \colhead{$\oNObsSurvEFF$} \\ & & \colhead{$\Hz$} & \colhead{$\GHz$} & \colhead{$\sec$} & \colhead{$\sec$} & \colhead{$\minute$} & & & & & \colhead{$\Watt$} & }
\startdata 
R12\tablenotemark{a}          & LBA          & $1950$  & $0.314$                  & $2400$               & $2400$               & $480$                  & $5$                    & $1$     & $1$ & $1$     & $1.4 \times 10^9$    & $7.2 \times 10^4$\\
H16\tablenotemark{b}          & ATA          & $0.7$    & $8$     & $1.5$                 & $93$ & $180$ & $6.5$ & $65$    & $1$ & $65$    & $8.8 \times 10^{16}$ & $5.5 \times 10^{9}$\\
                              &              & $0.7$    & $2.04$  & $1.5$                 & $93$ & $45$  & $6.5$ & $1959$ & $1$ & $1959$ & $5.5 \times 10^{16}$ & $4.3 \times 10^{10}$\\
                              &              & $0.7$    & $0.337$ & $1.5$                 & $93$ & $7.5$ & $6.5$ & $2822$ & $1$ & $2822$ & $3.8 \times 10^{14}$ & $1.0 \times 10^{10}$\\
                              &              & $0.7$    & $0.268$ & $1.5$                 & $93$ & $5.9$ & $6.5$ & $7459$ & $1$ & $7459$ & $1.1 \times 10^{15}$ & $2.1 \times 10^{10}$\\
BL (P20)                      & GBT (L)      & $2.79$   & $0.660$ & $18.25$               & $300$                 & $15$                   & $10$                   & $882$   & $1$ & $882$   & $2.1 \times 10^{12}$ & $1.9 \times 10^{9}$\\
BL (G21)                      & Parkes (UWL) & $3$      & $3.328$ & $18$                  & $600$                 & $120$\tablenotemark{c} & $20$                   & $7$     & $8.6 \times 10^6$\tablenotemark{d} & $6 \times 10^7$ & $4 \times 10^{18}$ & $3.9 \times 10^7$
\enddata
\tablecomments{Abbreviations: ATA -- Allen Telescope Array; BL -- Breakthrough Listen; GBT (L) -- Green Bank Telescope, L-band receiver; LBA -- Long Baseline Array; {and} UWL -- Ultra Wideband Low receiver.}
\tablenotetext{a}{R12 used {very long baseline interferometry} to observe {Gl} 581. I only consider broadcasts from that star, including when deriving the confusion limit. {Thus,} confusion for lines is determined by frequency resolution only, regardless of the increased angular resolution.}
\tablenotetext{b}{Both $\oDurationObs$ and ${\qSNThreshSurv}$ were lowered during the program, from $192$ to $93\ \sec$ and from $9$ to $6.5$, respectively.  In H16, the ATA only had an instantaneous bandwidth of $\oBandwidthTune = 70\ \MHz$.  Thus, many ``tunings'' at different frequencies were necessary to cover the entire survey bandwidth, with $\oDurationPoint$ calculated here as $\oDurationTune (\oBandwidthPoint/\oBandwidthTune)$.  The $\oBandwidthPoint$ are averages, based on the total ``star-MHz'' accumulated in each series of observations.}
\tablenotetext{c}{Actual observation times varied from $1$ to {$3\ \hr$}.}
\tablenotemark{d}{{When evaluating confusion, $\Mean{\hNObsGal} = 1$ is used, representing the nearest society's host star, as discussed in the text.}}
\tablereferences{R12 -- \citet{Rampadarath12}; H16 -- \citet{Harp16}; P20 -- \citet{Price20}; G21 -- \citet{Gajjar21}; M04 -- \citet{Meyer04}; N11 -- \citet{Norris11}}
\end{deluxetable*}

\begin{figure}
\centerline{\includegraphics[width=8.5cm]{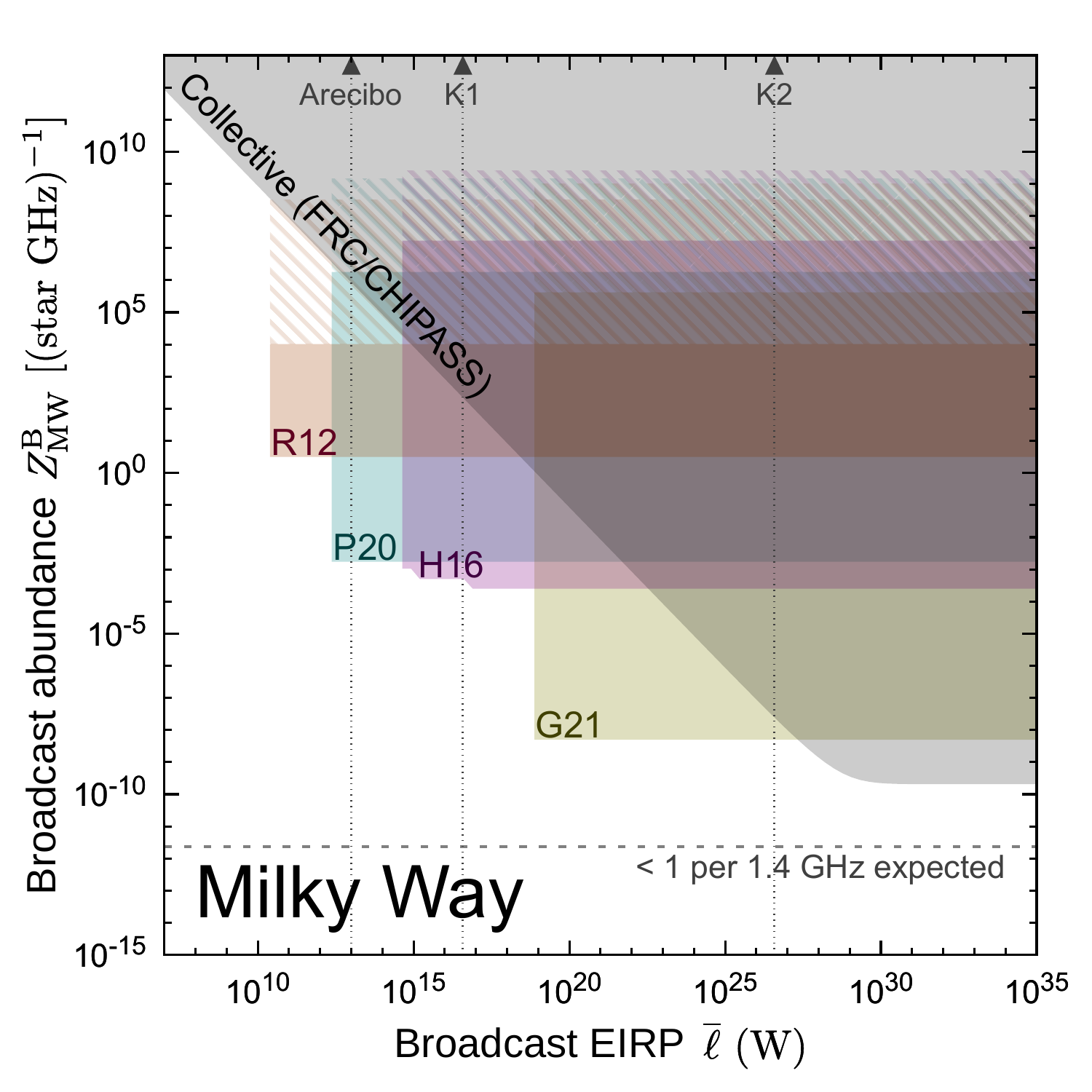}}
\figcaption{Order-of-magnitude constraints on ETI radio line broadcast luminosity and abundance in the Milky Way, assuming {all broadcasts have the same luminosity}. The collective bound from the assumption {that the Galaxy} lies on the FRC (grey) is shown {against} approximate {individualist} results from recent SETI surveys: R12 (red), P20 (green), H16 (violet), and G21 (gold). {For the SETI surveys, solid shading is used for the constraints when all drift rates are zero; if only noise confusion is present, the hatched regions are constrained as well.} These constraints assume {that} $\bDurationBox$ is longer than the observational programs. {The approximate EIRPs for the Arecibo planetary radar and Kardashev Type I (K1) and II (K2) broadcasts are marked.} \label{fig:MWLineDelta} }
\end{figure}

{T}he region{s} of $1.4\ \GHz$ {luminosity--abundance} parameter space constrained by SETI literature {are sketched} as the colored shaded regions in Figure~\ref{fig:MWLineDelta}. The bounds of these regions should be understood as order-of-magnitude estimates, as they assume a single characteristic distance for all stars in the sample and only include targeted stars.  Broadcasts from background stars are detectable if bright enough \citep{WlodarczykSroka20}, {increasing the mean number of sampled broadcasts at larger distances. Thus, the lower {boundaries of the shaded regions excluded by individualist programs} should actually bow downward at high $\bLisoBAR$. 

Unlike {when} we observe a single, distant metasociety, background societies within the Milky Way can have a wide range of distances. This spread in distances {widens} the fluence distribution to $\zPDF{\lFluenceEObs} \propto \lFluenceEObs^{-5/2}${, broad enough to prevent confusion (Section~\ref{sec:SNDefs})}. Confusion thus only sets in if the broadcasts from \emph{each} society are confused, regardless of whether we consider the targeted stars or the background stars.} {Seeing as each society is located at a single location, one star, I evaluate confusion as if $\Mean{\hNObsGal} = 1$ in equations~\ref{eqn:ConfusionThreshChord} and~\ref{eqn:NoiseConfusionThreshChord}, representing the host of this nearest society. Confusion also depends on the drift rate spread. It is most severe when all lines have zero drift (solid regions); the maximum abundances allowed, when only noise confusion is an issue, are $\sim 10^2 \endash 10^{4.5}$ times higher (hatched regions).}

Also shown is a collective bound that would arise if the Milky Way lies on the far-infrared--radio correlation (FRC; see section~\ref{sec:FRC}) for star-forming galaxies {(SFGs)}. {I employ a bandwidth of $64\ \MHz$, which is the bandwidth covered by CHIPASS, a recent 1.4 GHz southern sky survey used to estimate the Milky Way's radio emission (\citealt{Calabretta14}; see also \citealt{Zheng17}).} We cannot directly measure the radio luminosity of the Milky Way {since} we live inside it, but the assumption is validated by models of the Galaxy's synchrotron emission \citep{Strong10}. {The collective bound becomes more stringent for brighter broadcasts, of course, because fewer are needed to outshine the amount of radio emission that we actually expect from the Milky Way.} Nonetheless, these models are fit to observations of the Milky Way that only {cover} parts of the radio spectrum, leaving open the possibility that some radio transmitters could be hiding in the ``gaps'' \citep[consider the literature data in Table 1 of ][]{Zheng17}.  The {fact that we need to sample at least one broadcast to detect any artificial emission} sets a ${\bzAbundnu}$ threshold, below which the collective bounds fail to be constraining.\footnote{Values of $\oNuMidSurv {\bzAbundnu} = d^2 \Mean{{\bzNTime}}/[d\bTStart d\ln \bNuMid]$ much lower than one per Galaxy are meaningful because broadcasts turn on and off over time. The Galaxy could have $10^{-3}$ active transmitters on average, for example, if only one turns on per millennium and transmits for just a year.} Future work can set collective bounds along individual {sight lines} by comparing the sky density of stars with the Galactic radio background.

It is clear from Figure~\ref{fig:MWLineDelta} that the conventional individualist search strategy is highly effective in the Milky Way.  Collective bounds set some weak limits on very low luminosity broadcasts ($\bLisoBAR \la 10^{10}\ \Watt$), and rule out the possibility that extreme confusion is preventing detection of bright broadcasts, but the ${\bzAbundnu}$ values involved are {far-fetched}{:} more than one broadcast per star per {kilohertz} {for $\bLisoBAR$ up to $10^{13}\ \Watt$, the EIRP of our brightest transmitters, the planetary radars on facilities like Arecibo}. {Kardashev Type I radio transmitters are limited to about 200 per star per {gigahertz} through the collective bound, compared with the individualist limits of $\sim 10^{-3}$ per star per {gigahertz} from \citet{Harp16}. As we progress to higher broadcast EIRPs, surveys probing more distant regions of the Galaxy, like \citet{Gajjar21}, continue to be more restrictive than the collective bound because they cannot find even one to have a null result. For Kardashev Type II radio transmitters, the collective bound sets a limit of one per {$50$} million stars per {gigahertz}, while \citet{Gajjar21} {set} a limit about one {order of magnitude} stronger.}

Surprisingly, though, the collective bound is the tightest of those in Figure~\ref{fig:MWLineDelta} for the rarest, very bright transmitters {-- {reaching} about one per five billion stars per {gigahertz}}.  The plotted individualist SETI programs are targeted, covering only a small fraction of the Galaxy's stars.  There are also historical all-sky SETI surveys, and even continuum and HI radio surveys that could be brought to bear.  The problem is {that} these all have limited frequency ranges: if there is only one transmitter in the Galaxy somewhere in the frequency band $0 \endash 10\ \GHz$, it almost certainly falls outside the $0.4\endash 0.5\ \MHz$ covered by Big Ear or META \citep{Dixon85,Horowitz93}, or the $40 \endash 60$ MHz bandwidth of NVSS{,} HIPASS{, or CHIPASS} \citep{Condon98,Meyer04,Calabretta14}. Breakthrough Listen is surveying the bulk of the Galactic Plane over hundreds of {megahertz} \citep{Isaacson17}, partly closing this window.

\subsection{M31}
\label{sec:M31}
\begin{figure*}
\centerline{\includegraphics[width=8.5cm]{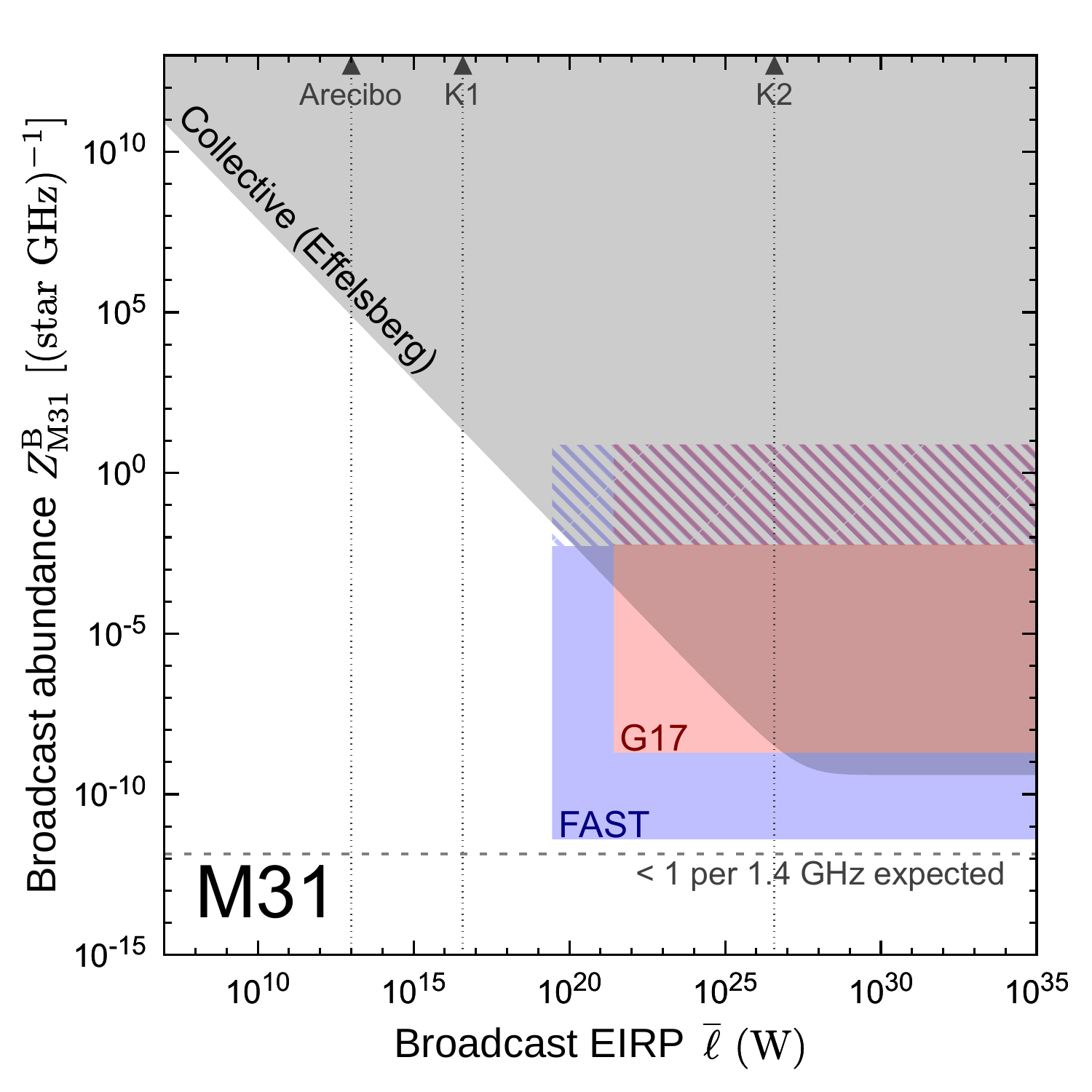}\,\includegraphics[width=8.5cm]{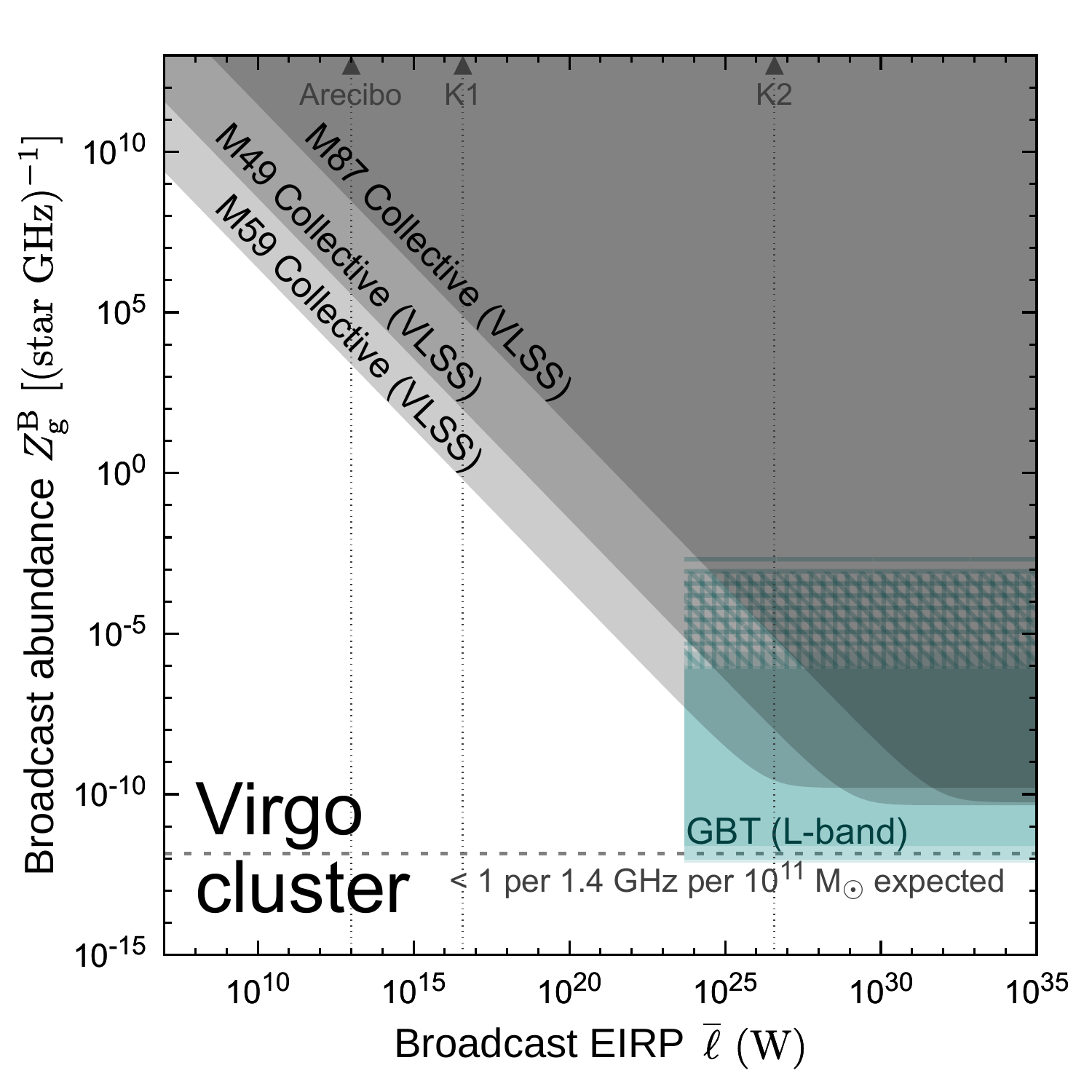}}
\figcaption{Current and forthcoming constraints on radio line broadcasts in M31 (left) and three example Virgo cluster ellipticals (right), assuming {that} {all broadcasts in each galaxy have the same EIRP $\bLisoBAR$}.  The G17 observations of M31 exclude the red shaded region, while example future observations with FAST and GBT are in blue and green, respectively. {Solid shading applies to individualist constraints when $\bzDriftRateBAR = 0$, and hatched {shading is} for the noise confusion limit at high $\bzDriftRateBAR$ (vertical for M49, diagonal for M31 and M87, horizontal for M59).} \label{fig:M31LineDelta}\label{fig:VirgoLineDelta}}
\end{figure*}

M31 is the nearest {large} spiral galaxy to our own, the only one besides the Milky Way and the {lower-mass} M33 with a {completed SETI survey since the turn of the century until very recently} (Table~\ref{table:M31VirgoIndivLines}).  \citet{Gray17} carried out a search for narrowband radio transmitters on the Jansky Very Large Array (JVLA). The Five-hundred-meter Aperture Spherical Radio Telescope (FAST) also will observe M31 in partnership with Breakthrough Listen \citep{Li20}.  Null results from these surveys are plotted with the collective bound from single-dish Effelsberg observations of total 1.4 GHz radio emission (\citealt{Beck98}; Table~\ref{table:CollectiveObs}) in Figure~\ref{fig:M31LineDelta}.

\begin{deluxetable*}{lcccccc}
\tabletypesize{\footnotesize}
\tablecolumns{7}
\tablewidth{0pt}
\tablecaption{Summary of past and proposed individualist searches for lines covering 1.4 GHz in M31 or Virgo Cluster ellipticals\label{table:M31VirgoIndivLines}}
\tablehead{ & & \multicolumn{3}{c}{M31} & \multicolumn{2}{c}{Virgo Ellipticals} \\ \colhead{Quantity} & \colhead{Unit} & \colhead{G17 (HI)} & \colhead{G17 (LSR)} & \colhead{FAST (HSR)} & \colhead{GBT (HSR)}}
\startdata
Instrument                  & \nodata             & JVLA                 & JVLA                  & FAST                 & GBT         \\
Type                        & \nodata             & Line                 & Line                  & Line                 & Line        \\
Completed?                  & \nodata             & \checkmark           & \checkmark            &                      &             \\
Reference                   & \nodata             & G17                  & G17                   & L20                  & {I17} \\    
$\oNuMidObs$                & $\GHz$              & $1.420$              & $1.421 \endash 1.423$ & $1.4$                & $1.4$       \\
$\ithetaFWHM$ (primary)     & $\arcmin$           & $31.7$               & $31.7$                & $1.5$                & $9$         \\
$\ithetaFWHM$ (synthesized) & $\arcsec$           & $4.3 \endash 14$     & $4.3 \endash 14$      & \nodata              & \nodata     \\
$\iNAntenna$                & \nodata             & $27$                 & $27$                  & $1$                  & $1$         \\
$\oNPointSurv$              & \nodata             & $5$                  & $5$                   & $84$                 & $1$         \\
$\oNBeamPoint$              & \nodata             & $\sim 2 \times 10^5$ & $\sim 2 \times 10^5$  & $19$                 & $1$         \\
$\oDurationDatum$           & $\sec$              & $1200$               & $300$                 & $0.25$\tablenotemark{a} & $18$     \\
$\oDurationObs$             & $\sec$              & $1200$               & $300$                 & $600$                & $300$       \\
$\oBandwidthObs$            & $\Hz$               & $122$                & $15$                  & $4$                  & $3$         \\
$\oDurationBeam$            & $\sec$              & $1200$               & $300$                 & $600$                & $900$       \\
$\oBandwidthBeam$           & $\MHz$              & $1$                  & $0.125$               & $500$                & $660$       \\
${\Mean{\hNBeamGal}}$      	        & {\nodata}   & $5.3 \times 10^8$ \tablenotemark{b} & $5.3 \times 10^8$ \tablenotemark{b} & $7.8 \times 10^7$ \tablenotemark{b} & $(0.6 \endash 2.1) \times 10^{12}$\tablenotemark{c}\\
${\qSNThreshSurv}$  & \nodata             & $7$                  & $7$                   & $10$                 & $10$        \\
$\bLisoBAR$ limit           & $\Watt$             & $2.2 \times 10^{21}$ & $1.5 \times 10^{21}$  & $2.3 \times 10^{19}$ & $(2.9 \endash 3.3) \times 10^{23}$
\enddata
\tablecomments{Abbreviations: LSR --{local standard of rest}; HSR -- {high spectral resolution.}}
\tablenotetext{a}{Minimum possible time; used for observations of GJ 273 reported in \citet{Li20}.}
\tablenotetext{b}{The projected sky density $d{\Mean{\hMAggSurvGal}}/d\SkyVar$ of M31's stars is roughly $2,700\ \Msun\ {\arcsec}^{-2}$ in the center of the M31N3 and M31S3 pointings \citep{Tamm12}. {The effective} number of stars in the beam for confusion noise is calculated using this density instead of the higher densities closer to the center, since a detection was not found in any pointing.}
\tablenotetext{c}{Stellar mass of M87, M49, and M59 unresolved by GBT; total {number of} stars in galaxy is used.}
\tablereferences{G17 -- \citet{Gray17}; L20 -- \citet{Li20}{; I17 -- \citet{Isaacson17}}}
\end{deluxetable*}

\citet{Gray17} and future FAST limits especially can already rule out a wide chunk of $\bLisoBAR$--${\bzAbundnu}$ parameter space, but they are much more limited than Milky Way observations.  At the bright end, the JVLA line search and the Effelsberg{-derived} collective bound only probe ${\bzAbundnu} \ga 100 \endash 1,000\ \GHz^{-1} \mathrm{M31}^{-1}$ because they have limited bandwidth, the same problem encountered with the Milky Way.  Because M31 covers a smaller region of the sky, it is a relatively simpl{e} matter for FAST to scan all of M31 with its wide bandwidth and close this window. The collective bound is the only constraint on line broadcasts with $\bLisoBAR \la 10^{19}\ \Watt$. {Since M31 is much fainter than the Milky Way is presumed to be in L-band despite having more stars, the collective bound is about {$10$} times stronger than for our Galaxy.} With the collective bound, we can rule out the existence of hundreds of Kardashev Type I transmitters per {gigahertz} around each star in M31, for example, {a population easily built by an advanced galactic metasociety.} Additionally, confusion sets in much sooner for M31 than in the Milky Way, when there is {$\sim 0.01 \endash 10$} line broadcast per star per {gigahertz}.

\begin{deluxetable*}{lcccccc}
\tabletypesize{\footnotesize}
\tablecolumns{6}
\tablewidth{0pt}
\tablecaption{Summary of galaxy properties and observations used for collective bounds\label{table:CollectiveObs}}
\tablehead{\colhead{Quantity} & \colhead{Units} & \colhead{{MW}} & \colhead{M31} & \colhead{M87} & \colhead{M49} & \colhead{M59}}
\startdata
$\yDistance$                            & $\Mpc$            & \nodata                 & $0.783$                      & $16.65$                      & $16.40$              & $15.45$\\
${\Mean{\hMAggSurvGal}}$               & $\Msun$           & $10^{10.78}$            & $10^{11.00}$            & $10^{11.53}$                 & $10^{11.62}$         & $10^{11.08}$\\
${\Mean{\hNSurvGal}}$\tablenotemark{a} & \nodata           & $3.0 \times 10^{11}$    & $4.5 \times 10^{11}$    & $1.7 \times 10^{12}$         & $2.1 \times 10^{12}$ & $6.0 \times 10^{11}$\\ 
\hline
Instrument                              & \nodata           & (FRC)                   & Effelsberg              & VLA                          & VLA                  & VLA\\
Radio reference                         & \nodata           & C14                     & B98                     & C98                          & C98                  & C98, B11\\
$\oNuMidObs$                            & $\GHz$            & $1.3945$\tablenotemark{b} & $1.465$                 & $1.4$                        & $1.4$                & $1.4$\\
$\dFluxENuObsGal$\tablenotemark{c}      & $\Jy$             & \nodata                 & $4.6 \pm 0.4$           & $138 \pm 5$\tablenotemark{d} & $0.220 \pm 0.008$    & $-0.0004 \pm 0.00045$\\   
$\dLnu$\tablenotemark{e}                & $\Watt\ \Hz^{-1}$ & $2.6 \times 10^{21}$\tablenotemark{f} & $4.0 \times 10^{20}$    & $4.9 \times 10^{24}$         & $7.6 \times 10^{21}$ & $1.4 \times 10^{19}$\\
$\dLnu / {\Mean{\hNTimeGal}}$ & $\Watt\ \GHz^{-1} \starUnit^{-1}$ & $8.7 \times 10^{18}$  & $8.8 \times 10^{17}$    & $2.9 \times 10^{21}$         & $3.6 \times 10^{18}$ & $2.4 \times 10^{16}$\\
$\ithetaFWHM$ (primary)                 & $\arcmin$         & \nodata                 & $9.35$\tablenotemark{g} & $31$                         & $31$                 & $31$\\
$\oNPointSurv$                          & \nodata           & \nodata                 & $68$\tablenotemark{h}   & $1$                          & $1$                  & $1$\\
$\oNBeamPoint$ (primary)                & \nodata           & \nodata                 & $1$                     & $1$                          & $1$                  & $1$\\
$\oDurationObs$                         & $\sec$            & \nodata                 & $0.67$\tablenotemark{i} & $23.3$                       & $23.3$               & $23.3$\\
$\oBandwidthObs$                        & $\MHz$            & $64$\tablenotemark{b}   & $20$                    & $42$                         & $42$                 & $42$
\enddata
\tablenotetext{a}{Assumes mean stellar mass of $0.2\ \Msun$ \citep{Chabrier03}.}
\tablenotetext{b}{{Values for the CHIPASS {southern sky} continuum radio survey.}}
\tablenotetext{c}{Reported spectral flux; $\dFluxENuObsGal = \dFluenceEObsGal / (\oDurationObs \oBandwidthObs)$.}
\tablenotetext{d}{By including single-dish data, \citet{Brown11} find a larger M87 radio flux of $210\ \Jy$.  M87's radio emission includes significant contributions from large-scale lobes and jets that are outside of the bulk of the stellar mass; if ETIs trace stellar mass, artificial radio emission would be more compact.  I thus use the VLSS measurement.}
\tablenotetext{e}{Uses $2\sigma$ flux upper bound on flux.}
\tablenotetext{f}{{Value derived from the {FRC} (equation~\ref{eqn:FRC}), using a star-formation rate of $1.65\ \Msun\,\yr^{-1}$ \citep{Licquia15}.}}
\tablenotetext{g}{Angular resolution of B98 map.}
\tablenotetext{h}{Effective number of pointings for M31 calculated as $4 {\Mean{{\hMAggObsGal}}} / (\pi \ithetaFWHM^2 d{\Mean{{\hMAggSurvGal}}}/d\SkyVar)$, where $d{\Mean{{\hMAggSurvGal}}}/d\SkyVar = 6,000\ \Msun\ {\arcsec}^{-2}$ is average over half-light isophote \citep{Tamm12}.}
\tablenotetext{i}{Calculated as $\oDurationObs = [2\ikTBack / (\iAeff {\SD{\qFluxENuGal}})]^2 / \oBandwidthObs$, from given rms noise of ${\SD{\qFluxENuGal}} = 5\ \mJy\ \mathrm{beam}^{-1}$, system temperature $26\ \Kelvin$, and notional effective area of $0.5 (\pi/4) (100\ \meter)^2$ for 100-meter aperture Effelsberg dish.} 
\tablereferences{Distances: {\citet{McConnachie12}} (M31), {\citet{Kashibadze20}} (M87, M49, M59); stellar masses: \citet{Licquia15} (MW), \citet{Tamm12} (M31), \citet{Jarrett19} (M87, M49, M59); radio data: B98 \citep{Beck98}, C98 \citep{Condon98}, B11 \citep{Brown11}, C14 \citep{Calabretta14}}
\end{deluxetable*}

Overall, the basic pattern is the same as for the Milky Way limits, but the collective bound is needed to rule out a much greater region of parameter space.

\subsection{Virgo cluster ellipticals}
\label{sec:VirgoEllipticals}
Of the galaxies in the {Breakthrough Listen nearby galaxies survey \citep{Isaacson17}}, the early-type Virgo Cluster galaxies are among the most massive and the most distant, which favor collective bounds. On the other hand, they also include radio galaxies, which are {radio-bright,} and thus {their collective bounds are weakened, allowing} a more luminous broadcast population.  Three of these galaxies {are} chosen {here} as representatives: M87, one of the brightest radio sources on the sky; M49, the largest and a much quieter radio galaxy; and M59, an elliptical with no detected 1.4 GHz radio emission at all (Table~\ref{table:CollectiveObs}). All three {have been} observed with the Green Bank Telescope (GBT) during Breakthrough Listen's survey (\citealt{Isaacson17}; Table~\ref{table:M31VirgoIndivLines}). The L-band beam width is much larger than the effective radii of these galaxies, so a single pointing will cover almost all of the stellar population.\footnote{The effective radii are comparable to the beam width of FAST at 1.4 GHz, or GBT at $\sim 10\ \GHz$, however.}

The line broadcast constraints in Figure~\ref{fig:VirgoLineDelta} show how collective bounds become progressively stronger as the galaxy's observed radio flux goes down.  In fact, collective bounds already rule out a substantial chunk of M59's {EIRP-abundance} parameter space {being} probed by the GBT. This follows from how faint it is in the radio, less than $\sim 40\ \Lsun\,\GHz^{-1}$. GBT's advantages in M59 are for a ``corner'' of rare, sub-Kardashev II transmitters ($10^{24} \endash 10^{27}\ \Watt$), and ultrarare bright transmitters, where the {need to detect at least one broadcast} limits the collective bound because of the {relatively} narrow bandwidth of VLSS ($42\ \MHz$; \citealt{Condon98}). Even the advantage for ultrarare lines could be eliminated by measuring M59's diffuse emission across a few {gigahertz} of bandwidth.  Of course, the issue is not just that the collective bounds are strong for the radio-silent M59, but {that} the individualist bounds are weak because of the great distances and insensitivity to faint broadcasts.  An isotropic narrowband transmitter in the Virgo Cluster would have to be about as luminous as Proxima Centauri to be detected by GBT.

Individualist bounds are more important for M49 and especially M87, since more abundant and/or brighter transmitters would be needed to match their radio luminosities. In all three galaxies, however, confusion sets in {no later than} when there is about one line broadcast per thousand stars per {gigahertz}{, and one per million stars per {gigahertz} for lines with zero drift}. Thus{,} confusion is a legitimate concern, as these abundances are plausible in a fully populated galaxy.

\section{General expectations for the collective bound}
\label{sec:GeneralCollective}

\subsection{The Far-infrared--Radio correlation and collective bounds}
\label{sec:FRC}
With the exception of some active galactic nuclei, galaxies are radio quiet. This makes the radio band {well suited} to collective bounds, because it would not take much power for ETIs to overwhelm the natural radio emission of a galaxy.  Low radio background in fact motivated radio transmitters as possible technosignatures \citep{Cocconi59}. In the range below $\sim 10\ \GHz$, the radio emission of most galaxies, if present, is dominated by synchrotron emission from {cosmic-ray} electrons and positrons, with $\qFluxENuGal \propto \FreqVar^{-\alpha}$ where $\alpha \sim 0.7 \endash 0.8$ for most normal {SFGs} \citep{Condon92}.  These are accelerated by phenomena associated with young massive stars and thus trace star-formation rate.  In fact, most {SFGs} lie on the FRC, with {a} ratio of far-infrared to 1.4 GHz radio luminosity ratio that is constant within a factor of about two at $\yRedshift \sim 0$ \citep{Yun01}.  For a Kroupa initial mass function, \citet{Murphy11} {relate} the mass star formation rate $\gSFRGal$ and radio brightness:
\begin{equation}
\label{eqn:FRC}
\left(\frac{\tAggLnuisoGal (1.4\ \GHz)}{\Watt\,\Hz^{-1}}\right) = {1.57 \times 10^{-21}} \left(\frac{\gSFRGal}{\Msun\,\yr^{-1}}\right),
\end{equation}
although dwarf galaxies have previously been found to be radio-quieter \citep{Bell03}. SFGs also become {somewhat} brighter in radio synchrotron emission at high redshift \citep{Delhaize17}.

Additionally, the majority of SFGs at a given redshift and SFR have similar stellar masses ${\Mean{\hMAggTimeGal}}$, a relation sometimes called the ``main sequence'' of SFGs. \citet{Speagle14} find a best-fit relation that can be expressed as
\begin{equation}
\left(\frac{\gSFRGal}{\Msun\,\yr^{-1}}\right) = 10^{{-0.13}} \left(\frac{{\Mean{\hMAggTimeGal}}}{{10^{10}}\ \Msun}\right)^{0.49},
\end{equation}
{at $\yRedshift = 0$, in terms of the Kroupa initial mass function.} The main sequence and FRC can be combined to estimate the radio emission of most SFGs at $\yRedshift = 0$:
\begin{equation}
\label{eqn:RadioVsMstar}
\left(\frac{\tAggLnuisoGal (1.4\ \GHz)}{\Watt\,\Hz^{-1}}\right) = 1.17 \times 10^{21} \left(\frac{{\Mean{\hMAggTimeGal}}}{10^{10} \Msun}\right)^{0.49} .
\end{equation}

In turn, this radio luminosity can be plugged into equations~{\ref{eqn:NEstimatedCollective},} \ref{eqn:FundamentalCollectiveBox}{,} and~\ref{eqn:FundamentalCollectiveChord} to yield {a} collective bound. {The population of artificial narrowband radio transmitters in {nonactive} $\yRedshift = 0$ galaxies is limited by the FRC to}:
\begin{multline}
{\bzAbundnu (\FreqVar)} \la 0.23\ \GHz^{-1}\,\starUnit^{-1}\ \left(\frac{{\Mean{\hMAggTimeGal}}}{10^{10} \Msun}\right)^{0.49} \\
{\cdot} \left(\frac{{\zMean{\bLiso}}}{10^{20}\ \Watt}\right)  \left(\frac{\FreqVar}{1.4\ \GHz}\right)^{-\alpha} \left(\frac{\MeanGal{\hM}}{0.2\ \Msun}\right),
\end{multline}
within a factor of a few{, as long as the broadcasts are individually much fainter than $\sim \oBandwidthObs \tAggLnuisoGal(1.4\ \GHz)$}.  For comparison, present SETI surveys of nearby galaxies are expected to be sensitive to broadcasts with $\bLiso \ga 10^{19} \endash 10^{20}\ \Watt$ \citep{Isaacson17}.  The FRC automatically implies that, in the vast majority of SFGs, most stars do not have {a transmitter that bright} in the examined frequency ranges.

A large fraction of stellar mass at low redshift is found in red early-type galaxies \citep{Moffett16}.  Although these galaxies can contain residual star formation and its attendant radio emission, by and large, most of them {are radio quiet} \citep{Nyland17}. Thus, collective bounds on radio emission can be even more powerful than for {SFGs} (section~\ref{sec:VirgoEllipticals}). Large early-type galaxies typically do have radio emission, however \citep{Sabater19}.

\subsection{Prospects for the collective bound for broadcasts at higher energies and other messengers}

The collective bound only lets us say that the ETI broadcast population is fainter than the galaxy as a whole. It thus favors wavebands where the galaxy's luminosity is known to be a small fraction of the bolometric luminosity. 

\paragraph{Infrared}
{SFGs} are bright in infrared because of the presence of thermal dust emission, with some residual contribution of stellar emission at short wavelengths \citep[e.g.,][]{Silva98}. Waste heat technosignature searches effectively use the collective bound \citep{Griffith15,Garrett15}, and they have focused on the mid-infrared because the classical Dyson sphere is ``habitable'' \citep{Dyson60}. {SFGs} are fortuitously fainter in mid-infrared, but nonetheless about $\sim 10\endash 20\%$ of the emission is released at $5 \endash 20\ \um$ (though in the form of narrow bands from small dust grains; e.g., \citealt{Dale01}), limiting these types of searches {\citep{Wright14-Search}}. Quiescent galaxies, however, including most ellipticals and lenticulars{,} are deficient in dust and infrared {emission,} though not entirely bereft of it \citep{Temi07,Smith12}, allowing more stringent collective bounds {\citep{Wright14-Search}}.

\paragraph{Optical}
Although optical transmissions are a long-sought technosignature, collective bounds are of dubious value for constraining them. Starlight is peaked in near-infrared, optical, and ultraviolet, so in order for broadcasts to contribute meaningfully to a galaxy's optical luminosity, they must practically outshine all the stars in the galaxy! Optical SETI's technosignatures are viable because we are not looking at the mean emission of a target {galaxy}, but for individualist rare fluctuations -- ultranarrowband lines or ultrashort pulses that stick out from the normal emission \citep{Schwartz61,Howard04,Abeysekara16}. Optical megastructure searches instead seek abnormal extinction of the starlight by the putative structures \citep{Annis99,Zackrisson15}.

Conceivably, the collective emission could be detectable if it was beamed, probably resulting in a smooth spectral component not found in nearby galaxies. Alternatively, broadcasts clustering around a ``magic frequency'' aggregate into an unnatural emission line in the galaxy's spectrum, which may stand out as being at an unusual wavelength or by apparently dominating the galaxy's luminosity if beamed without the usual signs of intense star formation or an active nucleus. Finally, the discreteness of extremely bright optical transmitters with $\bLiso \gg 1\ \Lsun$ might be detectable from  galactic surface brightness fluctuations.

\paragraph{High energy radiation}
Collective bounds should be moderately useful for constraining broadcasts of {high-energy} radiation. {X-ray SETI has been considered theoretically a few times in the literature, although observational results are scant \citep{Corbet97,Hippke17-XRay,Lacki20-LensFlare}. The same is true for neutrino SETI \citep{Subotowicz79,Learned94,Learned08,Lacki15-yEv}, and gamma-ray SETI is even more underdeveloped (for a rare exception, see \citealt{Harris86,Harris02}). The collective bound serves as a default upper limit.}

Most galaxies are X-ray emitters, as low-mass X-ray binaries alone contribute a minimum luminosity tracing stellar mass, about $\sim 10^{-4.5}\ \Lsun{/}\Msun$ at $z = 0$ and roughly {$10$} times higher at $z \sim 2 \endash 3$ \citep{Lehmer10,Lehmer16}. Star-forming galaxies also have soft X-ray emission from their interstellar medium and hard X-rays from high-mass X-ray binaries tracing star-formation rate \citep{Lehmer10,Mineo12-ISM}. X-ray telescopes are powerful enough to detect the comparatively low X-ray luminosities of normal galaxies out to cosmological distances, so the collective bound can be applied for a large number of galaxies \citep{Lehmer12}. Although not as stringent as the radio collective bounds, X-ray collective bounds are likely to be useful, especially given how underdeveloped {high-energy} SETI is.

In principle, {gamma}-ray and neutrino emission from galaxies is faint enough that the collective bound should be very useful. Observations of nearby {SFGs} indicate that typically $\sim 10^{-5}$ {to} $10^{-4}$ of the bolometric power is released as GeV {gamma rays} \citep{Ackermann12-SFGs}. Quiescent galaxies may be even fainter in {gamma rays} \citep[for theoretical discussion, see][]{Lien12}. Our sensitivity to {gamma rays} is comparatively weak, however: about a dozen star-forming galaxies beyond the Milky Way have been detected, only four of them (the Magellanic Clouds, M31, and M33) not intense starburst galaxies \citep{Abdollahi20,Ajello20}. We thus cannot rule out that many galaxies are abnormally {gamma}-ray bright from broadcasts. {\citet{Abdollahi20} {report} a current flux detection limit of $\sim 2 \times 10^{-12}\ \erg\,\cm^{-2}\,\sec^{-1}$ from \emph{Fermi-LAT} for GeV gamma-ray sources with an $\bEnuiso \propto \FreqVar^{-1}$ spectrum. The collective bound applied to narrowband sources suggests a maximum abundance of 
\begin{equation}
\FreqVar \bzAbundnu \sim 6 \times 10^{-5}\ \starUnit^{-1} \left[\frac{\zMean{\bLiso}}{1\ \Lsun} \frac{{\Mean{\hNObsGal}}}{10^{11}} \left(\frac{\yDistanceE}{10\ \Mpc}\right)^2\right]^{-1} .
\end{equation}
} Neutrino limits are {also quite poor.} {IceCube reports the ability to detect point sources in the northern TeV-PeV neutrino sky with fluxes below $\sim 3 \times 10^{-12}\ \erg\,\cm^{-2}\,\sec^{-1}$, assuming 1:1:1 flavor ratios and {an} $\EnergyVar^{-2}$ spectrum \citep{Aartsen17-PointSources}. The resulting limit is similar:
\begin{equation}
\FreqVar \bzAbundnu \sim 9 \times 10^{-5}\ \starUnit^{-1} \left[\frac{\zMean{\bLiso}}{1\ \Lsun} \frac{{\Mean{\hNObsGal}}}{10^{11}} \left(\frac{\yDistanceE}{10\ \Mpc}\right)^2\right]^{-1} .
\end{equation}
}

Still, given that we have essentially no {gamma}-ray or neutrino SETI constraints, even weak collective bounds from upper limits can serve as a starting point. {These limits are already sufficient to eliminate the possibility that {most big nearby} galaxies {like M81} are home to ETIs that convert $\ga 0.01\%$ of available starlight into GeV {gamma rays} or TeV neutrinos.}

{
\paragraph{Gravitational waves}
Gravitational waves have occasionally been suggested as a messenger for ETI broadcasts {\citep{Hippke18-Messenger}}. The difficulty of {gravitational-wave} SETI is the sheer weakness of gravity. LIGO and other gravitational wave observatories can detect black hole mergers from hundreds of {megaparsecs away} simply because they are the {most powerful} known events in the Universe \citep{Abbott19-GWTC1}. Still, LIGO and Virgo have been able to set limits on point sources of gravitational waves with flat spectra at around 25 Hz. The most recent limits are of order $10^{-10}$ {to} $10^{-9}\ \erg\,\cm^{-2}\,\sec^{{-1}}\,\Hz^{-1}$, depending on location on the sky \citep{Abbott21-PointSources}. For a population of persistent narrowband transmitters, the collective bound gives an abundance limit of 
\begin{equation}
\FreqVar \bzAbundnu \la 0.1 \endash 0.9\ \starUnit^{-1} \left[\frac{\zMean{\bLiso}}{1\ \Lsun} \frac{{\Mean{\hNObsGal}}}{10^{11}} \left(\frac{\yDistanceE}{10\ \Mpc}\right)^2\right]^{-1} .
\end{equation}
We can therefore reject the existence of Kardashev Type III {meta}societies broadcasting all of their power in $\sim 25\ \Hz$ gravitational waves in the nearest {large} galaxies. {This still leaves the rest of the {gravitational-wave} spectrum unexplored, {however,} and allows the existence of less powerful broadcasts of gravitational waves even in the nearest galaxies.}
}

\section{Conclusion}
\label{sec:Conclusion}

I consider two general types of constraints derived from the collection of measurements we make in a SETI survey{, the individualist signal-to-noise constraint and the collective bound}. Both types of limits are subject to {a} discreteness criterion: they cannot constrain models where broadcasts, or the transmitting societies, are sufficiently rare.

The individualist approach searches for single anomalous measurements that are incompatible with a natural background, a spike rising above the background {past} some signal-to-noise threshold. Most SETI {results} employ the individualist strategy. Only a single, sufficiently bright broadcast in the sample is necessary for a detection, which makes it useful if ETIs are rare. {But if there are too many broadcasts per observation, with too narrow a fluence distribution, the signal-to-noise ratio falls because they start to overlap}, ultimately resulting in the confusion limit where no detection is possible. {The sample variance sets an upper limit of
$$\qSNMeasureObsEST \la \left(\Median{\MaxSurv{\bzNWMeasureObsIONE}} - \Mean{\bzNWMeasureObsIONE}\right) / \sqrt{\Mean{\bzNWMeasureObsITWO}}$$
to the signal-to-noise {ratio} when the broadcasts are all about equally bright. {Sample confusion occurs when this ratio falls below a threshold $\qSNThreshSurv$ for detection; in some cases, the noise of the mutually interfering broadcasts can produce a noise confusion effect.} I present calculations about this degradation, including estimates for when confusion sets in for continuum {sources,}
$$\bzAbund \ga \left[\qSNThreshSurv^2 \Mean{\hNObsGal}\right]^{-1}$$
and narrowband radio {lines,}
$$\bzAbundnu \ga \left[\qSNThreshSurv^2 \Mean{\hNObsGal} \oBandwidthObs \bzFDriftObs (1 + \zMean{\bFCoherObs})\right]^{-1}$$
$${\bzAbundnu \ga \sqrt{\oNModeObs} \left[\qSNThreshSurv \Mean{\hNObsGal} \oBandwidthObs\right]^{-1}}$$
under standard assumptions (Table~\ref{table:StandardAssumptions}) with broadcasts of equal brightness. Confusion only happens for dense populations -- about one per channel per resolution element for narrowband observations -- but \emph{a priori} we cannot rule those out. As near as the Virgo Cluster, billions of stars and all their broadcasts can be blended into one radio beam, preventing detection if $\sim 1$ in {$1000$} stars has a {gigahertz} radio ``{beacon.}'' As we approach the confusion limit, more abundant broadcasts actually \emph{worsen} the prospects for detection.}

The collective bound{,}
$${\Mean{\bzNObs} \la \dbzNObs + \frac{C^{\prime 2}}{2} \left(1 + \sqrt{1 + \frac{4}{C^{\prime 2}} \dbzNObs}\right),}$$
follows from the aggregate emission, and is the simple observation that ETI broadcasts cannot outshine the {galaxy's observed emission} itself. It is unable to make a detection but is robust to confusion and uncertainty about the form of the broadcasts.  Generally speaking, the collective bound is more suitable for distant galaxies, because it only needs a measurement of the luminosity and because confusion is more severe at larger distances. The collective bound closes the window opened by confusion: we are not in fact missing vast populations of ETIs because their broadcasts are overlapping with each other. Collective bounds are most useful in wavebands where galaxies are faint and our instruments are sensitive, especially radio but also {possibly} X-rays. {Radio collective bounds of similar strength apply to all galaxies on the {FRC}, out to cosmological distances, and are even more constraining for radio-quiet quiescent galaxies.} {They set a limit of a few hundred Kardashev Type I radio transmitters per {gigahertz} in galaxies like the Milky Way {and} about a few hundred thousand Kardashev Type II GeV gamma-ray or TeV neutrino transmitters in nearby galaxies, and {they} constrain {Local Group} Kardashev Type III gravitational wave transmitters in the band observed by LIGO.}

Only the simplest properties of the aggregate emission of ETI populations have been considered. More powerful constraints could be found by exploiting the detailed statistical properties of this emission using the broadcast distributions. The underlying discrete nature of broadcasts {affects} the flux statistics \citep{Cordes97}, with an implied presence of Poisson fluctuations and spatial and temporal correlations. An example of this kind of approach can be found in the $P(D)$ method of constraining faint cosmic radio sources by examining fluctuations of the radio background \citep{Scheuer57,Condon74,Condon12,Vernstrom14} and extended to X-rays and gamma-rays \citep{Scheuer74,Malyshev11}. Likewise, measuring stellar Poisson fluctuations serves as a means of probing the stellar content and distance of other galaxies \citep[as in][]{Tonry88,Raimondo05}. Statistical methods allow us to glimpse deeper into the ETI luminosity function than we can hope with near-future targeted searches, sweeping another layer off the ``cosmic haystack'' \citep[c.f.,][]{Wright18}.

\acknowledgments
{{As with Paper I,} {I thank the referee, who{se dedicated efforts resulted in} detailed and helpful comments on this paper.} I {also} thank the Breakthrough Listen program for their support. Funding for \emph{Breakthrough Listen} research is sponsored by the Breakthrough Prize Foundation (\url{https://breakthroughprize.org/}).  In addition, I acknowledge the use of NASA's Astrophysics Data System and arXiv for this research.}

\appendix

\section{Derivations for the mean and sample variance of the aggregate emission}
\label{sec:AppendixAggregateEmission}
As before, we assume that all the broadcasts come from a single metasociety in a single galaxy (possibly the Milky Way). Suppose the {mean aggregate emission is the sum of the mean individual signals}.\footnote{\vphantom{T}{The total energy collected in radio observations is not generally the sum of the energies of each broadcast and the background, because they interfere with each other. But it can be shown that the mean energies do add linearly, so the condition still applies.}} Suppose the collected emission adds linearly. The sample-conditionalized mean tells us the mean emission that is collected during an observation from a fixed collection of societies and broadcasts:
\begin{equation}
\CsoMean{\qMeasureObs} = \Mean{\kMeasureObs} + \zSumSocObs \aSumBcObs \Mean{\lMeasureObs} = {\Mean{\kMeasureObs}} + \zSumSocObs \aSumBcObs \iAeff \iResponseMeasureObs(\lSkyLocation) \bEmissionisoObs {\lTransmittanceRObs} \lDilutionR .
\end{equation}

{In order to calculate the mean over all samples, we start by calculating the emission from a single society. Because a society is localized, all the broadcasts have the same dilution (distance). If the instrumental response has no dependence on frequency or time within the window $\ObsLabel$ (equation~\ref{eqn:InstrumentResponse}), all the $\iResponseMeasureObs$ terms will be the same as well, depending only on the sky position of the society $\aSkyLocation$.} Now, this is not precisely true: actual observations do not have perfectly sharp bandpasses, and beam size falls with frequency, for example, but this assumption should still suffice for typical SETI observations, which involve fine channels. Then, given the society's parameter tuple $\aTuple$, we have
\begin{equation}
\Mean{\maMeasureObs | \aTuple} = \iAeff \iResponseMeasureObs(\aSkyLocation) \lTransmittanceR \lDilutionR \Mean{\aSumBcObs {\lTransmittanceRObs} \bEmissionisoObs} = \iAeff \iResponseMeasureObs(\aSkyLocation) \lDilutionR \Mean{\baNObs} \aMean{\bEmissionisoObs} {\aMean{\lTransmittanceRObs}},
\end{equation}
where the second equality follows from Campbell's formula. 

Now this quantity can be interpreted as a random variable describing the society. We can therefore calculate the total mean with another application of Campbell's formula:
\begin{equation}
\Mean{\mzMeasureObs} = \Mean{\zSumSocObs\Mean{\maMeasureObs | \aTuple}} = \int_{\aHaystack} \Mean{\maMeasureObs | \aTuple} \azDistObs (\aTuple | \zTuple) d\aTuple .
\end{equation}
Plugging in the societal distribution from equation~\ref{eqn:SocDist}, we get
\begin{equation}
\label{eqn:MeanMeasureAppendix}
\Mean{\mzMeasureObs} = \azAbund \iAeff \zMean{\bEmissionisoObs} \zMean{\baNObs} \IntegralVolumeObs \iResponseMeasureObs({\hPosition}) \lDilutionR({\hPosition}) {\zMean{\lTransmittanceRObs | \hPosition}} \frac{d{\Mean{\hNObsGal}}}{d{\hPosition}} d{\hPosition} .
\end{equation}
This assumes that $\aMean{\bEmissionisoObs} = \zMean{\bEmissionisoObs}${, $\aMean{\lTransmittanceRObs} = \zMean{\lTransmittanceRObs | \hPosition}$} and $\Mean{\baNObs} = \zMean{\baNObs}$. 

The sample variance is slightly trickier, {because of the double sum over the broadcast and societal samples. To start, we note that the societies are a Poisson point process for a metasociety of given $\zTuple$, and so Campbell's second formula applies:}
\begin{equation}
\CsoVar{\Mean{\mzMeasureObs}} = \Var{\zSumSocObs \Mean{{\aSumBcObs \Mean{\lMeasureObs}}}} = \int_{\aHaystack} {\Mean{\Biggl(\aSumBcObs\Mean{\lMeasureObs}\Biggr)^2 \Bigg| \aTuple}} {\azDistObs}(\aTuple | \zTuple) d\aTuple
\end{equation}
{While the background has variance, the mean background is a constant and does not contribute to sample variance (the background is always ``sampled''). By the definition of variance,
\begin{equation}
\label{eqn:SampleVarMidStep}
\CsoVar{\Mean{\mzMeasureObs}} = \int_{\aHaystack} \left[\Var{{\aSumBcObs\Mean{\lMeasureObs}} \middle| \aTuple} +\Mean{{\aSumBcObs\Mean{\lMeasureObs}} \middle| \aTuple}^2\right] \azDistObs(\aTuple | \zTuple) d\aTuple .
\end{equation}
For the variance term in the integral, we again note that broadcasts are a Poisson point process for a society with given $\aTuple$, and so
\begin{equation}
\Var{\aSumBcObs\Mean{\lMeasureObs} \middle| \aTuple} = \Mean{\baNObs | \aTuple} \Mean{\aMean{\lMeasureObs^2} \middle| \aTuple} = \Mean{\baNObs} \iAeff^2 \iResponseMeasureObs(\aSkyLocation)^2 \aDilutionR^2  \aMean{\bEmissionisoObs^2} {\aMean{\lTransmittanceRObs^2}} .
\end{equation}}

{A difficulty arises if the transmittance is not a simple function of position: the first term in equation~\ref{eqn:SampleVarMidStep} uses $\aMean{\lTransmittanceRObs^2}${,} while the second uses $\aMean{\lTransmittanceRObs}^2$, and unlike the amount of emission, it is not independent of position. If the transmittance is solely a function of position, and not time or frequency, then it has a single value for each society and the two quantities are equal. {This, of course,} cannot be entirely accurate, but if we are dealing with observations of moderate bandwidths in the absence of absorption lines, we can approximate the extinction as constant over that bandwidth, and $\lTransmittanceRGen = \lTransmittanceR$.} {Again a}ssuming that $\aMean{\bEmissionisoObs} = \zMean{\bEmissionisoObs}$ and $\Mean{\baNObs} = \zMean{\baNObs}$,
\begin{equation}
\label{eqn:SampleVarMeasureAppendix}
\Var{\CsoMean{\mzMeasureObs}} = \iAeff^2 {\azAbund} \zMean{\baNObs} \left[\zMean{\bEmissionisoObs^2} + \zMean{\baNObs} \zMean{\bEmissionisoObs}^2\right] \IntegralVolumeObs [\iResponseMeasureObs(\hPosition) \lDilutionR(\hPosition) {\lTransmittanceR(\hPosition)}]^2 \frac{d{\Mean{\hNObsGal}}}{d\hPosition} d\hPosition .
\end{equation}

We can define the variable
\begin{equation}
\mzDilutionWMeasureGenIN = \zSumBcGen {[}\iResponseMeasureGen{(\lSkyLocation)} {\lDilutionR} {\lTransmittanceR} {]}^n {,}
\end{equation}
{for all $n$, in which case our assumptions give us
\begin{equation}
\Mean{\mzDilutionWMeasureGenIN} = \azAbund \zMean{\baNObs} \IntegralVolumeGen [\iResponseMeasureObs(\hPosition) \lDilutionR(\hPosition) {\lTransmittanceR(\hPosition)}]^n \frac{d{\Mean{\hNObsGal}}}{d\hPosition} d\hPosition {,}
\end{equation}
Equation~\ref{eqn:MeanMeasureAppendix} reduces to 
\begin{equation}
\label{eqn:MeanMeasureAppendixY}
\Mean{\mzMeasureObs} = \iAeff \zMean{\bEmissionisoObs} \Mean{\mzDilutionWMeasureGenIONE},
\end{equation}
while equation~\ref{eqn:SampleVarMeasureAppendix} can be written as
\begin{equation}
\label{eqn:SampleVarMeasureAppendixY}
\Var{\CsoMean{\mzMeasureObs}} = \iAeff^2 \left[\zMean{\bEmissionisoObs^2} + \zMean{\baNObs} \zMean{\bEmissionisoObs}^2\right] \Mean{\mzDilutionWMeasureObsITWO} .
\end{equation}

When the broadcasts are all part of a single distant galaxy} {and extinction is negligible or constant (e.g., from a smooth foreground screen like Earth's atmosphere)}, the dilution factor of each broadcast can be assumed to be the same{, allowing the substitution}
\begin{equation}
\label{eqn:FarGalaxySubstituion}
\zMean{\bEmissionisoGen^n} \Mean{\mjjDilutionWMeasureGenIN} \approx \zMean{\bEmissionisoGen^n} \zMeanGenNULL{{\lTransmittanceR^n} \lDilutionR^n} \Mean{\bzNWMeasureGenIN} \approx \zMean{\lFluenceGen^n} \Mean{\bzNWMeasureGenIN}.
\end{equation}
{It uses an a weighted number of broadcasts,
\begin{equation}
\bzNWMeasureGenIN = \zSumBcGen [\iResponseMeasureGen(\lSkyLocation)]^n,
\end{equation}
with a mean value of 
\begin{equation}
\Mean{\bzNWMeasureGenIN} = \azAbund \zMean{{\baNGen}} \IntegralVolumeGen [\iResponseMeasureGen(\lSkyLocation)]^n {\frac{d\Mean{\hNObsGal}}{d\hPosition} d\hPosition} .
\end{equation}
}

\section{Further details on calculation of radio energy statistics}
\subsection{Noise variance in radio detection}
\label{sec:RadioNoiseVar}

This {appendix} provides a summary of the mean noise variance calculation for radio broadcasts, $\Mean{\CsoVar{\qEnergyObs}}$. We start by describing the raw amplitudes with a complex phasor, $\lAmplitudeMode = |\lAmplitudeMode| \exp(i \lPhaseMode)$ (with $i = \sqrt{-1}$). {As in section~\ref{sec:Amplitudes},} I adopt a convention that $\lEnergyMode = |\lAmplitudeMode|^2$. The $\lEnergyMode$ may be fully randomized for incoherent signals, {be} constant for coherent signals, or display correlations for partially coherent signals. {The measured energy can be regarded as the sum of an unvarying mean component that is a measure of the mean energy output of the broadcast and a fluctuation component with zero mean (section~\ref{sec:Amplitudes}):}
\begin{equation}
\lEnergyMode = {\Mean{\lEnergyMode} + \fDelta \lEnergyMode =} \iAeff \iResponseEnergyMode(\lSkyLocation) \lFluenceEMode + {\fDelta \lEnergyMode} = \iAeff \iResponseEnergyMode(\lSkyLocation) \lDilutionE {\lTransmittanceEMode} \bEisoMode + {\fDelta \lEnergyMode}.
\end{equation}
{Likewise,} the phases may be linked by a relation, but individually they have a uniform distribution over $[-\pi, \pi)$ and I generally consider them independent between broadcasts and modes. These are summed together with a background of noise with $\kAmplitudeMode = |\kAmplitudeMode| \exp(i \kPhaseMode)$. The background amplitude has a complex Gaussian distribution, with $\Mean{\kEnergyMode} = \Mean{|\kAmplitudeMode|^2} = \ikTBack$. Furthermore, the background energy has an exponential ($\chi_2^2$) distribution, with $\Var{\kEnergyMode} = \Mean{\kEnergyMode}^2$. Thus, the measured amplitude for mode $\ModeLabel$ is
\begin{equation}
\qAmplitudeMode = |\kAmplitudeMode| \exp(i \kPhaseMode) + \SumBcObs |\lAmplitudeMode| \exp(i \lPhaseMode) {= \SumBcAdjoinedObs |\lAmplitudeMode| \exp(i \lPhaseMode)} .
\end{equation}
The {adjoined sample $\bSampleGenJADJOIN = \bSampleGenJ \cup \{{\kTuple}\}$ includes the} background noise {with amplitude} $\kAmplitudeMode${, which} is, for the purposes of these calculations, a special broadcast that is always present.

{Square-law} detectors measure the total energy as $\qEnergyMode = \qAmplitudeMode {\qAmplitudeMode}^{\ast}$:
\begin{equation}
\qEnergyMode = \SumBcAdjoinedMode \lEnergyMode + \DistinctDoubleSumBcAdjoinedMode |\lAmplitudeModeOne| |\lAmplitudeModeTwo| \cos(\lPhaseModeOne - \lPhaseModeTwo) . 
\end{equation}
{Then,} the energy from several modes can be added together to yield the energy measured by the observation. Let $\lEnergyObs = \SumModeInObs \lEnergyMode$:
\begin{equation}
\label{eqn:RadioEoFull}
\qEnergyObs = \SumBcAdjoinedObs \lEnergyObs + \SumModeInObs \DistinctDoubleSumBcAdjoinedObs |\lAmplitudeModeOne| |\lAmplitudeModeTwo| \cos(\lPhaseModeOne - \lPhaseModeTwo) . 
\end{equation}

The broadcasts in a sample have independent phases, giving us $\CsoMean{\cos(\lPhaseModeOne - \lPhaseModeTwo)} = 0$ if $\BcOneMark \ne \BcTwoMark$. Since $\CsoMean{\kEnergyObs} = \Mean{\kEnergyObs} = \oNModeObs \ikTBack$, $\CsoMean{\lEnergyObs} = {\Mean{\lEnergyObs} =} \iAeff \iResponseEnergyObs({\lSkyLocation}) {\lDilutionE {\lTransmittanceEObs} \bEisoObs}$, and the broadcasts have independent phases,
\begin{equation}
\CsoMean{\qEnergyObs} = \CsoMean{\kEnergyObs} + \SumBcObs \CsoMean{\lEnergyObs} = \oNModeObs \ikTBack + \SumBcObs \iAeff \iResponseEnergyObs(\lSkyLocation) {\lDilutionE {\lTransmittanceEObs} \bEisoObs},
\end{equation}
just as in equation~\ref{eqn:MeanEoRadio}. {It can then be shown that
\begin{equation}
\label{eqn:MeanFullRadio}
\Mean{\qEnergyObs} = \oNModeObs \ikTBack + \iAeff \zMean{\bEisoObs} \Mean{\mzDilutionWEnergyObsIONE} ,
\end{equation}
{when transmittance depends only on position,} just as for additive emission (equation~\ref{eqn:MeanMeasureAppendixY}).}

For the noise variance, we must calculate $\CsoVar{\qEnergyObs} = \CsoMean{{\qEnergyObs}^2} - \CsoMean{\qEnergyObs}^2$. To start,
\begin{multline}
{\qEnergyObs}^2 = \left(\SumBcAdjoinedObs \lEnergyObs\right)^2 + 2 \sum_{\bTupleONE \in \bSampleObsADJOIN} \lEnergyObsOne \SumModeInObs \sum_{\substack{\bTupleTWOONE, \bTupleTWOTWO \in \bSampleObsADJOIN \\ \bTupleTWOONE \ne \bTupleTWOTWO}} |\lAmplitudeModeTwoone| |\lAmplitudeModeTwotwo| \cos(\lPhaseModeTwoone - \lPhaseModeTwotwo) \\
+ \DoubleSumModeInObs \sum_{\substack{\bTupleONEONE, \bTupleONETWO \in \bSampleObsADJOIN \\ \bTupleONEONE \ne \bTupleONETWO}} \sum_{\substack{\bTupleTWOONE, \bTupleTWOTWO \in \bSampleObsADJOIN \\ \bTupleTWOONE \ne \bTupleTWOTWO}} |\lAmplitudeModeONEOneone| |\lAmplitudeModeONEOnetwo| |\lAmplitudeModeTWOTwoone| |\lAmplitudeModeTWOTwotwo| \cos(\lPhaseModeONEOneone - \lPhaseModeONEOnetwo) \cos(\lPhaseModeTWOTwoone - \lPhaseModeTWOTwotwo) .
\end{multline}
{Note that the final term includes in the summation {the} cases when $\ModeONELabel = \ModeTWOLabel$.} After averaging, the middle term vanishes. The final sum has nonzero terms when $\BcOneoneMark = \BcTwooneMark$ and $\BcOnetwoMark = \BcTwotwoMark$, or when $\BcOneoneMark = \BcTwotwoMark$ and $\BcOnetwoMark = \BcTwooneMark$, generally when $\ModeONELabel = \ModeTWOLabel$. {The other terms are taken to average to zero, under the assumption that the phases of the broadcasts are independent. If two different pairs of broadcasts are chosen for $\ModeONELabel$ and $\ModeTWOLabel$, then the phase differences between each pair should also be independent. If the same pair of broadcasts is chosen for $\ModeONELabel$ and $\ModeTWOLabel$, conceivably $\cos(\lPhaseModeONEOne - \lPhaseModeONETwo) = \cos(\lPhaseModeTWOOne - \lPhaseModeTWOTwo)$ if $\BcOneMark$ and $\BcTwoMark$ are {coherent and always} at the same exact frequency, but otherwise the phase differences should be scrambled by different frequencies and drift rates{, so those cases are assumed to average to zero too}.}

After proceeding through the algebra, we find
\begin{equation}
\label{eqn:RadioCsoVarEoFull}
\CsoVar{\qEnergyObs} = \SumModeInObs \DistinctDoubleSumBcAdjoinedObs \Mean{\lEnergyModeOne} \Mean{\lEnergyModeTwo} + \SumBcAdjoinedObs \DoubleSumModeInObs \Cov{\lEnergyModeONE, \lEnergyModeTWO} {.}
\end{equation}
{The covariance term contains all the information about the coherence of the signal.} {It is given by
\begin{equation}
\label{eqn:ModeCovariance}
\Cov{\lEnergyModeONE, \lEnergyModeTWO} = \Mean{\fDelta \lEnergyModeONE \fDelta \lEnergyModeTWO} - \Mean{\fDelta \lEnergyModeONE} \Mean{\fDelta \lEnergyModeTWO}.
\end{equation}
This property of the fluctuations is described by the $\bGTwo$ function (Appendix~\ref{sec:CoherentRadioVar}).}

The noise variance is the average of this quantity over all possible samples. Under the assumption that {the properties of the broadcast population and noise do not vary with mode,}
\begin{equation}
\Mean{\CsoVar{\qEnergyObs}} = \oNModeObs \left[\left(\ikTBack + \iAeff \Mean{\mzDilutionWEnergyModeIONE} \zMean{\bEisoMode}\right)^2 + \iAeff^2 \zMean{\baNMode} \Mean{\mzDilutionWEnergyModeITWO} \zMean{\bEisoMode}^2\right] + \zMean{\SumBcObs \DoubleSumModeInObs \Cov{{\lEnergyModeONE, \lEnergyModeTWO}}} 
\end{equation}
{for extinction that depends only on position.} {The average over covariances does \emph{not} include the background noise (i.e., the sample is not adjoined), which has already been accounted for under the assumption that it is incoherent.}

\subsection{Incoherent radio broadcasts}
\label{sec:IncoherentRadioVar}

Incoherent sources have complex Gaussian amplitudes, and their sum {is also} a complex Gaussian. {Thus,} we generally expect {the energy to be exponentially distributed, with} $\CsoVar{\qEnergyMode} = \CsoMean{\qEnergyMode}^2$. Furthermore, the noise should be independent between modes, so that $\CsoVar{\qEnergyObs} = \oNModeObs \CsoVar{\qEnergyMode}$. These conclusions are borne out from equation~\ref{eqn:RadioCsoVarEoFull}, if we adopt\footnote{Strictly speaking, ${\bGTwoPol}(|\oTStartModeONE -  \oTStartModeTWO|) - 1$ is not exactly zero for {nonzero} time delays -- its form is related to the Fourier transform of the (channel) bandpass-filtered signal \citep{Tan17}. Polyphase filterbanks use samples from several coarsely channelized spectra separated in time to suppress sidebands \citep{Price16}, so we do expect nonzero covariance. However, the covariance should be small, decreasing rapidly as the delay grows past $\oDurationMode$.}
\begin{equation}
\label{eqn:IncoherentCovariance}
{\Cov{\lEnergyModeONE, \lEnergyModeTWO}} = \begin{cases} 
				\Mean{\lEnergyMode}^2 & \text{if}~\ModeLabel = \ModeONELabel = \ModeTWOLabel \\
        0                     & \text{if}~\ModeLabel = \ModeONELabel \ne \ModeTWOLabel
	\end{cases} .
\end{equation}

According to equation~\ref{eqn:IncoherentCovariance}, the only nonzero covariance terms are the variance terms, so the noise variance is
\begin{equation}
\Mean{\CsoVar{\qEnergyObs}} = \oNModeObs \left[\left(\ikTBack + \iAeff \Mean{\mzDilutionWEnergyModeIONE} \zMean{\bEisoMode}\right)^2 + \iAeff^2 \Mean{\mzDilutionWEnergyModeITWO} \left(\zMean{\bEisoMode^2} + \zMean{\baNMode} \zMean{\bEisoMode}^2\right) \right]  .
\end{equation}

When expanded, there are five terms in the noise variance: (1) a constant term from system noise and natural background emission; (2) a term linear in $\Mean{\mzDilutionWEnergyModeIONE} \zMean{\bEisoMode}$, describing interference between the noise and the broadcasts; (3) a $\Mean{\mzDilutionWEnergyModeIONE}^2 \zMean{\bEisoMode}^2$ term describing the interference between the broadcasts among all societies; (4) a $\Mean{\mzDilutionWEnergyModeITWO} \zMean{\baNMode} \zMean{\bEisoMode}^{{2}}$ term for the interference between different broadcasts from the same society; and (5) a $\Mean{\mzDilutionWEnergyModeITWO} \zMean{\bEisoMode^2}$ term describing the wave noise from self-interference. {The last {term} is the mean self-noise of an individual broadcast \citep[e.g.,][]{Kulkarni89}, a noise that is intrinsic to the source even in the absence of all other noise. Terms three and four might also be included as ``self-noise'' of the broadcast population as a whole.}

\begin{figure*}
\centerline{\includegraphics[width=16cm]{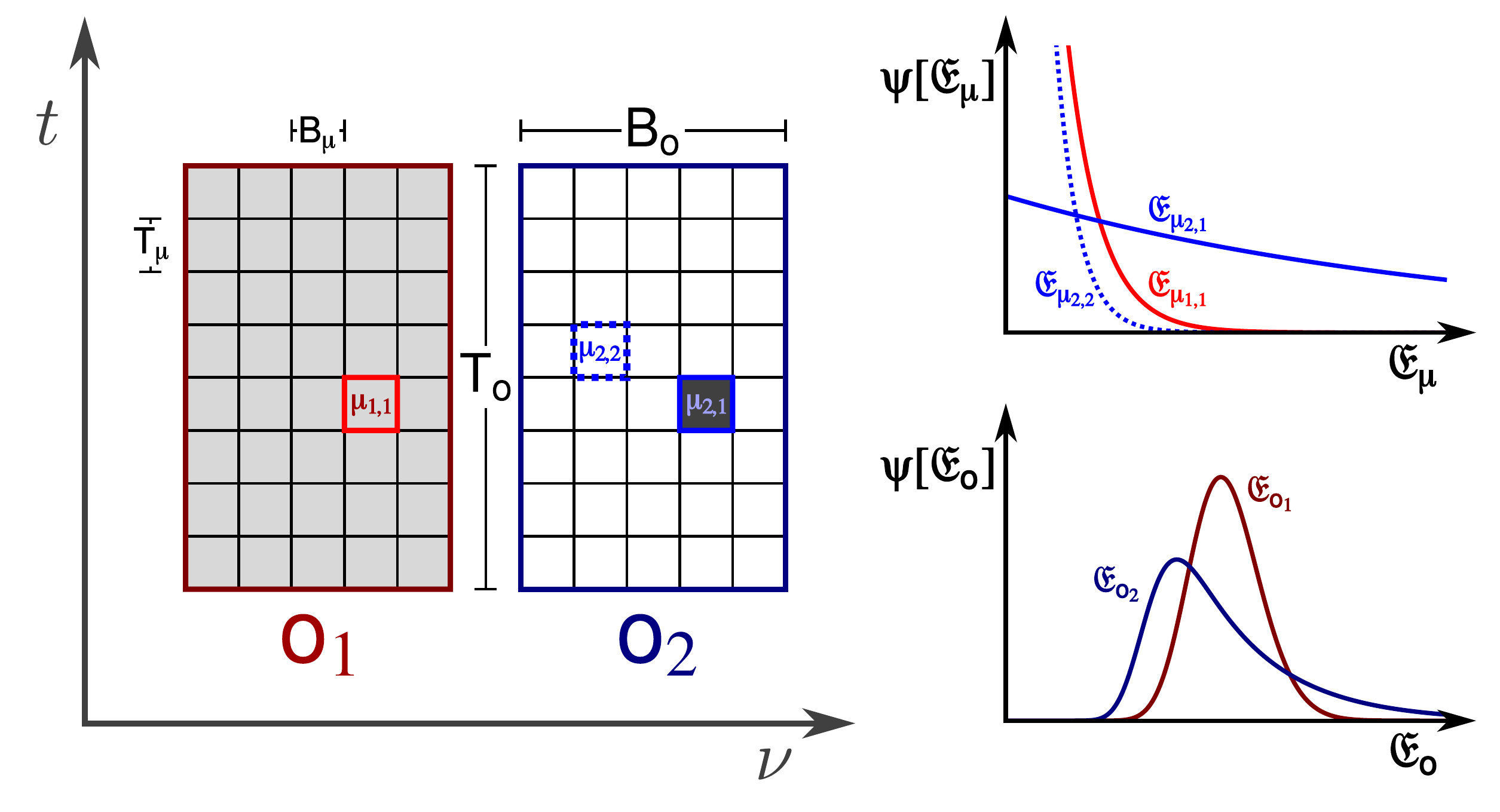}}
\figcaption{Radio observations ($\ObsONELabel$ and $\ObsTWOLabel$) measure collected energy by summing over the power in a number of modes {(amplitudes)}. {A spectrogram on the left shows two broadcasts, one for each observation.} The mean total energy {intercepted from the shown broadcasts {is}} the same in $\ObsONELabel$ and $\ObsTWOLabel$, but although the {$\ObsONELabel$} broadcast's energy is spread out among all cells (light shading), {the} $\ObsTWOLabel$ {broadcast's} energy is concentrated in one cell (dark shading) and the other cells only have background (white). {On the right, the resulting probability densities for the energy per mode (top) and energy per observation (bottom) {are shown}. T}he noise variance of ${\qEnergyObsTWO}$ is greater than that of ${\qEnergyObsTWO}$ despite having the same mean{, a result of the heavier PDF tail}.\label{fig:CellVariance}}
\end{figure*}

The noise variance depends on how each broadcast's energy is apportioned into modes. If the energy is clumped into a small fraction of the modes summed in an observation, the noise variance of incoherent broadcasts is increased: the $\zVar{\lEnergyMode}$ for occupied cells is $\propto ({\oNModeObs})^{-2}$, while the number of occupied cells is only proportional to $\oNModeObs$ (Figure~\ref{fig:CellVariance}).

The full variance for incoherent broadcasts is
\begin{multline}
\Var{\qEnergyObs}  = \oNModeObs \left[\left(\ikTBack + \iAeff \Mean{\mzDilutionWEnergyModeIONE} \zMean{\bEisoMode}\right)^2 + \iAeff^2 \Mean{\mzDilutionWEnergyModeITWO} \left(\zMean{\bEisoMode^2} + \zMean{\baNMode} \zMean{\bEisoMode}^2\right) \right]  \\
+ \iAeff^2 \Mean{\mzDilutionWEnergyObsITWO} \left(\zMean{\bEisoObs^2} + \zMean{\baNObs} \zMean{\bEisoObs}^2\right) .
\end{multline}
For convenience and intuition's sake when calculating the effective signal-to-noise {ratio}, we can express this in terms of $\lSNEnergyObsEST$ {(equation~\ref{eqn:lSNDef})} to find the generalization of equation~\ref{eqn:VarERadioIncoherent}:
\begin{multline}
\label{eqn:VarERadioIncoherentFull}
\frac{\Var{\qEnergyObs}}{\oNModeObs (\ikTBack)^2}  =  \left(1 + \lSNEnergyObsEST \sqrt{\oNModeObs} \frac{\Mean{\mzDilutionWEnergyModeIONE}}{\zMeanObs{\lDilutionE {\lTransmittanceE}}} \frac{\zMean{\bEisoMode}}{\zMean{\bEisoObs}}\right)^2 + \lSNEnergyObsEST^2 \oNModeObs \frac{\Mean{\mzDilutionWEnergyModeITWO}}{\zMeanObs{\lDilutionE {\lTransmittanceE}}^2} \left(\frac{\zMean{\bEisoMode^2}}{\zMean{\bEisoObs}^2} + \zMean{\baNMode} \frac{\zMean{\bEisoMode}^2}{\zMean{\bEisoObs}^2}\right) \\
+ \lSNEnergyObsEST^2 \frac{\Mean{\mzDilutionWEnergyObsITWO}}{\zMeanObs{\lDilutionE {\lTransmittanceE}}^2} \left(\frac{\zMean{\bEisoObs^2}}{\zMean{\bEisoObs}^2} + \zMean{\baNObs}\right) .
\end{multline}

\subsubsection{The box model and incoherent broadcasts} 
In the box model, the modes form a contiguous window in time and frequency, with $\oNModeObs = \oNumPolObs \oDurationObs \oBandwidthObs$. The mean measured energy $\Mean{\qEnergyObs}$ can be found with equation~\ref{eqn:MeanFullRadio}. 

The full expression for the variance is complicated even with the simplifying approximations of the box model. Narrowband emission in the form of lines {is} coherent and likely to have frequency drift{; they} are treated {with} the chord model (Appendix~\ref{sec:BoxChordModelVarE}).  The box model itself is very useful for wideband broadcasts. If $\bBandwidth \gg \oBandwidthObs$, the broadcasts are necessarily incoherent. 

I scale $\zMean{\baNGen}$ and ${\Mean{\mzDilutionWEnergyGenIN}}$ in equation~\ref{eqn:VarERadioIncoherentFull} under the assumption that the observation only includes one beam, covering the same sky field in all observations. {Thus,} ${\Mean{\mzDilutionWEnergyGenIN}} = {\Mean{\mzDilutionWEnergyObsIN}} \Mean{{\bzNGen}}/\Mean{{\bzNObs}}$. The equations include means of the polarization factor $\bFPolObs \equiv \sum_{\PolVar \in \oPolSetObs} \bFPol (\PolVar; \bPolQuantity)$ and its square, with $\zMean{\bFPolObs} = \oNumPolObs/2$.

For incoherent pulses ($\bDuration \ll \oDurationObs$, $\bBandwidth \gg \oBandwidthMode$),
\begin{multline}
\label{eqn:VarRadioPulse}
\frac{\Var{\qEnergyObs}}{\oNModeObs (\ikTBack)^2} \approx \left(1 + \frac{\lSNEnergyObsEST}{\sqrt{\oNModeObs}} \frac{\Mean{\mzDilutionWEnergyObsIONE}}{\zMeanObs{\lDilutionE {\lTransmittanceE}}}\right)^2 
+ \lSNEnergyObsEST^2 \frac{\Mean{\mzDilutionWEnergyObsITWO}}{\zMeanObs{\lDilutionE {\lTransmittanceE}}^2} \left[\frac{\zMean{\bEnuiso^2}}{\zMean{\bEnuiso}^2} \frac{4 \zMean{\bFPolObs^2}}{\oNumPolObs^2} \left(1 + {\frac{\zMean{\bFPolMode^2}}{\zMean{\bFPolObs^2}}} \frac{\oNumPolObs^2 \oDurationObs}{\oNModeObs \oDurationMode}\right) \right.\\
\left. + \zMean{\baNObs} \left(1 + \frac{1}{\oNModeObs}\right)\right] .
\end{multline}

Hisses ($\bDuration \gg \oDurationObs$, $\bBandwidth \gg \oBandwidthObs$) include steady continuum sources and have
\begin{multline}
\label{eqn:VarRadioHiss}
\frac{\Var{\qEnergyObs}}{\oNModeObs (\ikTBack)^2} \approx \left(1 + \frac{\lSNEnergyObsEST}{\sqrt{\oNModeObs}} \frac{\Mean{\mzDilutionWEnergyObsIONE}}{\zMeanObs{\lDilutionE {\lTransmittanceE}}}\right)^2 
+ \lSNEnergyObsEST^2 \frac{\Mean{\mzDilutionWEnergyObsITWO}}{\zMeanObs{\lDilutionE {\lTransmittanceE}}^2} \left[\frac{\zMean{\bLnuiso^2}}{\zMean{\bLnuiso}^2} \frac{4 \zMean{\bFPolObs^2}}{\oNumPolObs^2} \left(1 + {\frac{\zMean{\bFPolMode^2}}{\zMean{\bFPolObs^2}}} \frac{\oNumPolObs^2}{\oNModeObs}\right) \right.\\
\left.+ \zMean{\baNObs} \left(1 + \frac{1}{\oNModeObs}\right)\right] .
\end{multline}

Lines with $\bBandwidth \ll \oBandwidthObs$ necessarily have a coherence time ${\bTCoher} \gg \oBandwidthObs^{-1} \ge \oDurationMode$ (see section~\ref{sec:PartiallyCoherentRadioNoise}).\footnote{No signal can meaningfully have $\bDuration \ll \oDurationMode$ and $\bBandwidth \ll \oBandwidthMode$ simultaneously, so there is no strict blip regime for radio broadcasts.}

\subsection{Coherent radio broadcasts}
\label{sec:AppendixChord}
\label{sec:CoherentRadioVar}
The mean energy in coherent broadcasts is correlated {between different times}. This leads to a prolonged ``plateau'' in the covariance where it takes on values of order the variance. In a fully coherent broadcast, there is no variance and thus no covariance. A partially coherent broadcast can be viewed as band-limited Gaussian white noise, the sum of a continuum of small random sinusoids with slightly different frequencies, and so fluctuates with wave noise on long {timescales}.

{\subsubsection{Dedrifting, energy conservation, and the chord model}}
\label{sec:ChordConservation}
{Coherent radio broadcasts are narrowband,} {immediately suggesting the use of the chord model. The details of how we look for putative signals with incoherent dedrifting complicates this relatively simple picture, however.}

{A line drifts across $|\bDriftRate| \oDurationObs$ of bandwidth over an observation, a long-term smearing effect.} If this is greater than $\oBandwidthObs$ and $\oDurationObs/\oDurationDatum > 1$, the center of the line moves into different frequency channels in different dynamic spectra. {Summing along one channel results in the loss of power.} {T}he modes are first summed into intermediate data blocks {($\DatumLabel$), each with the same bandwidth as the channel as a whole ($\oBandwidthDatum = \oBandwidthObs$). Each data block $\DatumJLabel$ {is} then shifted in frequency by an amount $-(\oTStartDatumJ - \oTStartObs) \oDriftRateObs$, ``straightening out'' the skewed lines on the spectrogram that represent drifting signals. A broadcast that starts out at} drift rate $\bDriftRate$ \emph{in the inertial frame} {then is treated as though it has a new drift rate} $\bDriftRatePRIME = \bDriftRate - \oDriftRateObs$. {This simple translate-and-add method splits the window ``box'' into several boxes strung along a chord with $\oDriftRateObs$ in the inertial-frame (raw) spectrogram.} {As such, the chord model does not actually apply after dedrifting -- we have a ``pseudochord'' model for these observations, as opposed to the ``true chord'' model that applies when $\oDriftRateObs = 0$.}

{The amount of time that a broadcast spends in the reconstructed ``box'' in the dedrifted frame can be shorter or longer than the naive prediction based on $\bDriftRatePRIME$, a result of ``shadowing'' effects of the different inertial-frame boxes} and the chord running out of the constructed observation window and then {reentering} it. {The effects of the dedrifting can be described by a parameter
\begin{equation}
\bFDurationChordObs \equiv \frac{\zMeanObs{\bDurationObs | \bDriftRate}}{\zMeanObs{\bDurationObsCHORD | \bDriftRate = \bDriftRatePRIME}} = \zMeanObs{\bDurationObs | \bDriftRate} \frac{1 + |\bDriftRatePRIME| \oDurationObs / \oBandwidthObs}{\oDurationObs},
\end{equation}
where $\bDurationObsCHORD$ is the duration that would apply if we were working in a ``true'' chord model {with $\bDriftRate = \bDriftRatePRIME$} {and $\oDriftRateObs = 0$}.}

\begin{figure*}
\centerline{\includegraphics[width=18cm]{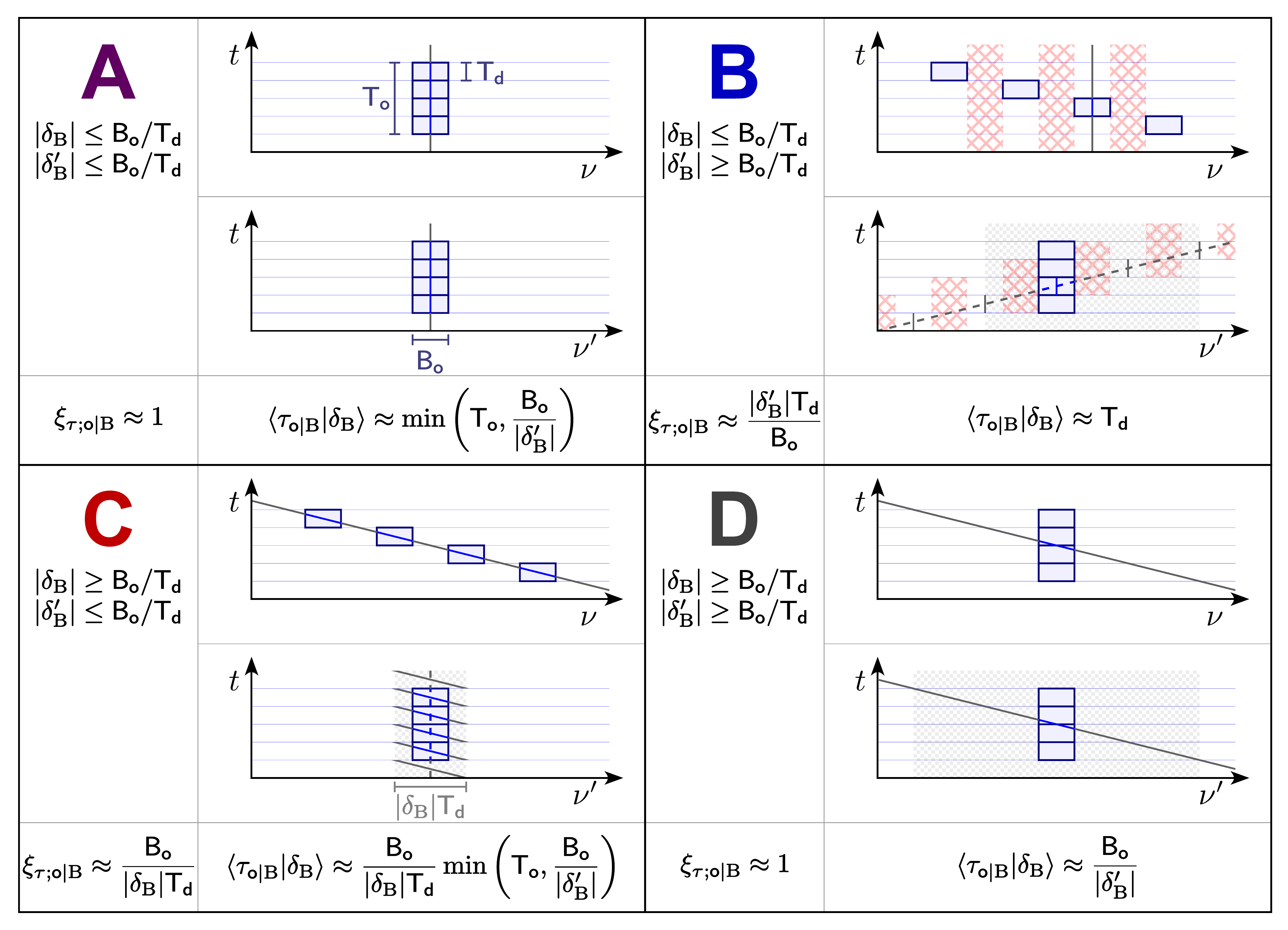}}
\figcaption{{The total energy from a narrowband broadcast sampled by a window, {as well as} the mean number of broadcasts sampled, depends on both the intrinsic drift rate $\bDriftRate$ and the ``observed'' drift rate $\bDriftRatePRIME$. Dedrifting slices up spectrograms into data bins of duration $\oDurationDatum$, applying a frequency shift linear in time, inducing an apparent drift rate $\bDriftRatePRIME$ (dotted lines). Within each data bin, however, there is no frequency shift, so the broadcast behaves as if it has the original drift rate $\bDriftRate$ (solid lines). Examples from each of the four regimes are shown. Within each regime, the observation window is constructed from data {subwindows} (blue boxes). The top spectrogram in each regime sketches the spectrogram before dedrifting is applied; the bottom {one} shows the spectrogram constructed by dedrifting. In regimes B, C, and D, the line is smeared (grey checkerboard shading). The pink hatched shading for the regime B spectrograms {shows} where a line is missed by the observation, falling through the gaps despite falling in the nominal bandwidth of the dedrifted channel. A line in one of these gaps would be picked up in one of the neighboring channels. The $\bFDurationChordObs$ factor expresses the ratio between the time the chord spends in the dedrifted observation and the time predicted in the true chord model.} \label{fig:Dedrift}}
\end{figure*}

{There are four basic regimes of behavior, as shown in Figure~\ref{fig:Dedrift}. Regime A is closest to the simple chord model, where a chord is not likely to pass straight through the bandwidth of a channel during a single data window. Regime D is also fairly simple -- in both the inertial frame and the dedrifted frame, the broadcast has high drift rate, passing through the observation window once and never returning. Regime C represents a relatively successful dedrifting of a {high-$\bDriftRate$} broadcast. A well-known issue with this dedrifting method is that the data block's duration imposes a short-term smearing in frequency: the broadcast enters and leaves the datum's bandwidth before the next datum can correct, spreading the energy over many channels. Like long-term smearing, this reduces sensitivity. Regime B, however, is not as familiar, because it represents a case where dedrifting \emph{ruins} a broadcast already with low drift rate. In this case, the data block's duration imposes a minimum ``exposure'' {that} is much longer than naively expected from $\bDriftRatePRIME$ -- a kind of apparent short-term desmearing. However, it is also possible for a broadcast ostensibly in the frequency range of the dedrifted channel to be ``missed'' as the data blocks skip over wide intervals in frequency (pink shading in the figure). Although regime B does not apply to a detection, it still comes up when considering the sample variance of the broadcast population.}

{Both the mean number and typical time the broadcast spend in an observation are affected by the dedrifting process. The product of the two is constant, a fundamental result that is needed for energy conservation. Suppose we had a population of broadcasts, all with the same drift rate and the same luminosity. The mean amount of energy intercepted per window is proportional to $\Mean{\bzNGen} \Mean{\bEisoGen} = \Mean{\bzNGen} \Mean{\bLiso} \Mean{\bFPolGen} \Mean{\bDurationGen} \propto \Mean{\bzNGen} \Mean{\bDurationGen}$. Now suppose that the instantaneous bandwidth of the observation is $\oBandwidthGen$ at all times, even though which frequencies are covered can shift from one moment to the next. W}hen averaged over all samples, the mean aggregate emission of a population is unaffected by drift rate, as long as the abundance of lines is approximately constant over {the entire bandwidth spanned by the observation}. This is a consequence of frequency translation symmetry and conservation of energy. If a larger spread of drift rate resulted in $\Mean{\mzFluenceGen}$ going down for one channel, then it should do so for the next one, and the one after that, and the one after that, and so on. That would mean that the summed energy over the bandwidth $\ga 2 \oBandwidthGen \bzDriftRateBAR$ has also decreased. But where could the energy have gone? The broadcasts are still shining as {brightly;} they merely have shifted around within that bandwidth. Additionally, for every broadcast that leaves the window, we expect one broadcast to enter on average.

Thus, as a general principle, {
\begin{equation}
\frac{d\Mean{\bzNGen}}{d\bDriftRate} \zMean{\bDurationGen | \bDriftRate} = \Mean{\bzNGen} \zPDFGen{\bDriftRate} \zMean{\bDurationGen | \bDriftRate} = \Mean{\hNGenGal} \bzAbundnu \oBandwidthGen \oDurationGen \zPDF{\bDriftRate}.
\end{equation}
Integrating over drift rate also gives us
\begin{equation}
\Mean{\bzNGen} \zMean{\bDurationGen} = \Mean{\bzNObsZD} \oDurationGen = \Mean{\hNGenGal} \bzAbundnu \oBandwidthGen \oDurationGen .
\end{equation}
} {Finally, putting the luminosity back in, a robust result of energy conservation is
\begin{equation}
\label{eqn:ChordEConservation}
\Mean{\bzNGen} \zMean{{\bEisoGen}} = \Mean{\bzNGenZD} \zMean{\bLiso} \zMean{\bFPolGen} \oDurationGen = \Mean{\bzNObsZD} \zMean{\bLiso} \frac{\oNumPolGen}{2} \frac{\oBandwidthGen}{\oBandwidthObs} \oDurationGen .
\end{equation}
This average applies also to weighted numbers, like $\Mean{\bzNWMeasureGenIN}$ and even $\Mean{\mzDilutionWEnergyGenIN}$:} {
\begin{equation}
\Mean{\mzDilutionWEnergyGenIN} \zMean{\bEisoGen} = \Mean{\mzDilutionWEnergyGenZDIN} \zMean{\bLiso} \zMean{\bFPolGen} \oDurationGen~\text{and}~\Mean{\bzNWEnergyGenIN} \zMean{\bEisoGen} =\Mean{\bzNWEnergyGenINZD} \zMean{\bLiso} \zMean{\bFPolGen} \oDurationGen,
\end{equation}
which can be seen by applying the energy conservation principle to the population in each small region of the galaxy and then adding.}

\subsubsection{Covariance for coherent broadcasts}
\label{sec:CoherentCovariance}
For the covariance, I adopt the form ({see} equation~\ref{eqn:ModeCovariance})
\begin{multline}
\label{eqn:FullCoherentCov}
{\Cov{\lEnergyModeONE, \lEnergyModeTWO}} = {[\iAeff \iResponseEnergyMode(\lSkyLocation) \lDilutionE \bLiso]^2 {\cdot} {\lTransmittanceEModeONE} \bDurationModeONE \bFPolModeONE {\cdot} {\lTransmittanceEModeTWO} \bDurationModeTWO  \bFPolModeTWO} \\
{{\cdot} \left[{\bGTwoPol} (|\oTStartModeONE - \oTStartModeTWO|) - 1\right]  {\cdot} \begin{cases}
	1         & \text{if}~\PolVar(\ModeONELabel) = \PolVar(\ModeTWOLabel)\\
	\bFPolpol & \text{if}~\PolVar(\ModeONELabel) \ne \PolVar(\ModeTWOLabel)
	\end{cases}.}
\end{multline}
The coherence function ${\bGTwoPol} (\fDelta \TimeVar) = \MeanAll{|\lAmplitude(\TimeVar)|^2 |\lAmplitude(\TimeVar + \fDelta \TimeVar)|^2} / \MeanAll{|\lAmplitude(\TimeVar)|}^2$ is well-known in quantum optics, describing the covariance between the intensity of electromagnetic radiation at two different times separated by $\fDelta \TimeVar$ {in a single polarization} \citep[e.g.,][]{Foellmi09,Tan14,Guerin17,Tan17,Hippke21}.\footnote{The averaging over the amplitudes and their product is defined to be over a sufficiently long timescale $\gg \bTCoher$, such that the amplitudes fluctuate chaotically. If the averaging is only done over $\ll \bTCoher$, the fluctuations will not be accounted for, and ${\bGTwoPol}$ will appear to be $1$ for all sampled $\fDelta \TimeVar$, just as for a perfectly coherent source.} As an approximation, I adopt a step function:
\begin{equation}
\label{eqn:gTwo}
{\bGTwoPol}(\fDelta \TimeVar) - 1 = \begin{cases}
                                 1 & \text{if}~\fDelta \TimeVar \le \bTCoher/2~\text{and not perfectly coherent}\\
									               0 & \text{if}~\fDelta \TimeVar > \bTCoher/2~\text{or perfectly coherent},
											           \end{cases}
\end{equation}
although the exact form depends on the spectrum of the broadcast \citep{Tan17} and is identically zero for a perfectly coherent radiation source. The instantaneous bandwidth $\bBandwidthTime$ of the chord is approximately $\bTCoher^{-1}$. In order for the chord model to be valid, the line needs a coherence time $\ga \oBandwidthObs^{-1}$, though this still leaves open the possibility that $\bTCoher \ll \oDurationObs$.

The polarization bears some special discussion. Coherent light is always polarized, and the polarization state remains approximately constant over time intervals shorter than the coherence timescale \citep{Hecht98}. {In this sense, coherent broadcasts are always polarized, when viewed on short enough timescales.} On {timescales longer than $\sim \bTCoher$, the polarization state may or may not fluctuate, depending on whether the fluctuations in each polarization are correlated. This is what the $\bFPolpol$ factor describes. If the broadcast is unpolarized on long {timescales} (like a thermal source with a narrow bandpass filter applied), then the fluctuations in one polarization are independent of those in the other on any delay timescale. This is clear when we consider thermal emission, where the two independent polarizations are essentially different messengers. We then set $\bFPolpol = 0$. But if the broadcast has a constant polarization on long timescales (e.g., a linear polarizer is placed in front of the transmitter, and we observe both linear polarizations), then the fluctuations are correlated. For example, if the broadcast is fully linear polarized at an angle $45^{\circ}$, the amplitudes in both polarizations will increase and decrease in tandem. For fully polarized broadcasts, $\bFPolpol = 1$.}

{Now we must sum over all the modes the broadcast crosses. The modes generally form a grid in time and frequency, with matching points in each polarization. It is clearer to sum along different polarizations first, to bring out all the polarization dependence, before summing along time and frequency. We start by picking one of the polarizations from $\oPolSetObs$, dubbed $\overline{\PolVar}$ here, and stripping the observation down to that polarization, for a new window $\ObsPolLabel(\overline{\PolVar})$. The modes in this new window still cover all the time-frequency pairs, and we can use it to define a sum over the time and frequency regions covered by the observation. Equation~\ref{eqn:FullCoherentCov} then is written as
\begin{multline}
\DoubleSumModeInObs \Cov{\lEnergyModeONE, \lEnergyModeTWO} = [\iAeff \iResponseEnergyMode(\lSkyLocation) \lDilutionE \bLiso]^2 \DoubleSumModeInObsPolOVER  {\lTransmittanceEModeONE} \bDurationModeONE {\cdot} {\lTransmittanceEModeTWO} \bDurationModeTWO \left[{\bGTwoPol} (|\oTStartModeONE - \oTStartModeTWO|) - 1\right] \\
\cdot \sum_{\PolVar_1, \PolVar_2 \in \oPolSetObs} \bFPolModeONE \bFPolModeTWO (1 \cdot \IndicatorOf{\PolVar_1 = \PolVar_2} + \bFPolpol \cdot \IndicatorOf{\PolVar_1 \ne \PolVar_2}).
\end{multline}
Remember, $\bFPolModeONE${,} $\bFPolModeTWO$, and $\bFPolpol$ describe the \emph{mean} emission of the broadcast, not the fluctuations -- they should be constants in the chord model, regardless of the instantaneous polarization state.} {The sum over polarizations can be solved explicitly, noting that $\bFPolTWO = 1 - \bFPolONE$ for orthogonal polarizations $\PolVar_1$ and $\PolVar_2$:
\begin{equation}
\bFPolpolObs \equiv \sum_{\PolVar_1, \PolVar_2 \in \oPolSetObs} \bFPolModeONE \bFPolModeTWO (1 \cdot \IndicatorOf{\PolVar_1 = \PolVar_2} + \bFPolpol \cdot \IndicatorOf{\PolVar_1 \ne \PolVar_2}) = \bFPolObs^2 - 2 \bFPolOVER (1 - \bFPolOVER) (1 - \bFPolpol) \IndicatorOf{\oNumPolObs = 1} .
\end{equation}

I write the total sum over all modes as}
\begin{equation}
\DoubleSumModeInObs \Cov{\lEnergyModeONE, \lEnergyModeTWO} = [\iAeff \iResponseEnergyMode(\lSkyLocation) {\lDilutionE} {\lTransmittanceE} {\bEisoObs}]^2 {\bFCoherObs}, 
\end{equation}
{with
\begin{equation}
\bFCoherObs \equiv \bFPolpolObs \frac{1}{\bDurationObs^2 \bFPolObs^2} \DoubleSumModeInObsPolOVER  \bDurationModeONE \bDurationModeTWO \left[{\bGTwoPol} (|\oTStartModeONE - \oTStartModeTWO|) - 1\right] .
\end{equation}
}
{As in previous sections, I assume {that} transmittance solely depends on position. All of the complications in the chord running in and out of the window while competing against the finite coherence time are stuffed into this sum. To do a full calculation for all possible cases is beyond the scope of the paper, because of all the different regimes shown in Figure~\ref{fig:Dedrift}. Its value can be calculated in some cases:}
\begin{itemize}
\item {If the broadcast is perfectly coherent, there are no fluctuations to contribute to the covariance, and thus $\bFCoherObs = 0$.}
\item {If the coherence time is much longer than $\bDurationObs$ (which may be longer than the naive $\bDurationObsCHORD$, however), then $\bGTwoPol {- 1}$ is always $1$ whenever the ``chord'' is in the observation window, regardless of how dedrifting is affecting things. Then we are simply summing all the $\bDurationMode$ twice, once for $\bDurationModeONE$ and {once for} $\bDurationModeTWO$. Within any chord-like model, $\SumModeInObs \bDurationMode = \bDurationObs$, so $\bFCoherObs = \bFPolpolObs / \bFPolObs^2$.}
\item {When $\oDriftRateObs = 0$, then we are working in the true chord model,} {and the broadcast enters and leaves the window once. The inner summation over $\ModeTWOLabel$ essentially integrates for a duration $\bTCoher$, subject to additional cutoffs when the chord {enters and} leaves the observational window:
\begin{equation}
\SumModeTWOInObsPolOVER \bDurationModeTWO [{\bGTwoPol} (|\oTStartModeONE - \oTStartModeTWO|) - 1] \approx \min(\bTStartObs + \bDurationObs, \oTStartModeONE + \bTCoher/2) - \max(\bTStartObs, \oTStartModeONE - \bTCoher/2) \approx \min(\bTCoher, \bDurationObs) .
\end{equation}
{Here $\bTStartObs$ is when the chord enters the observation window, either because the window itself begins or the chord has drifted into the window's bandpass; $\bTStartObs + \bDurationObs$ is when the chord leaves, either because the window ends or it drifts out of the frequency range.} There are two natural limits: when $\bDurationMode \ll \bTCoher \ll \bDurationObs$, this sum is usually $\sim \bTCoher$, and when $\bDurationObs \ll \bTCoher$, it is $\bDurationObs$. The sum over $\ModeONELabel$ integrates the time the chord spends in the observational window, which is $\bDurationObs$. Thus,
\begin{equation}
\bFCoherObs \approx \frac{\bFPolpolObs}{\bFPolObs^2} \min\left(1, \frac{\bTCoher}{\bDurationObs}\right).
\end{equation}}
\item {If $\bDurationMode \ll \bTCoher \ll \oDurationDatum$ for a typical ``datum'' window $\DatumLabel$ used for dedrifting, then the inner sum over $\ModeTWOLabel$ is nonzero usually within only one such datum. {Hence} we have
\begin{equation}
\bFCoherObs \approx \frac{\bFPolpolObs}{\bFPolObs^2} \min\left(1, \frac{\bTCoher}{\bDurationDatum}\right).
\end{equation}}
\end{itemize}

\subsubsection{\texorpdfstring{Variance for coherent broadcast {populations}}{Variance for coherent broadcast populations}}
\label{sec:BoxChordModelVarE}
{The full expression for the energy variance is found by taking the weighted mean sum of the covariance terms for all the broadcasts in the sample.} Applying Campbell's theorem, the average over broadcast samples for a single society is:
\begin{equation}
\Mean{\aSumBcObs \DoubleSumModeInObs {\Cov{\lEnergyModeONE, \lEnergyModeTWO}} \middle| \aTuple} \approx \Mean{\baNObs} [\iAeff \aDilutionE {\aTransmittanceE} \iResponseEnergyObs(\aSkyLocation)]^2 \aMean{{\bEisoObs^2}} {\aMean{\bFCoherObs}} \IndicatorOf{\aTuple \in \aHaystackObs} .
\end{equation}
Another application of Campbell's theorem for the societal distribution, assuming {that} societies are interchangeable, gives us the average over all samples,
\begin{equation}
\Mean{\SumBcObs \DoubleSumModeInObs {\Cov{\lEnergyModeONE, \lEnergyModeTWO}}} \approx \iAeff^2 \Mean{\mzDilutionWEnergyObsITWO} \zMean{{\bEisoObs^2}} {\zMean{\bFCoherObs}} .
\end{equation}

Finally, we arrive at the total variance:
\begin{multline}
\label{eqn:CoherentRadioFullVar}
\Var{\qEnergyObs} \approx \oNModeObs \left[\left(\ikTBack + \iAeff \Mean{\mzDilutionWEnergyModeIONE} {\zMean{\bEisoMode}}\right)^2 + \iAeff^2 \Mean{\mzDilutionWEnergyModeITWO} \zMean{\baNMode} \zMean{\bEisoMode}^2\right] \\
+ \iAeff^2 \Mean{\mzDilutionWEnergyObsITWO} \left(\zMean{\bEisoObs^2} {\left({1} + {\zMean{\bFCoherObs}}\right)} + \zMean{\baNObs} \zMean{\bEisoObs}^2\right) .
\end{multline}
All of the dependence of the variance on the drift rate distribution comes from the {term proportional to {$\zMean{\bEisoObs^2} (1 + \zMean{\bFCoherObs})$}}. {This sample variance term decreases as the spread of drift rates increases: while more broadcasts are intercepted, the amount of energy caught falls because they cross the window more quickly, and the quadratic dependence on $\bEisoObs$ is the stronger effect.}

The terms constant and linear in $\zMean{\bLiso} \zMean{\bDurationMode}$ in equation~\ref{eqn:CoherentRadioFullVar} are known \citep[as in][]{Cordes97}. One noise variance term in the incoherent expression is missing, reflecting the lack of wave noise in a perfectly coherent signal. Despite the lack of self-interference in individual broadcasts, the mutual interference when there is more than one broadcast adds an additional kind of wave noise. Whenever ${\Mean{\bzNMode}} \gg \zMean{\bLiso^2}/\zMean{\bLiso}^2$, equation~\ref{eqn:VarERadioPartCoherent} converges to the incoherent limit of equation~\ref{eqn:VarERadioIncoherent}. This {is,} of {course,} what we expect from the central limit theorem. Furthermore, the partially coherent case (${\zMean{\bFCoherObs} \sim} 1$) adds a term that effectively magnifies the sample variance, and this term can lead to much greater variance than in the incoherent case.

To understand intuitively how the variance is affected by drift rate, it is helpful to write it in terms of the expected number of broadcasts and individual broadcast signal-to-noise {ratio} if the drift rate is zero {($\lSNEnergyObsESTZD$; equation~\ref{eqn:lSNZeroDef})}. {Most of the terms simplify through the application of conservation of energy (equation~\ref{eqn:ChordEConservation}).}
\begin{multline}
\label{eqn:VarRadioChord}
\frac{\Var{\qEnergyObs}}{\oNModeObs (\ikTBack)^2} \approx \left(1 + \frac{\lSNEnergyObsESTZD}{\sqrt{\oNModeObs}} \frac{\Mean{\mzDilutionWEnergyObsZDIONE}}{\zMeanObs{\lDilutionE {\lTransmittanceE}}}\right)^2
 + \lSNEnergyObsESTZD^2 \frac{\Mean{\mzDilutionWEnergyObsZDITWO}}{\zMeanObs{\lDilutionE {\lTransmittanceE}}^2} \left[\frac{4 {\zMean{\bFPol^2}}}{\oNumPolObs^2} \frac{\zMean{\bLiso^2}}{\zMean{\bLiso}^2} {\bzFDriftObs}\right.\\
\left. \left({1 + \zMean{\bFCoherObs}}\right)  + \zMean{\baNObsZD} \left(1 + \frac{1}{\oNModeObs}\right)\right] ,
\end{multline}
{where $\bzFDriftObs \equiv \zMean{\bDurationObs^2}/[\zMean{\bDurationObs} \oDurationObs]$. Paper I presented calculations for $\zMean{\bDurationObs^2}$ in the true chord model ($\oDriftRateObs = 0$)}

\subsubsection{\texorpdfstring{{Serendipitous lines in the chord model}}{Serendipitous lines in the chord model}}
\label{sec:ChordSNFull}

Whether or not {we have found a line through a deliberate dedrifting search}, there {is} also be a background of ``serendipitous'' lines. If all lines have the same intrinsic {flux}, the ``brightest'' line will be the one that happens to be closest to being dedrifted by accident for whatever $\oDriftRateObs$ is used{, the one with the smallest $|\bDriftRatePRIME|$ if $\oDurationObs \gg \oDurationDatum$}. This can be roughly estimated by calculating the minimum drift rate {magnitude} expected in a survey. {Using the results of Paper I, the minimum drift rate magnitude in a population sampled by window $\GenLabel$ is found as
\begin{equation}
\zCDFGen{|\bDriftRatePRIME|}(\bzDriftRateLOGEN) = 1 - \frac{1}{\Mean{\bzNGen}} \ln\left[\frac{1}{2}\left(e^{\Mean{\bzNGen}} + 1\right)\right] .
\end{equation}
if $\bzNGen$ is Poissonian {(which in general requires the diffuse approximation)}. 

{Let us suppose the uniform drift rate distribution holds, with the distribution centered at $\bzDriftRateMid$ and having a width $\bzDriftRateBAR$ (equation~\ref{eqn:UniformDriftDist}). To simplify things further, assume} $|\bzDriftRateMid - \oDriftRateObs| \le \bzDriftRateBAR$, and $\bzDriftRateLOGEN \le \bzDriftRateBAR - |\bzDriftRateMid|$, {which excludes} the possibility that $\bzDriftRateLOGEN$ spills off the edges of the uniform distribution. It can be shown that
\begin{equation}
\label{eqn:LowestDriftRate}
\frac{\bzDriftRateLOGEN}{\oBandwidthGen / \oDurationGen}  = \sqrt{\left(\bzDriftNatBARGen^2 + 2 \bzDriftNatBARGen + (\bzDriftNatMidPRIMEGen)^2\right) \left(1 - \frac{1}{\Mean{\bzNGen}} \ln\left[\frac{1}{2}\left(e^{\Mean{\bzNGen}} + 1\right)\right]\right) + 1} - 1 .
\end{equation}
The {scaled} quantities $\bzDriftNatBARGen \equiv \bzDriftRateBAR \oDurationGen / \oBandwidthGen$ and $\bzDriftNatMidPRIMEGen \equiv (\bzDriftRateMid - \oDriftRateObs) \oDurationGen / \oBandwidthGen$}  express drift rates in ``natural units'' of $\oBandwidthGen/\oDurationGen$.

{To calculate the ``brightest'' expected signal from serendipitous lines, we want to consider all lines caught by all observations of the target {galaxy}. Thus, the appropriate window to use in equation~\ref{eqn:LowestDriftRate} covers each point of the sky for the entire dwell time and the entire bandwidth. Now, the bandwidth in a pointing is generally very large in a radio SETI search, at least a few hundred {kilohertz} if not many {megahertz} or {gigahertz}. The frequency window is large enough that broadcasts drift in or out only at the very edge, and ${\bzDriftNatBARSurv, |\bzDriftNatMidPRIMESurv|} \ll 1$. When $\Mean{\bzNSurv} \ga 1$, 
\begin{equation}
\bzDriftRateLOSURV \approx \frac{\bzDriftRateBAR \ln 2}{\Mean{\bzNSurv}} , 
\end{equation}
}{with one line with $|\bDriftRatePRIME| \approx \bzDriftRateLOSURV$ expected. As the number of background lines increases, the minimum drift rate magnitude falls proportionally. Eventually, we expect some to be dedrifted by chance. For a fixed $\oDriftRateObs$ with $\oDurationObs \gg \oDurationDatum$, a line is dedrifted when the long-term smearing effect is smaller than the bandwidth of a channel or the short-term smearing effect, $\bDriftRatePRIME \le \max(|\oDriftRateObs| \oDurationDatum/\oDurationObs, \oBandwidthObs / \oDurationObs)$. On average, this is achieved when}
\begin{equation}
{\Mean{\bzNSurv} \ga \frac{\bzDriftRateBAR \ln 2}{\max(|\oDriftRateObs| \oDurationDatum / \oDurationObs, \oBandwidthObs / \oDurationObs)}} {~\text{implying}~\Mean{\bzNObsZD} \ga \frac{\bzDriftNatBARObs \ln 2}{\max(|\oDriftRateObs| \oDurationDatum / \oBandwidthObs, 1)} \frac{\oBandwidthObs}{\oBandwidthSurv} \frac{\Mean{\hNObsGal}}{\Mean{\hNSurvGal}}} .
\end{equation}
The {serendipitous contribution to $\bzNObsEFF$} then proceeds through the sparse and confusion regimes for these ``brightest'' dedrifted lines.

Although lines may be sparse enough that the typical trial $\oDriftRateObs$ has no dedrifted lines, the optimal signal-to-noise is found for a specific $\oDriftRateObs$ {that} matches {the $\bDriftRate$ of any specific line.} This improvement in performance is modeled with the ${\bzIndicatorObs}$ variable set to $1$. This line, by careful {dedrifting}, achieves {an effective bandwidth of $\dbBandwidthTimeOFObsMIN \approx \max(\oBandwidthObs, |\oDriftRateObs| \oDurationDatum)$}. But if {an observation} has intercepted multiple lines{, all of equal flux}, it is unlikely that any one dedrifted line happens to coincide with the maximum fluctuations in the background of lines. Thus, the optimal signal-to-noise {ratio} depends on whether the signal-to-noise from a{n} intentionally dedrifted line rises above that of the serendipitous lines.

From these considerations, we have
\begin{equation}
{\bzNObsEFF} \approx \max\left(\frac{{\bzIndicatorObs}}{\max(1, |\oDriftRateObs|\oDurationDatum/\oBandwidthObs)}, \fDelta {\bzNObsEFF}\right)
\end{equation}
with the serendipitous contribution
{
\begin{equation}
\fDelta{\bzNObsEFF} = \begin{cases}
		   0 & \text{if}~\Median{{\bzNSurv}} = 0\\
                   \displaystyle \zMaxSurvNULL{\frac{\oBandwidthObs}{{\dbBandwidthTimeOFObs}}} \approx \min\left(\frac{\Mean{\bzNSurv} \oBandwidthObs}{\bzDriftRateBAR \oDurationObs (\ln 2)}, 1\right) & \text{if}~\Median{{\bzNSurv}} \ge 1~\text{and}~\bzDriftRateLOSURV \ge \displaystyle \max\left(|\oDriftRateObs|, \frac{\oBandwidthObs}{\oDurationObs}\right)\\
                   \frac{\displaystyle \Median{\MaxSurv{\bzNObs({|\bDriftRatePRIME| \le \dbBandwidthTimeOFObsMIN})}} - \Mean{{\bzNObs}({|\bDriftRatePRIME| \le \dbBandwidthTimeOFObsMIN})}}{\displaystyle \max(1, |\oDriftRateObs|\oDurationDatum/\oBandwidthObs)} & \text{if}~\bzDriftRateLOSURV < \displaystyle \max\left(|\oDriftRateObs|, \frac{\oBandwidthObs}{\oDurationObs}\right)
                  \end{cases}
\end{equation}
when $\oDurationObs \gg \oDurationDatum$. {When no dedrifting correction is applied ($\oDriftRateObs = 0$, $\dbBandwidthTimeOFObsMIN = \oBandwidthObs$), and $\bzDriftNatBARObs \ge 1$, it can be shown with the results of Paper I, Appendix C that
\begin{equation}
\Mean{\bzNObs(|\bDriftRatePRIME| \le \dbBandwidthTimeOFObsMIN)} = \Mean{\bzNObsZD} \frac{3}{2 \bzDriftNatBARObs} ,
\end{equation}
keeping $\fDelta \bzNObsEFF \sim 1$ well past the point {at which} a {low drift} rate population would be confused.} These effective numbers are defined relative to the brightness of a zero-drift line.}

\subsubsection{\texorpdfstring{Evaluating {sample} confusion and sensitivity in surveys {for drifting lines}}{Evaluating sample confusion and sensitivity in surveys for drifting lines}}
\label{sec:ChordSensitivityDetails}
To estimate when {sample} confusion sets in for a survey, I presuppose {$\bzIndicatorObs = 1$} and ignore the serendipitous contribution. The sample variance gives an upper limit on the signal-to-noise {ratio}. Under {the usual} assumptions (diffuse approximation, identical luminosities, two polarizations observed, distant {galaxy, $\oDurationObs \gg \oDurationDatum$}), 
\begin{equation}
{\qSNEnergyObsEST \la \frac{1}{\max(1, |\oDriftRateObs| \oDurationDatum/\oBandwidthObs)} \left[\Mean{\bzNWEnergyObsITWOZD} {\bzFDriftObs} (1 + {\zMean{\bFCoherObs}})\right]^{-1/2} .}
\end{equation}
Thus, confusion necessarily sets in for this particular dedrifted observation when
\begin{equation}
\label{eqn:ConfusionFoundLine}
\Mean{\bzNWEnergyObsITWOZD} > \left[{\qSNThreshSurv}^2 \max\left(1, \frac{|\oDriftRateObs|^2 \oDurationDatum^2}{\oBandwidthObs^2}\right) \bzFDriftObs \left(1 + {\zMean{\bFCoherObs}}\right)\right]^{-1}
\end{equation}
{Applying equation~\ref{eqn:LowestDriftRate} to the observation window, it can be shown that this roughly corresponds to there being $[\qSNThreshSurv^2 \max(1, |\oDriftRateObs| \oDurationDatum / \oBandwidthObs)]^{-1}$ serendipitously dedrifted lines per observation.}

The conventional assumption is that broadcasts are extremely rare. {Thus,} we will need to dedrift observations by an amount $\oDriftRateObs \approx \bDriftRate$, typically {by} $\sim {\zMean{|\bDriftRate|}}$, a value that may be high for broad drift rate distributions. Yet when there are many broadcasts, some will happen to have a drift rate near $0$ by chance, as long as the drift rate distribution extends to zero and the lines have linear drift.

This has two consequences. First, {sample} confusion only occurs when there are a lot of broadcasts.  Because surveys (generally) are made of many observations, there is likely to be a fortuitously dedrifted broadcast for a $\oDriftRateObs = 0$ observation well before confusion sets in. Confusion only prevents detection if it prevents detection \emph{at every tried drift rate, including those near zero}. The ``confusion limits'' on our sensitivities to broadcasts should therefore be evaluated for $\oDriftRateObs = 0$, 
\begin{equation}
\label{eqn:ConfusionBest}
{\Mean{\bzNWEnergyObsITWOZD} > \left[\qSNThreshSurv^2 \bzFDriftObs \left(1 + {\zMean{\bFCoherObs}}\right)\right]^{-1} .}
\end{equation}
Accordingly, {wider} drift rate {distribution}s actually preserve our sensitivity against confusion, because more broadcasts are summed into the background of any one observation.

Second, the {EIRP} sensitivity of a SETI survey is sensitive to smearing. For narrowband observations with incoherent dedrifting, the sensitivity loss can be significant \citep{Margot21} -- for a \emph{typical} line. But for high enough values of $\Mean{{\bzNObs}}$, a survey is expected to find lines with zero drift, for which the smearing penalty does not apply. Thus, sensitivity loss from drift is dependent on the mean number of broadcasts that are intercepted by the survey. {This is basically for the same reason that a survey is more likely to make a detection if there are many broadcasts with a broad luminosity distribution: with so many broadcasts, some will be so far on the {high-luminosity} tail that they should be detectable.}

Lines with significant curvature in their drifts are subject to more smearing, as they do not remain at zero drift for long. This problem would plague even coherent dedrifting and would require more advanced techniques that fit {higher-order} terms to the drifts. Nonetheless, the same basic points should apply if some lines lack curvature by chance.

{
\section{Derivation of photon counting variance}
\label{sec:PhotonDerivation}
This {appendix} presents a short derivation of the noise variance when counting photons from broadcasts. From equation~\ref{eqn:SampleVarMeasureAppendixY}, we find the sample variance
\begin{equation}
\Var{\CsoMean{\qPhotonObs}} = \iAeff^2 \left[\zMean{\bPhotonisoObs^2} + \zMean{\baNObs} \zMean{\bPhotonisoObs}^2\right] \Mean{\mzDilutionWMeasureGenITWO} ,
\end{equation}
where $\mzDilutionWPhotonObsITWO$ employs the photon distance $\yDistanceQ = \yDistanceM$. For the noise variance, we start by noting that the total number of collected photons is simply a linear sum, with no cross-interference:
\begin{equation}
\qPhotonObs = \kPhotonObs + \zSumSocObs \aSumBcObs \lPhotonObs .
\end{equation}
Furthermore, the broadcasts of a society are independent of each {other,} and the aggregate emission of the societies in the metasociety are also independent; all are independent of the background noise. {Hence,} it follows from independence that the individual variances add and
\begin{equation}
\CsoVar{\qPhotonObs} = \Mean{\kPhotonObs} + \zSumSocObs \aSumBcObs \Var{\lPhotonObs}.
\end{equation}
The background noise is almost certainly Poissonian, but it is possible that the variance in the photon broadcasts is not. Sub-Poissonian photon statistics are possible for artificial transmitters, while artificial modulation can greatly increase the variance of individual broadcasts. But if the broadcasts each have Poissonian photon statistics {(as unmodulated lasers do)},
\begin{equation}
\CsoVar{\qPhotonObs} = \Mean{\kPhotonObs} + \zSumSocObs \aSumBcObs \Mean{\lPhotonObs}.
\end{equation}

The noise variance is then the mean over all samples. First, {the results of Appendix~\ref{sec:AppendixAggregateEmission} apply; equation~\ref{eqn:MeanMeasureAppendix} for photons is}
\begin{equation}
\Mean{\zSumSocObs \aSumBcObs \Mean{\lPhotonObs}} = \azAbund  \iAeff \zMean{\bPhotonisoObs} \zMean{\baNObs}  \IntegralVolumeObs \iResponsePhotonObs({\hPosition}) \lDilutionQ({\hPosition}) {\lTransmittanceQ(\hPosition)} \frac{d\Mean{{\hNObsGal}}}{d{\hPosition}} d{\hPosition} ,
\end{equation}
as long as the societies are interchangeable {and extinction depends only on position}. {S}ome algebra gives us
\begin{equation}
\label{eqn:FullPhotonVar}
\Var{\qPhotonObs} = \Mean{\kPhotonObs} + \iAeff \zMean{\bPhotonisoObs} \Mean{\mzDilutionWPhotonObsIONE} + \iAeff^2 \left[ \zMean{\bPhotoniso^2} + \zMean{\baNObs} \zMean{\bPhotonisoObs}^2\right] \Mean{\mzDilutionWPhotonObsITWO} ,
\end{equation}
which can also be expressed as
\begin{equation}
\label{eqn:FullPhotonVarAlt}
\frac{\Var{\qPhotonObs}}{\Mean{\kPhotonObs}} = 1 + \frac{\lSNPhotonObsEST}{\sqrt{\Mean{\kPhotonObs}}} \frac{\Mean{\mzDilutionWPhotonObsIONE}}{\zMean{\lDilutionQ}} + \lSNPhotonObsEST^2 \left(\frac{\zMean{\bPhotonisoObs^2}}{\zMean{\bPhotonisoObs}^2} + \zMean{\baNObs}\right) \frac{\Mean{\mzDilutionWPhotonObsITWO}}{\zMean{\lDilutionQ}^2} 
\end{equation}
for convenience when calculating signal-to-noise {ratio}.
}

\section{Maximum of many Poisson random variables}
\label{sec:MaxPoisson}

Suppose we have a fixed number ${\oNObs}$ of independent samples, each containing ${\jjN}$ events, and we would like to estimate the maximum ${\jjN}$ among those ${\oNObs}$ samples, ${\Max{\jjN}}$. The median $\Median{{\Max{\jjN}}}$ provides a typical estimate. Each ${\jjN}$ has a Poisson distribution with a mean $\Mean{{\jjN}}$.  In particular, for {equally bright} broadcasts and insignificant background, the maximum $\qSN$ achieved is determined by the difference between the maximum number of broadcasts intercepted by an instrument and the mean. 

The cumulative mass function (CMF) of the ${\jjN}$ is given by the regularized upper incomplete gamma function, $\fQgammaReg(n + 1, \Mean{{\jjN}})$.  Then, by extreme value theory (\citealt{Gumbel58}{; \citealt{Castillo05}}; see Paper I), the maximum of ${\oNObs}$ independent realizations has $\fpP(n \le \Mean{{\jjN}}) = \lfloor \fQgammaReg(n + 1, \Mean{{\jjN}}) \rfloor^{{\oNObs}}$. There is no analytic expression for $\Mean{{\Max{\jjN}}}$ that I am aware of, but the median is defined by
\begin{equation}
\label{eqn:NBMaxObs}
\left\lfloor \fQgammaReg({\Median{\Max{\jjN}}} + 1, \Mean{{\jjN}}) \right\rfloor = e^{-(\ln 2)/{\oNObs}} \approx 1 - \frac{\ln 2}{{\oNObs}} ,
\end{equation}
with the approximation becoming more precise when ${\oNObs} \gg 1$.   

The Poisson distribution function is discrete, leading to jumps in the CMF. {Note also} that $\fQgammaReg(n + 1, \Mean{{\jjN}})$ includes the probability that ${\jjN} = \lfloor n \rfloor$. I use a continuous probability distribution that treats ${\jjN}$ as a continuous quantity.

\subsection{\texorpdfstring{Sparse limit: $\Mean{{N^\mathrm{J}}} \ll 1$}{Sparse limit: <N> << 1}}
\begin{figure}
\centerline{\includegraphics[width=8.5cm]{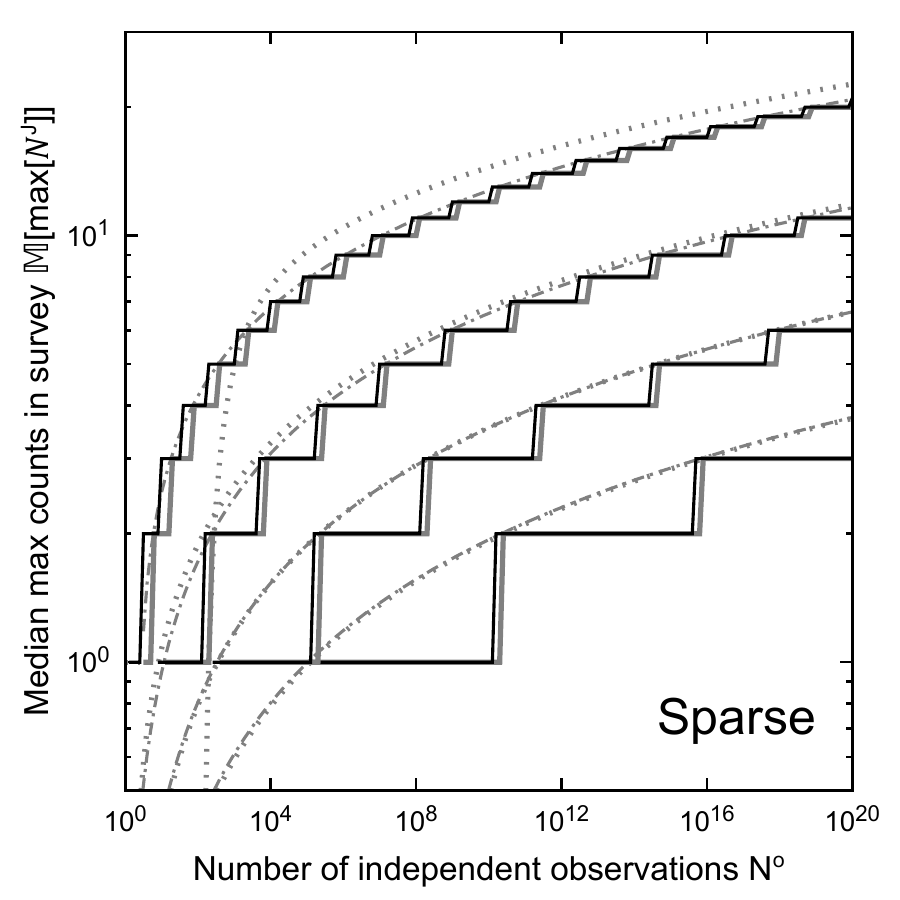}\includegraphics[width=8.5cm]{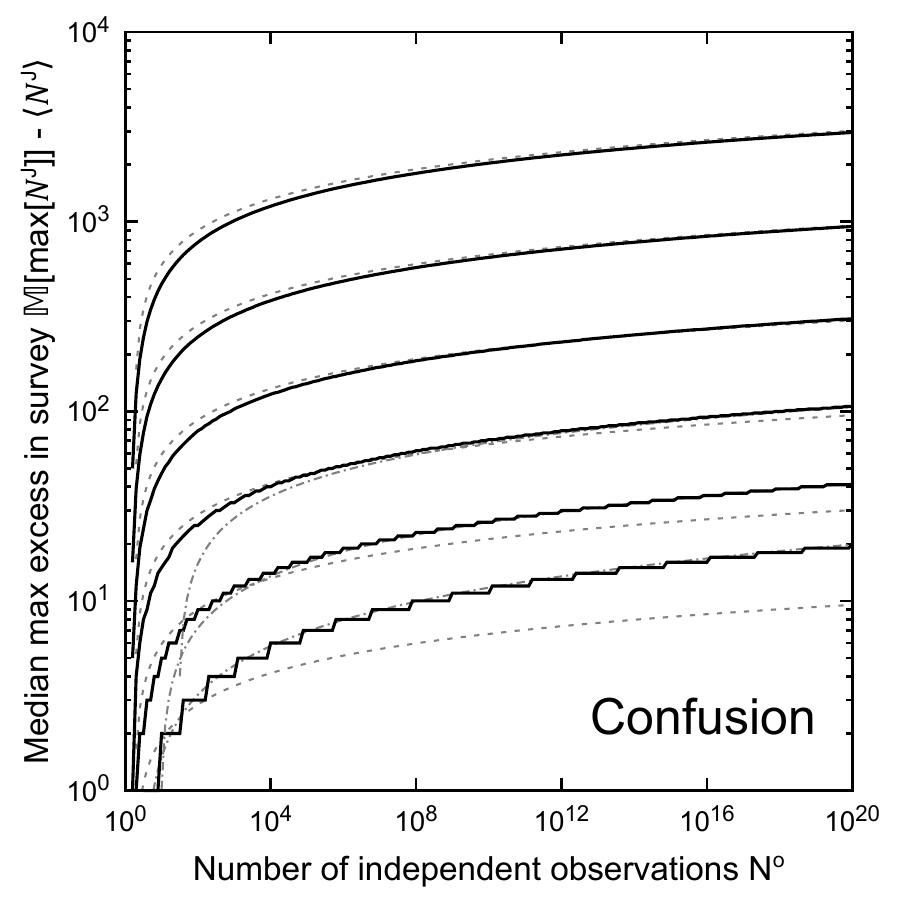}}
\figcaption{Approximations to ${\Median{\Max{\jjN}}}$ in the sparse (left) and confusion (right) regimes.  On left, the numerically computed value (solid black; equation~\ref{eqn:NBMaxObs}) is compared to estimates ({gray}): equation~\ref{eqn:SparseNBMaxIterable} ({gray} solid; without rounding down, {dashed-dotted}) and equation~\ref{eqn:SparseNBMaxMain} without rounding down (grey, dotted). On right, the numerical result (solid black) compared to equation~\ref{eqn:ConfusionNBMaxSimple} ({gray} dashed) and equation~\ref{eqn:SparseNBMaxMain} without rounding ({gray dashed-dotted}). From top to bottom, $\Mean{{\jjN}} = 1, 10^{-1}, 10^{-2.5}, 10^{-5}$ on {the} left and $\Mean{{\jjN}} = 10^5, 10^4, 10^3, 10^2, 10^1, 1$ on {the} right. \label{fig:PoissonNMaxApprox}} 
\end{figure}

When the typical observation is expected to be empty, the Poisson probability falls off exponentially {with $\Mean{{\jjN}}$}.  The {CMF} starts out near $1$ for ${\jjN} = 0$, and each time ${\jjN}$ is incremented, virtually all of the remainder is eliminated. Thus, to a good approximation,
\begin{equation}
\label{eqn:QPoissonSparseApprox}
\fpP({\jjN} < n; \Mean{{\jjN}}) = \fQgammaReg(n, \Mean{{\jjN}}) \approx 1 - \frac{\Mean{{\jjN}}^n}{n!} .
\end{equation}
By taking advantage of Stirling's approximation, $n! \approx (n/e)^{n} \sqrt{2 \pi n}$, some algebra gives us 
\begin{equation}
\label{eqn:SparseNBMaxIterable}
{\Median{\Max{\jjN}}} \approx \left\lfloor \Mean{{\jjN}} \exp\left[1 + \fWLambertZero\left(\frac{1}{\Mean{{\jjN}} e} \ln \left[\frac{{\oNObs} e^{-\Mean{{\jjN}}-1}}{\ln 2 \sqrt{2\pi \Mean{{\jjN}}}} \left(\frac{{\Median{\Max{\jjN}}} + 1/2}{{\Median{\Max{\jjN}}}}\right)^{{\Median{\Max{\jjN}}}+1/2}\right]\right)\right] - 1/2 \right\rfloor
\end{equation}
using the principal branch of the Lambert $\fWLambert$ function, $\fWLambertZero$, and after rounding down because ${\jjN}$ is a discrete variable. The right-hand side depends on ${\Median{\Max{\jjN}}}$, but the dependence is fairly weak for ${\Median{\Max{\jjN}}} \ga 1$, approaching $\sqrt{e}$, giving us
\begin{equation}
\label{eqn:SparseNBMaxMain}
{\Median{\Max{\jjN}}} \approx \left\lfloor \Mean{{\jjN}} \exp\left[1 + \fWLambertZero\left(\frac{1}{\Mean{{\jjN}} e} \ln \left[\frac{{\oNObs} e^{-\Mean{{\jjN}}-1/2}}{\ln 2 \sqrt{2\pi \Mean{{\jjN}}}} \right]\right)\right] - 1/2 \right\rfloor .
\end{equation}
Equation~\ref{eqn:SparseNBMaxMain} is an excellent approximation when ${\oNObs} \gg 1$. The argument for $\fWLambertZero$ is generally quite small, but an adequate approximation is $\fWLambertZero (x) \approx \ln x + \ln \ln x {\cdot} (1/\ln x - 1)$ \citep{Roy10}.  Its argument must be $\ge -1/e$ to be valid.  Several terms are needed, however; taking $\fWLambertZero (x) \to \ln x$ leads to underestimates of ${\Median{\Max{\jjN}}}$ (Figure~\ref{fig:PoissonNMaxApprox}).

\subsection{\texorpdfstring{Confusion limit: $\Mean{{N^\mathrm{J}}} \gg 1$}{Confusion limit: <N> >> 1}}
In the confusion limit, the Poisson distribution approaches a normal distribution with mean and variance $\Mean{{\jjN}}$.  Although this approximation does not necessarily hold far out on the tails of the distribution, it suggests that
\begin{equation}
{\Median{\Max{\jjN}}} = \Mean{{\jjN}} + C \sqrt{\Mean{{\jjN}}},
\end{equation}
with $C \sim 1$, because the normal distribution falls off with each standard deviation and outliers will be rare.  

Now, for $x = a + \sqrt{2a} C^{\prime}$,
\begin{equation}
1 - \fQgammaReg(a + 1, x) \approx \frac{1}{2} \erfc(-C^{\prime}) - \frac{1}{3} \sqrt{\frac{2}{\pi a}} (1 + {C^{\prime}}^2) e^{-{C^{\prime}}^2}
\end{equation}
when $a$ is large, with $\erfc$ referring to the complementary error function \citep{Paris10}.  In this case, $a \to {\Median{\Max{\jjN}}}$ and $x \to \Mean{{\jjN}}$, which gives us $C \approx {C^{\prime}}/\sqrt{2}$. {Furthermore,} the second term on the {right-hand side} is small because ${\Median{\Max{\jjN}}} \approx \Mean{{\jjN}}$ is large by assumption, and it is suppressed rapidly as $u$ increases past $1$.  Thus,
\begin{equation}
u \approx \sqrt{2} \erf^{-1} \left[2 \left(\frac{1}{2}\right)^{1/{\oNObs}} - 1\right] .
\end{equation}
For large ${\oNObs}$, this can be written as $C^2 \approx \fWLambertZero[{\oNObs}^2 / (2 \pi (\ln 2)^2)]$.  The first term of $\fWLambertZero(x)$ gives us the relatively simple
\begin{equation}
\label{eqn:ConfusionNBMaxSimple}
C \sim \sqrt{2 \ln \frac{{\oNObs}}{\sqrt{2\pi} \ln 2}}.
\end{equation}
This is a poor approximation when $\Mean{{\jjN}} \sim 1$, but it does well for large $\Mean{{\jjN}}$ (Figure~\ref{fig:PoissonNMaxApprox}, dashed lines).

Remarkably, equation~\ref{eqn:SparseNBMaxMain} also does well in the confusion regime when ${\oNObs} \gg 1$ (dash-dotted lines), despite equation~\ref{eqn:QPoissonSparseApprox} being a poor approximation.  More accurately, $\fQgammaReg(n + 1, \Mean{{\jjN}}) \approx 1 - \Mean{{\jjN}}^{n + 1}/(n + 1)! {\cdot} C \sqrt{\Mean{{\jjN}}}$ where $C \le \sqrt{\pi/2}$ decreases with ${\oNObs}$, but the ${\Median{\Max{\jjN}}}$ only has a logarithmic dependence on the additional factor.

{
\subsection{Applicability to populations of broadcasts}
These approximations are used in this paper to estimate the fluctuations in the number of broadcasts for the signal-to-noise ratio. Broadcasts are not generally Poisson; they are clustered into societies. We can ignore this when the diffuse approximation is applicable ($\Mean{\baNObs} \ll 1$). {Then,} $\baNObs$ is basically a Bernoulli variable flagging the rare societies that are broadcasting in the $\ObsLabel$ window; the societies of a metasociety are a Poisson point process, and a Poisson sum of Bernoulli variables is itself Poisson. The approximations also apply if the number of societies is fixed: perhaps we posit a single society for a survey covering a single star that we are trying to constrain, much like the single metasociety assumption applying for a target galaxy in this paper. 

The approximations also apply in a way for the opposite limit, when $\zMean{\baNObs} \gg \Mean{\azNObs}$ and $\gg 1$, if all societies are interchangeable (with $\Mean{\baNObs} = \zMean{\baNObs}$). {Then,} there are small deviations in the number of {broadcasts} per society, but very nearly all societies have the same number. The fluctuations in the signal-to-noise {ratio} are mostly the result of the varying number of societies, with $\Median{\maxOfObsOverSurv{\bzNObs}} \approx \Median{\maxOfObsOverSurv{\azNObs}} \zMean{\baNObs}$. Note, however, that the number of effective observations is much smaller, roughly the number of {nonoverlapping} pointings -- measurements in different channels and likely even different epochs sample the same societies.
}

\bibliographystyle{aasjournal}
\bibliography{ETIPopulations_2_Collective_arXiv}

\begin{thebibliography}{}
\expandafter\ifx\csname natexlab\endcsname\relax\def\natexlab#1{#1}\fi
\providecommand{\url}[1]{\href{#1}{#1}}
\providecommand{\dodoi}[1]{doi:~\href{http://doi.org/#1}{\nolinkurl{#1}}}
\providecommand{\doeprint}[1]{\href{http://ascl.net/#1}{\nolinkurl{http://ascl.net/#1}}}
\providecommand{\doarXiv}[1]{\href{https://arxiv.org/abs/#1}{\nolinkurl{https://arxiv.org/abs/#1}}}

\bibitem[{{Aartsen} {et~al.}(2017){Aartsen}, {Abraham}, {Ackermann}, {Adams},
  {Aguilar}, {Ahlers}, {Ahrens}, {Altmann}, {Andeen}, {Anderson}, {Ansseau},
  {Anton}, {Archinger}, {Arg{\"u}elles}, {Auffenberg}, {Axani}, {Bai},
  {Barwick}, {Baum}, {Bay}, {Beatty}, {Becker Tjus}, {Becker}, {BenZvi},
  {Berley}, {Bernardini}, {Bernhard}, {Besson}, {Binder}, {Bindig}, {Bissok},
  {Blaufuss}, {Blot}, {Bohm}, {B{\"o}rner}, {Bos}, {Bose}, {B{\"o}ser},
  {Botner}, {Braun}, {Brayeur}, {Bretz}, {Bron}, {Burgman}, {Carver}, {Casier},
  {Cheung}, {Chirkin}, {Christov}, {Clark}, {Classen}, {Coenders}, {Collin},
  {Conrad}, {Cowen}, {Cross}, {Day}, {de Andr{\'e}}, {De Clercq}, {del Pino
  Rosendo}, {Dembinski}, {De Ridder}, {Desiati}, {de Vries}, {de Wasseige}, {de
  With}, {DeYoung}, {D{\'\i}az-V{\'e}lez}, {di Lorenzo}, {Dujmovic}, {Dumm},
  {Dunkman}, {Eberhardt}, {Ehrhardt}, {Eichmann}, {Eller}, {Euler}, {Evenson},
  {Fahey}, {Fazely}, {Feintzeig}, {Felde}, {Filimonov}, {Finley}, {Flis},
  {F{\"o}sig}, {Franckowiak}, {Friedman}, {Fuchs}, {Gaisser}, {Gallagher},
  {Gerhardt}, {Ghorbani}, {Giang}, {Gladstone}, {Glauch}, {Gl{\"u}senkamp},
  {Goldschmidt}, {Golup}, {Gonzalez}, {Grant}, {Griffith}, {Haack}, {Haj
  Ismail}, {Hallgren}, {Halzen}, {Hansen}, {Hansmann}, {Hanson}, {Hebecker},
  {Heereman}, {Helbing}, {Hellauer}, {Hickford}, {Hignight}, {Hill}, {Hoffman},
  {Hoffmann}, {Holzapfel}, {Hoshina}, {Huang}, {Huber}, {Hultqvist}, {In},
  {Ishihara}, {Jacobi}, {Japaridze}, {Jeong}, {Jero}, {Jones}, {Jurkovic},
  {Kappes}, {Karg}, {Karle}, {Katz}, {Kauer}, {Keivani}, {Kelley},
  {Kheirandish}, {Kim}, {Kintscher}, {Kiryluk}, {Kittler}, {Klein}, {Kohnen},
  {Koirala}, {Kolanoski}, {Konietz}, {K{\"o}pke}, {Kopper}, {Kopper},
  {Koskinen}, {Kowalski}, {Krings}, {Kroll}, {Kr{\"u}ckl}, {Kr{\"u}ger},
  {Kunnen}, {Kunwar}, {Kurahashi}, {Kuwabara}, {Labare}, {Lanfranchi},
  {Larson}, {Lauber}, {Lennarz}, {Lesiak-Bzdak}, {Leuermann}, {Lu},
  {L{\"u}nemann}, {Madsen}, {Maggi}, {Mahn}, {Mancina}, {Mandelartz},
  {Maruyama}, {Mase}, {Maunu}, {McNally}, {Meagher}, {Medici}, {Meier}, {Meli},
  {Menne}, {Merino}, {Meures}, {Miarecki}, {Mohrmann}, {Montaruli}, {Moulai},
  {Nahnhauer}, {Naumann}, {Neer}, {Niederhausen}, {Nowicki}, {Nygren},
  {Obertacke Pollmann}, {Olivas}, {O'Murchadha}, {Palczewski}, {Pandya},
  {Pankova}, {Peiffer}, {Penek}, {Pepper}, {P{\'e}rez de los Heros}, {Pieloth},
  {Pinat}, {Price}, {Przybylski}, {Quinnan}, {Raab}, {R{\"a}del}, {Rameez},
  {Rawlins}, {Reimann}, {Relethford}, {Relich}, {Resconi}, {Rhode}, {Richman},
  {Riedel}, {Robertson}, {Rongen}, {Rott}, {Ruhe}, {Ryckbosch}, {Rysewyk},
  {Sabbatini}, {Sanchez Herrera}, {Sandrock}, {Sandroos}, {Sarkar},
  {Satalecka}, {Schlunder}, {Schmidt}, {Schoenen}, {Sch{\"o}neberg},
  {Schumacher}, {Seckel}, {Seunarine}, {Soldin}, {Song}, {Spiczak}, {Spiering},
  {Stanev}, {Stasik}, {Stettner}, {Steuer}, {Stezelberger}, {Stokstad},
  {St{\"o}ssl}, {Str{\"o}m}, {Strotjohann}, {Sullivan}, {Sutherland},
  {Taavola}, {Taboada}, {Tatar}, {Tenholt}, {Ter-Antonyan}, {Terliuk},
  {Te{\v{s}}i{\'c}}, {Tilav}, {Toale}, {Tobin}, {Toscano}, {Tosi},
  {Tselengidou}, {Turcati}, {Unger}, {Usner}, {Vandenbroucke}, {van
  Eijndhoven}, {Vanheule}, {van Rossem}, {van Santen}, {Veenkamp}, {Vehring},
  {Voge}, {Vogel}, {Vraeghe}, {Walck}, {Wallace}, {Wallraff}, {Wandkowsky},
  {Weaver}, {Weiss}, {Wendt}, {Westerhoff}, {Whelan}, {Wickmann}, {Wiebe},
  {Wiebusch}, {Wille}, {Williams}, {Wills}, {Wolf}, {Wood}, {Woolsey},
  {Woschnagg}, {Xu}, {Xu}, {Xu}, {Yanez}, {Yodh}, {Yoshida}, {Zoll}, \&
  {IceCube Collaboration}}]{Aartsen17-PointSources}
{Aartsen}, M.~G., {Abraham}, K., {Ackermann}, M., {et~al.} 2017, \apj, 835,
  151, \dodoi{10.3847/1538-4357/835/2/151}

\bibitem[{{Abbott} {et~al.}(2019){Abbott}, {Abbott}, {Abbott}, {Abraham},
  {Acernese}, {Ackley}, {Adams}, {Adhikari}, {Adya}, {Affeldt}, {Agathos},
  {Agatsuma}, {Aggarwal}, {Aguiar}, {Aiello}, {Ain}, {Ajith}, {Allen},
  {Allocca}, {Aloy}, {Altin}, {Amato}, {Ananyeva}, {Anderson}, {Anderson},
  {Angelova}, {Antier}, {Appert}, {Arai}, {Araya}, {Areeda}, {Ar{\`e}ne},
  {Arnaud}, {Arun}, {Ascenzi}, {Ashton}, {Aston}, {Astone}, {Aubin}, {Aufmuth},
  {AultONeal}, {Austin}, {Avendano}, {Avila-Alvarez}, {Babak}, {Bacon},
  {Badaracco}, {Bader}, {Bae}, {Baker}, {Baldaccini}, {Ballardin}, {Ballmer},
  {Banagiri}, {Barayoga}, {Barclay}, {Barish}, {Barker}, {Barkett}, {Barnum},
  {Barone}, {Barr}, {Barsotti}, {Barsuglia}, {Barta}, {Bartlett}, {Bartos},
  {Bassiri}, {Basti}, {Bawaj}, {Bayley}, {Bazzan}, {B{\'e}csy}, {Bejger},
  {Belahcene}, {Bell}, {Beniwal}, {Berger}, {Bergmann}, {Bernuzzi}, {Bero},
  {Berry}, {Bersanetti}, {Bertolini}, {Betzwieser}, {Bhandare}, {Bidler},
  {Bilenko}, {Bilgili}, {Billingsley}, {Birch}, {Birney}, {Birnholtz},
  {Biscans}, {Biscoveanu}, {Bisht}, {Bitossi}, {Bizouard}, {Blackburn},
  {Blackman}, {Blair}, {Blair}, {Blair}, {Bloemen}, {Bode}, {Boer}, {Boetzel},
  {Bogaert}, {Bondu}, {Bonilla}, {Bonnand}, {Booker}, {Boom}, {Booth}, {Bork},
  {Boschi}, {Bose}, {Bossie}, {Bossilkov}, {Bosveld}, {Bouffanais}, {Bozzi},
  {Bradaschia}, {Brady}, {Bramley}, {Branchesi}, {Brau}, {Briant}, {Briggs},
  {Brighenti}, {Brillet}, {Brinkmann}, {Brisson}, {Brockill}, {Brooks},
  {Brown}, {Brunett}, {Buikema}, {Bulik}, {Bulten}, {Buonanno}, {Buskulic},
  {Bustamante Rosell}, {Buy}, {Byer}, {Cabero}, {Cadonati}, {Cagnoli},
  {Cahillane}, {Calder{\'o}n Bustillo}, {Callister}, {Calloni}, {Camp},
  {Campbell}, {Canepa}, {Cannon}, {Cao}, {Cao}, {Capocasa}, {Carbognani},
  {Caride}, {Carney}, {Carullo}, {Casanueva Diaz}, {Casentini}, {Caudill},
  {Cavagli{\`a}}, {Cavalier}, {Cavalieri}, {Cella}, {Cerd{\'a}-Dur{\'a}n},
  {Cerretani}, {Cesarini}, {Chaibi}, {Chakravarti}, {Chamberlin}, {Chan},
  {Chao}, {Charlton}, {Chase}, {Chassande-Mottin}, {Chatterjee}, {Chaturvedi},
  {Chatziioannou}, {Cheeseboro}, {Chen}, {Chen}, {Chen}, {Cheng}, {Cheong},
  {Chia}, {Chincarini}, {Chiummo}, {Cho}, {Cho}, {Cho}, {Christensen}, {Chu},
  {Chua}, {Chung}, {Chung}, {Ciani}, {Ciobanu}, {Ciolfi}, {Cipriano}, {Cirone},
  {Clara}, {Clark}, {Clearwater}, {Cleva}, {Cocchieri}, {Coccia}, {Cohadon},
  {Cohen}, {Colgan}, {Colleoni}, {Collette}, {Collins}, {Cominsky},
  {Constancio}, {Conti}, {Cooper}, {Corban}, {Corbitt}, {Cordero-Carri{\'o}n},
  {Corley}, {Cornish}, {Corsi}, {Cortese}, {Costa}, {Cotesta}, {Coughlin},
  {Coughlin}, {Coulon}, {Countryman}, {Couvares}, {Covas}, {Cowan}, {Coward},
  {Cowart}, {Coyne}, {Coyne}, {Creighton}, {Creighton}, {Cripe}, {Croquette},
  {Crowder}, {Cullen}, {Cumming}, {Cunningham}, {Cuoco}, {Canton}, {D{\'a}lya},
  {Danilishin}, {D'Antonio}, {Danzmann}, {Dasgupta}, {Da Silva Costa},
  {Datrier}, {Dattilo}, {Dave}, {Davier}, {Davis}, {Daw}, {DeBra},
  {Deenadayalan}, {Degallaix}, {De Laurentis}, {Del{\'e}glise}, {Del Pozzo},
  {DeMarchi}, {Demos}, {Dent}, {De Pietri}, {Derby}, {De Rosa}, {De Rossi},
  {DeSalvo}, {de Varona}, {Dhurandhar}, {D{\'\i}az}, {Dietrich}, {Di Fiore},
  {Di Giovanni}, {Di Girolamo}, {Di Lieto}, {Ding}, {Di Pace}, {Di Palma}, {Di
  Renzo}, {Dmitriev}, {Doctor}, {Donovan}, {Dooley}, {Doravari}, {Dorrington},
  {Downes}, {Drago}, {Driggers}, {Du}, {Ducoin}, {Dupej}, {Dwyer}, {Easter},
  {Edo}, {Edwards}, {Effler}, {Ehrens}, {Eichholz}, {Eikenberry}, {Eisenmann},
  {Eisenstein}, {Essick}, {Estelles}, {Estevez}, {Etienne}, {Etzel}, {Evans},
  {Evans}, {Fafone}, {Fair}, {Fairhurst}, {Fan}, {Farinon}, {Farr}, {Farr},
  {Fauchon-Jones}, {Favata}, {Fays}, {Fazio}, {Fee}, {Feicht}, {Fejer}, {Feng},
  {Fernandez-Galiana}, {Ferrante}, {Ferreira}, {Ferreira}, {Ferrini},
  {Fidecaro}, {Fiori}, {Fiorucci}, {Fishbach}, {Fisher}, {Fishner},
  {Fitz-Axen}, {Flaminio}, {Fletcher}, {Flynn}, {Fong}, {Font}, {Forsyth},
  {Fournier}, {Frasca}, {Frasconi}, {Frei}, {Freise}, {Frey}, {Frey},
  {Fritschel}, {Frolov}, {Fulda}, {Fyffe}, {Gabbard}, {Gadre}, {Gaebel},
  {Gair}, {Gammaitoni}, {Ganija}, {Gaonkar}, {Garcia},
  {Garc{\'\i}a-Quir{\'o}s}, {Garufi}, {Gateley}, {Gaudio}, {Gaur}, {Gayathri},
  {Gemme}, {Genin}, {Gennai}, {George}, {George}, {Gergely}, {Germain},
  {Ghonge}, {Ghosh}, {Ghosh}, {Ghosh}, {Giacomazzo}, {Giaime}, {Giardina},
  {Giazotto}, {Gill}, {Giordano}, {Glover}, {Godwin}, {Goetz}, {Goetz},
  {Goncharov}, {Gonz{\'a}lez}, {Gonzalez Castro}, {Gopakumar}, {Gorodetsky},
  {Gossan}, {Gosselin}, {Gouaty}, {Grado}, {Graef}, {Granata}, {Grant}, {Gras},
  {Grassia}, {Gray}, {Gray}, {Greco}, {Green}, {Green}, {Gretarsson}, {Groot},
  {Grote}, {Grunewald}, {Gruning}, {Guidi}, {Gulati}, {Guo}, {Gupta}, {Gupta},
  {Gustafson}, {Gustafson}, {Haegel}, {Halim}, {Hall}, {Hall}, {Hamilton},
  {Hammond}, {Haney}, {Hanke}, {Hanks}, {Hanna}, {Hannam}, {Hannuksela},
  {Hanson}, {Hardwick}, {Haris}, {Harms}, {Harry}, {Harry}, {Haster},
  {Haughian}, {Hayes}, {Healy}, {Heidmann}, {Heintze}, {Heitmann}, {Hello},
  {Hemming}, {Hendry}, {Heng}, {Hennig}, {Heptonstall}, {Hernandez Vivanco},
  {Heurs}, {Hild}, {Hinderer}, {Hoak}, {Hochheim}, {Hofman}, {Holgado},
  {Holland}, {Holt}, {Holz}, {Hopkins}, {Horst}, {Hough}, {Howell}, {Hoy},
  {Hreibi}, {Huang}, {Huerta}, {Huet}, {Hughey}, {Hulko}, {Husa}, {Huttner},
  {Huynh-Dinh}, {Idzkowski}, {Iess}, {Ingram}, {Inta}, {Intini}, {Irwin},
  {Isa}, {Isac}, {Isi}, {Iyer}, {Izumi}, {Jacqmin}, {Jadhav}, {Jani},
  {Janthalur}, {Jaranowski}, {Jenkins}, {Jiang}, {Johnson}, {Johnson-McDaniel},
  {Jones}, {Jones}, {Jones}, {Jonker}, {Ju}, {Junker}, {Kalaghatgi},
  {Kalogera}, {Kamai}, {Kandhasamy}, {Kang}, {Kanner}, {Kapadia}, {Karki},
  {Karvinen}, {Kashyap}, {Kasprzack}, {Katsanevas}, {Katsavounidis}, {Katzman},
  {Kaufer}, {Kawabe}, {Keerthana}, {K{\'e}f{\'e}lian}, {Keitel}, {Kennedy},
  {Key}, {Khalili}, {Khan}, {Khan}, {Khan}, {Khan}, {Khazanov}, {Khursheed},
  {Kijbunchoo}, {Kim}, {Kim}, {Kim}, {Kim}, {Kim}, {Kim}, {Kimball}, {King},
  {King}, {Kinley-Hanlon}, {Kirchhoff}, {Kissel}, {Kleybolte}, {Klika},
  {Klimenko}, {Knowles}, {Koch}, {Koehlenbeck}, {Koekoek}, {Koley},
  {Kondrashov}, {Kontos}, {Koper}, {Korobko}, {Korth}, {Kowalska}, {Kozak},
  {Kringel}, {Krishnendu}, {Kr{\'o}lak}, {Kuehn}, {Kumar}, {Kumar}, {Kumar},
  {Kumar}, {Kuo}, {Kutynia}, {Kwang}, {Lackey}, {Lai}, {Lam}, {Landry}, {Lane},
  {Lang}, {Lange}, {Lantz}, {Lanza}, {Lartaux-Vollard}, {Lasky}, {Laxen},
  {Lazzarini}, {Lazzaro}, {Leaci}, {Leavey}, {Lecoeuche}, {Lee}, {Lee}, {Lee},
  {Lee}, {Lee}, {Lee}, {Lehmann}, {Lenon}, {Leroy}, {Letendre}, {Levin}, {Li},
  {Li}, {Li}, {Li}, {Lin}, {Linde}, {Linker}, {Littenberg}, {Liu}, {Liu}, {Lo},
  {Lockerbie}, {London}, {Longo}, {Lorenzini}, {Loriette}, {Lormand},
  {Losurdo}, {Lough}, {Lousto}, {Lovelace}, {Lower}, {L{\"u}ck}, {Lumaca},
  {Lundgren}, {Lynch}, {Ma}, {Macas}, {Macfoy}, {MacInnis}, {Macleod},
  {Macquet}, {Maga{\~n}a-Sandoval}, {Maga{\~n}a Zertuche}, {Magee}, {Majorana},
  {Maksimovic}, {Malik}, {Man}, {Mandic}, {Mangano}, {Mansell}, {Manske},
  {Mantovani}, {Marchesoni}, {Marion}, {M{\'a}rka}, {M{\'a}rka}, {Markakis},
  {Markosyan}, {Markowitz}, {Maros}, {Marquina}, {Marsat}, {Martelli},
  {Martin}, {Martin}, {Martynov}, {Mason}, {Massera}, {Masserot}, {Massinger},
  {Masso-Reid}, {Mastrogiovanni}, {Matas}, {Matichard}, {Matone}, {Mavalvala},
  {Mazumder}, {McCann}, {McCarthy}, {McClelland}, {McCormick}, {McCuller},
  {McGuire}, {McIver}, {McManus}, {McRae}, {McWilliams}, {Meacher}, {Meadors},
  {Mehmet}, {Mehta}, {Meidam}, {Melatos}, {Mendell}, {Mercer}, {Mereni},
  {Merilh}, {Merzougui}, {Meshkov}, {Messenger}, {Messick}, {Metzdorff},
  {Meyers}, {Miao}, {Michel}, {Middleton}, {Mikhailov}, {Milano}, {Miller},
  {Miller}, {Millhouse}, {Mills}, {Milovich-Goff}, {Minazzoli}, {Minenkov},
  {Mishkin}, {Mishra}, {Mistry}, {Mitra}, {Mitrofanov}, {Mitselmakher},
  {Mittleman}, {Mo}, {Moffa}, {Mogushi}, {Mohapatra}, {Montani}, {Moore},
  {Moraru}, {Moreno}, {Morisaki}, {Mours}, {Mow-Lowry}, {Mukherjee},
  {Mukherjee}, {Mukherjee}, {Mukund}, {Mullavey}, {Munch}, {Mu{\~n}iz},
  {Muratore}, {Murray}, {Nagar}, {Nardecchia}, {Naticchioni}, {Nayak},
  {Neilson}, {Nelemans}, {Nelson}, {Nery}, {Neunzert}, {Ng}, {Ng}, {Nguyen},
  {Nichols}, {Nielsen}, {Nissanke}, {Nitz}, {Nocera}, {North}, {Nuttall},
  {Obergaulinger}, {Oberling}, {O'Brien}, {O'Dea}, {Ogin}, {Oh}, {Oh}, {Ohme},
  {Ohta}, {Okada}, {Oliver}, {Oppermann}, {Oram}, {O'Reilly}, {Ormiston},
  {Ortega}, {O'Shaughnessy}, {Ossokine}, {Ottaway}, {Overmier}, {Owen}, {Pace},
  {Pagano}, {Page}, {Pai}, {Pai}, {Palamos}, {Palashov}, {Palomba},
  {Pal-Singh}, {Pan}, {Pang}, {Pang}, {Pankow}, {Pannarale}, {Pant},
  {Paoletti}, {Paoli}, {Papa}, {Parida}, {Parker}, {Pascucci}, {Pasqualetti},
  {Passaquieti}, {Passuello}, {Patil}, {Patricelli}, {Pearlstone}, {Pedersen},
  {Pedraza}, {Pedurand}, {Pele}, {Penn}, {Perego}, {Perez}, {Perreca},
  {Pfeiffer}, {Phelps}, {Phukon}, {Piccinni}, {Pichot}, {Piergiovanni},
  {Pillant}, {Pinard}, {Pirello}, {Pitkin}, {Poggiani}, {Pong}, {Ponrathnam},
  {Popolizio}, {Porter}, {Powell}, {Prajapati}, {Prasad}, {Prasai}, {Prasanna},
  {Pratten}, {Prestegard}, {Privitera}, {Prodi}, {Prokhorov}, {Puncken},
  {Punturo}, {Puppo}, {P{\"u}rrer}, {Qi}, {Quetschke}, {Quinonez}, {Quintero},
  {Quitzow-James}, {Raab}, {Radkins}, {Radulescu}, {Raffai}, {Raja}, {Rajan},
  {Rajbhandari}, {Rakhmanov}, {Ramirez}, {Ramos-Buades}, {Rana}, {Rao},
  {Rapagnani}, {Raymond}, {Razzano}, {Read}, {Regimbau}, {Rei}, {Reid},
  {Reitze}, {Ren}, {Ricci}, {Richardson}, {Richardson}, {Ricker},
  {Riemenschneider}, {Riles}, {Rizzo}, {Robertson}, {Robie}, {Robinet},
  {Rocchi}, {Rolland}, {Rollins}, {Roma}, {Romanelli}, {Romano}, {Romel},
  {Romie}, {Rose}, {Rosi{\'n}ska}, {Rosofsky}, {Ross}, {Rowan}, {R{\"u}diger},
  {Ruggi}, {Rutins}, {Ryan}, {Sachdev}, {Sadecki}, {Sakellariadou}, {Salafia},
  {Salconi}, {Saleem}, {Salemi}, {Samajdar}, {Sammut}, {Sanchez}, {Sanchez},
  {Sanchis-Gual}, {Sandberg}, {Sanders}, {Santiago}, {Sarin}, {Sassolas},
  {Sathyaprakash}, {Saulson}, {Sauter}, {Savage}, {Schale}, {Scheel},
  {Scheuer}, {Schmidt}, {Schnabel}, {Schofield}, {Sch{\"o}nbeck}, {Schreiber},
  {Schulte}, {Schutz}, {Schwalbe}, {Scott}, {Scott}, {Seidel}, {Sellers},
  {Sengupta}, {Sennett}, {Sentenac}, {Sequino}, {Sergeev}, {Setyawati},
  {Shaddock}, {Shaffer}, {Shahriar}, {Shaner}, {Shao}, {Sharma}, {Shawhan},
  {Shen}, {Shink}, {Shoemaker}, {Shoemaker}, {ShyamSundar}, {Siellez},
  {Sieniawska}, {Sigg}, {Silva}, {Singer}, {Singh}, {Singhal}, {Sintes},
  {Sitmukhambetov}, {Skliris}, {Slagmolen}, {Slaven-Blair}, {Smith}, {Smith},
  {Somala}, {Son}, {Sorazu}, {Sorrentino}, {Souradeep}, {Sowell}, {Spencer},
  {Srivastava}, {Srivastava}, {Staats}, {Stachie}, {Standke}, {Steer},
  {Steinke}, {Steinlechner}, {Steinlechner}, {Steinmeyer}, {Stevenson},
  {Stocks}, {Stone}, {Stops}, {Strain}, {Stratta}, {Strigin}, {Strunk},
  {Sturani}, {Stuver}, {Sudhir}, {Summerscales}, {Sun}, {Sunil}, {Suresh},
  {Sutton}, {Swinkels}, {Szczepa{\'n}czyk}, {Tacca}, {Tait}, {Talbot},
  {Talukder}, {Tanner}, {T{\'a}pai}, {Taracchini}, {Tasson}, {Taylor}, {Thies},
  {Thomas}, {Thomas}, {Thondapu}, {Thorne}, {Thrane}, {Tiwari}, {Tiwari},
  {Tiwari}, {Toland}, {Tonelli}, {Tornasi}, {Torres-Forn{\'e}}, {Torrie},
  {T{\"o}yr{\"a}}, {Travasso}, {Traylor}, {Tringali}, {Trovato}, {Trozzo},
  {Trudeau}, {Tsang}, {Tse}, {Tso}, {Tsukada}, {Tsuna}, {Tuyenbayev}, {Ueno},
  {Ugolini}, {Unnikrishnan}, {Urban}, {Usman}, {Vahlbruch}, {Vajente},
  {Valdes}, {van Bakel}, {van Beuzekom}, {van den Brand}, {Van Den Broeck},
  {Vander-Hyde}, {van Heijningen}, {van der Schaaf}, {van Veggel}, {Vardaro},
  {Varma}, {Vass}, {Vas{\'u}th}, {Vecchio}, {Vedovato}, {Veitch}, {Veitch},
  {Venkateswara}, {Venugopalan}, {Verkindt}, {Vetrano}, {Vicer{\'e}}, {Viets},
  {Vine}, {Vinet}, {Vitale}, {Vo}, {Vocca}, {Vorvick}, {Vyatchanin}, {Wade},
  {Wade}, {Wade}, {Walet}, {Walker}, {Wallace}, {Walsh}, {Wang}, {Wang},
  {Wang}, {Wang}, {Wang}, {Ward}, {Warden}, {Warner}, {Was}, {Watchi},
  {Weaver}, {Wei}, {Weinert}, {Weinstein}, {Weiss}, {Wellmann}, {Wen},
  {Wessel}, {We{\ss}els}, {Westhouse}, {Wette}, {Whelan}, {White}, {Whiting},
  {Whittle}, {Wilken}, {Williams}, {Williamson}, {Willis}, {Willke}, {Wimmer},
  {Winkler}, {Wipf}, {Wittel}, {Woan}, {Woehler}, {Wofford}, {Worden},
  {Wright}, {Wu}, {Wysocki}, {Xiao}, {Yamamoto}, {Yancey}, {Yang}, {Yap},
  {Yazback}, {Yeeles}, {Yu}, {Yu}, {Yuen}, {Yvert}, {Zadro{\.Z}ny}, {Zanolin},
  {Zappa}, {Zelenova}, {Zendri}, {Zevin}, {Zhang}, {Zhang}, {Zhang}, {Zhao},
  {Zhou}, {Zhou}, {Zhu}, {Zimmerman}, {Zlochower}, {Zucker}, {Zweizig}, {LIGO
  Scientific Collaboration}, \& {Virgo Collaboration}}]{Abbott19-GWTC1}
{Abbott}, B.~P., {Abbott}, R., {Abbott}, T.~D., {et~al.} 2019, Physical Review
  X, 9, 031040, \dodoi{10.1103/PhysRevX.9.031040}

\bibitem[{{Abbott} {et~al.}(2021){Abbott}, {Abbott}, {Abraham}, {Acernese},
  {Ackley}, {Adams}, {Adams}, {Adhikari}, {Adya}, {Affeldt}, {Agarwal},
  {Agathos}, {Agatsuma}, {Aggarwal}, {Aguiar}, {Aiello}, {Ain}, {Ajith},
  {Akutsu}, {Aleman}, {Allen}, {Allocca}, {Altin}, {Amato}, {Anand},
  {Ananyeva}, {Anderson}, {Anderson}, {Ando}, {Angelova}, {Ansoldi}, {Antelis},
  {Antier}, {Appert}, {Arai}, {Arai}, {Arai}, {Araki}, {Araya}, {Araya},
  {Areeda}, {Ar{\`e}ne}, {Aritomi}, {Arnaud}, {Aronson}, {Asada}, {Asali},
  {Ashton}, {Aso}, {Aston}, {Astone}, {Aubin}, {Aufmuth}, {Aultoneal},
  {Austin}, {Babak}, {Badaracco}, {Bader}, {Bae}, {Bae}, {Baer}, {Bagnasco},
  {Bai}, {Baiotti}, {Baird}, {Bajpai}, {Ball}, {Ballardin}, {Ballmer}, {Bals},
  {Balsamo}, {Baltus}, {Banagiri}, {Bankar}, {Bankar}, {Barayoga}, {Barbieri},
  {Barish}, {Barker}, {Barneo}, {Barone}, {Barr}, {Barsotti}, {Barsuglia},
  {Barta}, {Bartlett}, {Barton}, {Bartos}, {Bassiri}, {Basti}, {Bawaj},
  {Bayley}, {Baylor}, {Bazzan}, {B{\'e}csy}, {Bedakihale}, {Bejger},
  {Belahcene}, {Benedetto}, {Beniwal}, {Benjamin}, {Bennett}, {Bentley},
  {Benyaala}, {Bergamin}, {Berger}, {Bernuzzi}, {Bersanetti}, {Bertolini},
  {Betzwieser}, {Bhandare}, {Bhandari}, {Bhattacharjee}, {Bhaumik}, {Bidler},
  {Bilenko}, {Billingsley}, {Birney}, {Birnholtz}, {Biscans}, {Bischi},
  {Biscoveanu}, {Bisht}, {Biswas}, {Bitossi}, {Bizouard}, {Blackburn},
  {Blackman}, {Blair}, {Blair}, {Blair}, {Bobba}, {Bode}, {Boer}, {Bogaert},
  {Boldrini}, {Bondu}, {Bonilla}, {Bonnand}, {Booker}, {Boom}, {Bork},
  {Boschi}, {Bose}, {Bose}, {Bossilkov}, {Boudart}, {Bouffanais}, {Bozzi},
  {Bradaschia}, {Brady}, {Bramley}, {Branch}, {Branchesi}, {Brau}, {Breschi},
  {Briant}, {Briggs}, {Brillet}, {Brinkmann}, {Brockill}, {Brooks}, {Brooks},
  {Brown}, {Brunett}, {Bruno}, {Bruntz}, {Bryant}, {Buikema}, {Bulik},
  {Bulten}, {Buonanno}, {Buscicchio}, {Buskulic}, {Byer}, {Cadonati}, {Caesar},
  {Cagnoli}, {Cahillane}, {Cain}, {Bustillo}, {Callaghan}, {Callister},
  {Calloni}, {Camp}, {Canepa}, {Cannavacciuolo}, {Cannon}, {Cao}, {Cao}, {Cao},
  {Capocasa}, {Capote}, {Carapella}, {Carbognani}, {Carlin}, {Carney},
  {Carpinelli}, {Carullo}, {Carver}, {Diaz}, {Casentini}, {Castaldi},
  {Caudill}, {Cavagli{\`a}}, {Cavalier}, {Cavalieri}, {Cella},
  {Cerd{\'a}-Dur{\'a}n}, {Cesarini}, {Chaibi}, {Chakravarti}, {Champion},
  {Chan}, {Chan}, {Chan}, {Chan}, {Chandra}, {Chanial}, {Chao}, {Charlton},
  {Chase}, {Chassande-Mottin}, {Chatterjee}, {Chaturvedi}, {Chen}, {Chen},
  {Chen}, {Chen}, {Chen}, {Chen}, {Chen}, {Chen}, {Chen}, {Cheng}, {Cheong},
  {Cheung}, {Chia}, {Chiadini}, {Chiang}, {Chierici}, {Chincarini}, {Chiofalo},
  {Chiummo}, {Cho}, {Cho}, {Choate}, {Choudhary}, {Choudhary}, {Christensen},
  {Chu}, {Chu}, {Chu}, {Chua}, {Chung}, {Ciani}, {Ciecielag}, {Cie{\'s}lar},
  {Cifaldi}, {Ciobanu}, {Ciolfi}, {Cipriano}, {Cirone}, {Clara}, {Clark},
  {Clark}, {Clarke}, {Clearwater}, {Clesse}, {Cleva}, {Coccia}, {Cohadon},
  {Cohen}, {Cohen}, {Colleoni}, {Collette}, {Colpi}, {Compton}, {Constancio},
  {Conti}, {Cooper}, {Corban}, {Corbitt}, {Cordero-Carri{\'o}n}, {Corezzi},
  {Corley}, {Cornish}, {Corre}, {Corsi}, {Cortese}, {Costa}, {Cotesta},
  {Coughlin}, {Coughlin}, {Coulon}, {Countryman}, {Cousins}, {Couvares},
  {Covas}, {Coward}, {Cowart}, {Coyne}, {Coyne}, {Creighton}, {Creighton},
  {Criswell}, {Croquette}, {Crowder}, {Cudell}, {Cullen}, {Cumming},
  {Cummings}, {Cuoco}, {Cury{\l}o}, {Canton}, {D{\'a}lya}, {Dana},
  {Daneshgaranbajastani}, {D'Angelo}, {Danilishin}, {D'Antonio}, {Danzmann},
  {Darsow-Fromm}, {Dasgupta}, {Datrier}, {Dattilo}, {Dave}, {Davier}, {Davies},
  {Davis}, {Daw}, {Dean}, {Debra}, {Deenadayalan}, {Degallaix}, {de Laurentis},
  {Del{\'e}glise}, {Del Favero}, {de Lillo}, {de Lillo}, {Del Pozzo},
  {Demarchi}, {de Matteis}, {D'Emilio}, {Demos}, {Dent}, {Depasse}, {de
  Pietri}, {De Rosa}, {de Rossi}, {Desalvo}, {de Simone}, {Dhurandhar},
  {D{\'\i}az}, {Diaz-Ortiz}, {Didio}, {Dietrich}, {di Fiore}, {di Fronzo}, {di
  Giorgio}, {di Giovanni}, {di Girolamo}, {di Lieto}, {Ding}, {di Pace}, {di
  Palma}, {di Renzo}, {Divakarla}, {Dmitriev}, {Doctor}, {D'Onofrio},
  {Donovan}, {Dooley}, {Doravari}, {Dorrington}, {Drago}, {Driggers}, {Drori},
  {Du}, {Ducoin}, {Dupej}, {Durante}, {D'Urso}, {Duverne}, {Dwyer}, {Easter},
  {Ebersold}, {Eddolls}, {Edelman}, {Edo}, {Edy}, {Effler}, {Eguchi},
  {Eichholz}, {Eikenberry}, {Eisenmann}, {Eisenstein}, {Ejlli}, {Enomoto},
  {Errico}, {Essick}, {Estell{\'e}s}, {Estevez}, {Etienne}, {Etzel}, {Evans},
  {Evans}, {Ewing}, {Fafone}, {Fair}, {Fairhurst}, {Fan}, {Farah}, {Farinon},
  {Farr}, {Farr}, {Farrow}, {Fauchon-Jones}, {Favata}, {Fays}, {Fazio},
  {Feicht}, {Fejer}, {Feng}, {Fenyvesi}, {Ferguson}, {Fernandez-Galiana},
  {Ferrante}, {Ferreira}, {Fidecaro}, {Figura}, {Fiori}, {Fishbach}, {Fisher},
  {Fittipaldi}, {Fiumara}, {Flaminio}, {Floden}, {Flynn}, {Fong}, {Font},
  {Fornal}, {Forsyth}, {Franke}, {Frasca}, {Frasconi}, {Frederick}, {Frei},
  {Freise}, {Frey}, {Fritschel}, {Frolov}, {Fronz{\'e}}, {Fujii}, {Fujikawa},
  {Fukunaga}, {Fukushima}, {Fulda}, {Fyffe}, {Gabbard}, {Gadre}, {Gaebel},
  {Gair}, {Gais}, {Galaudage}, {Gamba}, {Ganapathy}, {Ganguly}, {Gao},
  {Gaonkar}, {Garaventa}, {Garc{\'\i}a-N{\'u}{\~n}ez},
  {Garc{\'\i}a-Quir{\'o}s}, {Garufi}, {Gateley}, {Gaudio}, {Gayathri}, {Ge},
  {Gemme}, {Gennai}, {George}, {Gergely}, {Gewecke}, {Ghonge}, {Ghosh},
  {Ghosh}, {Ghosh}, {Ghosh}, {Ghosh}, {Giacomazzo}, {Giacoppo}, {Giaime},
  {Giardina}, {Gibson}, {Gier}, {Giesler}, {Giri}, {Gissi}, {Glanzer},
  {Gleckl}, {Godwin}, {Goetz}, {Goetz}, {Gohlke}, {Goncharov}, {Gonz{\'a}lez},
  {Gopakumar}, {Gosselin}, {Gouaty}, {Grace}, {Grado}, {Granata}, {Granata},
  {Grant}, {Gras}, {Grassia}, {Gray}, {Gray}, {Greco}, {Green}, {Green},
  {Gretarsson}, {Gretarsson}, {Griffith}, {Griffiths}, {Griggs}, {Grignani},
  {Grimaldi}, {Grimes}, {Grimm}, {Grote}, {Grunewald}, {Gruning}, {Guerrero},
  {Guidi}, {Guimaraes}, {Guix{\'e}}, {Gulati}, {Guo}, {Guo}, {Gupta}, {Gupta},
  {Gupta}, {Gustafson}, {Gustafson}, {Guzman}, {Ha}, {Haegel}, {Hagiwara},
  {Haino}, {Halim}, {Hall}, {Hamilton}, {Hammond}, {Han}, {Haney}, {Hanks},
  {Hanna}, {Hannam}, {Hannuksela}, {Hansen}, {Hansen}, {Hanson}, {Harder},
  {Hardwick}, {Haris}, {Harms}, {Harry}, {Harry}, {Hartwig}, {Hasegawa},
  {Haskell}, {Hasskew}, {Haster}, {Hattori}, {Haughian}, {Hayakawa}, {Hayama},
  {Hayes}, {Healy}, {Heidmann}, {Heintze}, {Heinze}, {Heinzel}, {Heitmann},
  {Hellman}, {Hello}, {Helmling-Cornell}, {Hemming}, {Hendry}, {Heng},
  {Hennes}, {Hennig}, {Hennig}, {Vivanco}, {Heurs}, {Hild}, {Hill}, {Himemoto},
  {Hines}, {Hiranuma}, {Hirata}, {Hirose}, {Hochheim}, {Hofman}, {Hohmann},
  {Holgado}, {Holland}, {Hollows}, {Holmes}, {Holt}, {Holz}, {Hong}, {Hopkins},
  {Hough}, {Howell}, {Hoy}, {Hoyland}, {Hreibi}, {Hsieh}, {Hsu}, {Huang},
  {Huang}, {Huang}, {Huang}, {Huang}, {Huang}, {H{\"u}bner}, {Huddart},
  {Huerta}, {Hughey}, {Hui}, {Hui}, {Husa}, {Huttner}, {Huxford}, {Huynh-Dinh},
  {Ide}, {Idzkowski}, {Iess}, {Ikenoue}, {Imam}, {Inayoshi}, {Inchauspe},
  {Ingram}, {Inoue}, {Intini}, {Ioka}, {Isi}, {Isleif}, {Ito}, {Itoh}, {Iyer},
  {Izumi}, {Jaberianhamedan}, {Jacqmin}, {Jadhav}, {Jadhav}, {James}, {Jan},
  {Jani}, {Janssens}, {Janthalur}, {Jaranowski}, {Jariwala}, {Jaume},
  {Jenkins}, {Jeon}, {Jeunon}, {Jia}, {Jiang}, {Jin}, {Johns}, {Jones},
  {Jones}, {Jones}, {Jones}, {Jones}, {Jonker}, {Ju}, {Jung}, {Jung}, {Junker},
  {Kaihotsu}, {Kajita}, {Kakizaki}, {Kalaghatgi}, {Kalogera}, {Kamai},
  {Kamiizumi}, {Kanda}, {Kandhasamy}, {Kang}, {Kanner}, {Kao}, {Kapadia},
  {Kapasi}, {Karat}, {Karathanasis}, {Karki}, {Kashyap}, {Kasprzack},
  {Kastaun}, {Katsanevas}, {Katsavounidis}, {Katzman}, {Kaur}, {Kawabe},
  {Kawaguchi}, {Kawai}, {Kawasaki}, {K{\'e}f{\'e}lian}, {Keitel}, {Key},
  {Khadka}, {Khalili}, {Khan}, {Khan}, {Khazanov}, {Khetan}, {Khursheed},
  {Kijbunchoo}, {Kim}, {Kim}, {Kim}, {Kim}, {Kim}, {Kim}, {Kimball}, {Kimura},
  {King}, {Kinley-Hanlon}, {Kirchhoff}, {Kissel}, {Kita}, {Kitazawa},
  {Kleybolte}, {Klimenko}, {Knee}, {Knowles}, {Knyazev}, {Koch}, {Koekoek},
  {Kojima}, {Kokeyama}, {Koley}, {Kolitsidou}, {Kolstein}, {Komori},
  {Kondrashov}, {Kong}, {Kontos}, {Koper}, {Korobko}, {Kotake}, {Kovalam},
  {Kozak}, {Kozakai}, {Kozu}, {Kringel}, {Krishnendu}, {Kr{\'o}lak}, {Kuehn},
  {Kuei}, {Kumar}, {Kumar}, {Kumar}, {Kumar}, {Kume}, {Kuns}, {Kuo}, {Kuo},
  {Kuromiya}, {Kuroyanagi}, {Kusayanagi}, {Kwak}, {Kwang}, {Laghi}, {Lalande},
  {Lam}, {Lamberts}, {Landry}, {Lane}, {Lang}, {Lange}, {Lantz}, {La Rosa},
  {Lartaux-Vollard}, {Lasky}, {Laxen}, {Lazzarini}, {Lazzaro}, {Leaci},
  {Leavey}, {Lecoeuche}, {Lee}, {Lee}, {Lee}, {Lee}, {Lee}, {Lee}, {Lehmann},
  {Lema{\^\i}tre}, {Leon}, {Leonardi}, {Leroy}, {Letendre}, {Levin}, {Leviton},
  {Li}, {Li}, {Li}, {Li}, {Li}, {Li}, {Lin}, {Lin}, {Lin}, {Lin}, {Lin},
  {Linde}, {Linker}, {Linley}, {Littenberg}, {Liu}, {Liu}, {Liu}, {Liu},
  {Llorens-Monteagudo}, {Lo}, {Lockwood}, {Lollie}, {London}, {Longo}, {Lopez},
  {Lorenzini}, {Loriette}, {Lormand}, {Losurdo}, {Lough}, {Lousto}, {Lovelace},
  {L{\"u}ck}, {Lumaca}, {Lundgren}, {Luo}, {Macas}, {Macinnis}, {MacLeod},
  {MacMillan}, {Macquet}, {Hernandez}, {Maga{\~n}a-Sandoval}, {Magazz{\`u}},
  {Magee}, {Maggiore}, {Majorana}, {Makarem}, {Maksimovic}, {Maliakal},
  {Malik}, {Man}, {Mandic}, {Mangano}, {Mango}, {Mansell}, {Manske},
  {Mantovani}, {Mapelli}, {Marchesoni}, {Marchio}, {Marion}, {Mark},
  {M{\'a}rka}, {M{\'a}rka}, {Markakis}, {Markosyan}, {Markowitz}, {Maros},
  {Marquina}, {Marsat}, {Martelli}, {Martin}, {Martin}, {Martinez}, {Martinez},
  {Martinovic}, {Martynov}, {Marx}, {Masalehdan}, {Mason}, {Massera},
  {Masserot}, {Massinger}, {Masso-Reid}, {Mastrogiovanni}, {Matas},
  {Mateu-Lucena}, {Matichard}, {Matiushechkina}, {Mavalvala}, {McCann},
  {McCarthy}, {McClelland}, {McClincy}, {McCormick}, {McCuller}, {McGhee},
  {McGuire}, {McIsaac}, {McIver}, {McManus}, {McRae}, {McWilliams}, {Meacher},
  {Mehmet}, {Mehta}, {Melatos}, {Melchor}, {Mendell}, {Menendez-Vazquez},
  {Menoni}, {Mercer}, {Mereni}, {Merfeld}, {Merilh}, {Merritt}, {Merzougui},
  {Meshkov}, {Messenger}, {Messick}, {Meyers}, {Meylahn}, {Mhaske}, {Miani},
  {Miao}, {Michaloliakos}, {Michel}, {Michimura}, {Middleton}, {Milano},
  {Miller}, {Millhouse}, {Mills}, {Milotti}, {Milovich-Goff}, {Minazzoli},
  {Minenkov}, {Mio}, {Mir}, {Mishkin}, {Mishra}, {Mishra}, {Mistry}, {Mitra},
  {Mitrofanov}, {Mitselmakher}, {Mittleman}, {Miyakawa}, {Miyamoto},
  {Miyazaki}, {Miyo}, {Miyoki}, {Mo}, {Mogushi}, {Mohapatra}, {Mohite},
  {Molina}, {Molina-Ruiz}, {Mondin}, {Montani}, {Moore}, {Moraru}, {Morawski},
  {More}, {Moreno}, {Moreno}, {Mori}, {Morisaki}, {Moriwaki}, {Mours},
  {Mow-Lowry}, {Mozzon}, {Muciaccia}, {Mukherjee}, {Mukherjee}, {Mukherjee},
  {Mukherjee}, {Mukund}, {Mullavey}, {Munch}, {Mu{\~n}iz}, {Murray},
  {Musenich}, {Nadji}, {Nagano}, {Nagano}, {Nakamura}, {Nakano}, {Nakano},
  {Nakashima}, {Nakayama}, {Nardecchia}, {Narikawa}, {Naticchioni}, {Nayak},
  {Nayak}, {Negishi}, {Neil}, {Neilson}, {Nelemans}, {Nelson}, {Nery},
  {Neunzert}, {Ng}, {Ng}, {Nguyen}, {Nguyen}, {Nguyen}, {Quynh}, {Ni},
  {Nichols}, {Nishizawa}, {Nissanke}, {Nocera}, {Noh}, {Norman}, {North},
  {Nozaki}, {Nuttall}, {Oberling}, {O'Brien}, {Obuchi}, {O'Dell}, {Ogaki},
  {Oganesyan}, {Oh}, {Oh}, {Oh}, {Ohashi}, {Ohishi}, {Ohkawa}, {Ohme}, {Ohta},
  {Okada}, {Okutani}, {Okutomi}, {Olivetto}, {Oohara}, {Ooi}, {Oram},
  {O'Reilly}, {Ormiston}, {Ormsby}, {Ortega}, {O'Shaughnessy}, {O'Shea},
  {Oshino}, {Ossokine}, {Osthelder}, {Otabe}, {Ottaway}, {Overmier}, {Pace},
  {Pagano}, {Page}, {Pagliaroli}, {Pai}, {Pai}, {Palamos}, {Palashov},
  {Palomba}, {Pan}, {Panda}, {Pang}, {Pang}, {Pankow}, {Pannarale}, {Pant},
  {Paoletti}, {Paoli}, {Paolone}, {Parisi}, {Park}, {Parker}, {Pascucci},
  {Pasqualetti}, {Passaquieti}, {Passuello}, {Patel}, {Patricelli}, {Payne},
  {Pechsiri}, {Pedraza}, {Pegoraro}, {Pele}, {Arellano}, {Penn}, {Perego},
  {Pereira}, {Pereira}, {Perez}, {P{\'e}rigois}, {Perreca}, {Perri{\`e}s},
  {Petermann}, {Petterson}, {Pfeiffer}, {Pham}, {Phukon}, {Piccinni}, {Pichot},
  {Piendibene}, {Piergiovanni}, {Pierini}, {Pierro}, {Pillant}, {Pilo},
  {Pinard}, {Pinto}, {Piotrzkowski}, {Piotrzkowski}, {Pirello}, {Pitkin},
  {Placidi}, {Plastino}, {Pluchar}, {Poggiani}, {Polini}, {Pong}, {Ponrathnam},
  {Popolizio}, {Porter}, {Powell}, {Pracchia}, {Pradier}, {Prajapati},
  {Prasai}, {Prasanna}, {Pratten}, {Prestegard}, {Principe}, {Prodi},
  {Prokhorov}, {Prosposito}, {Prudenzi}, {Puecher}, {Punturo}, {Puosi},
  {Puppo}, {P{\"u}rrer}, {Qi}, {Quetschke}, {Quinonez}, {Quitzow-James},
  {Raab}, {Raaijmakers}, {Radkins}, {Radulesco}, {Raffai}, {Rail}, {Raja},
  {Rajan}, {Ramirez}, {Ramirez}, {Ramos-Buades}, {Rana}, {Rapagnani}, {Rapol},
  {Ratto}, {Raymond}, {Raza}, {Razzano}, {Read}, {Rees}, {Regimbau}, {Rei},
  {Reid}, {Reitze}, {Relton}, {Renzini}, {Rettegno}, {Ricci}, {Richardson},
  {Richardson}, {Richardson}, {Ricker}, {Riemenschneider}, {Riles}, {Rizzo},
  {Robertson}, {Robie}, {Robinet}, {Rocchi}, {Rocha}, {Rodriguez},
  {Rodriguez-Soto}, {Rolland}, {Rollins}, {Roma}, {Romanelli}, {Romano},
  {Romano}, {Romel}, {Romero}, {Romero-Shaw}, {Romie}, {Rose}, {Rosi{\'n}ska},
  {Rosofsky}, {Ross}, {Rowan}, {Rowlinson}, {Roy}, {Roy}, {Rozza}, {Ruggi},
  {Ryan}, {Sachdev}, {Sadecki}, {Sadiq}, {Sago}, {Saito}, {Saito}, {Sakai},
  {Sakai}, {Sakellariadou}, {Sakuno}, {Salafia}, {Salconi}, {Saleem}, {Salemi},
  {Samajdar}, {Sanchez}, {Sanchez}, {Sanchez}, {Sanchis-Gual}, {Sanders},
  {Sanuy}, {Saravanan}, {Sarin}, {Sassolas}, {Satari}, {Sato}, {Sato},
  {Sauter}, {Savage}, {Savant}, {Sawada}, {Sawant}, {Sawant}, {Sayah},
  {Schaetzl}, {Scheel}, {Scheuer}, {Schindler-Tyka}, {Schmidt}, {Schnabel},
  {Schneewind}, {Schofield}, {Sch{\"o}nbeck}, {Schulte}, {Schutz}, {Schwartz},
  {Scott}, {Scott}, {Seglar-Arroyo}, {Seidel}, {Sekiguchi}, {Sekiguchi},
  {Sellers}, {Sergeev}, {Sengupta}, {Sennett}, {Sentenac}, {Seo}, {Sequino},
  {Setyawati}, {Shaffer}, {Shahriar}, {Shams}, {Shao}, {Sharifi}, {Sharma},
  {Sharma}, {Shawhan}, {Shcheblanov}, {Shen}, {Shibagaki}, {Shikauchi},
  {Shimizu}, {Shimoda}, {Shimode}, {Shink}, {Shinkai}, {Shishido}, {Shoda},
  {Shoemaker}, {Shoemaker}, {Shukla}, {Shyamsundar}, {Sieniawska}, {Sigg},
  {Singer}, {Singh}, {Singh}, {Singha}, {Sintes}, {Sipala}, {Skliris},
  {Slagmolen}, {Slaven-Blair}, {Smetana}, {Smith}, {Smith}, {Somala}, {Somiya},
  {Son}, {Soni}, {Soni}, {Sorazu}, {Sordini}, {Sorrentino}, {Sorrentino},
  {Sotani}, {Soulard}, {Souradeep}, {Sowell}, {Spagnuolo}, {Spencer}, {Spera},
  {Srivastava}, {Srivastava}, {Staats}, {Stachie}, {Steer}, {Steinlechner},
  {Steinlechner}, {Stops}, {Stover}, {Strain}, {Strang}, {Stratta}, {Strunk},
  {Sturani}, {Stuver}, {S{\"u}dbeck}, {Sudhagar}, {Sudhir}, {Sugimoto}, {Suh},
  {Summerscales}, {Sun}, {Sun}, {Sunil}, {Sur}, {Suresh}, {Sutton}, {Suzuki},
  {Suzuki}, {Swinkels}, {Szczepa{\'n}czyk}, {Szewczyk}, {Tacca}, {Tagoshi},
  {Tait}, {Takahashi}, {Takahashi}, {Takamori}, {Takano}, {Takeda}, {Takeda},
  {Talbot}, {Tanaka}, {Tanaka}, {Tanaka}, {Tanaka}, {Tanaka}, {Tanasijczuk},
  {Tanioka}, {Tanner}, {Tao}, {Tapia}, {Martin}, {Martin}, {Tasson}, {Telada},
  {Tenorio}, {Terkowski}, {Test}, {Thirugnanasambandam}, {Thomas}, {Thomas},
  {Thompson}, {Thondapu}, {Thorne}, {Thrane}, {Tiwari}, {Tiwari}, {Tiwari},
  {Toland}, {Tolley}, {Tomaru}, {Tomigami}, {Tomura}, {Tonelli},
  {Torres-Forn{\'e}}, {Torrie}, {E Melo}, {T{\"o}yr{\"a}}, {Trapananti},
  {Travasso}, {Traylor}, {Tringali}, {Tripathee}, {Troiano}, {Trovato},
  {Trozzo}, {Trudeau}, {Tsai}, {Tsai}, {Tsang}, {Tsang}, {Tsao}, {Tse}, {Tso},
  {Tsubono}, {Tsuchida}, {Tsukada}, {Tsuna}, {Tsutsui}, {Tsuzuki}, {Turconi},
  {Tuyenbayev}, {Ubhi}, {Uchikata}, {Uchiyama}, {Udall}, {Ueda}, {Uehara},
  {Ueno}, {Ueshima}, {Ugolini}, {Unnikrishnan}, {Uraguchi}, {Urban}, {Ushiba},
  {Usman}, {Utina}, {Vahlbruch}, {Vajente}, {Vajpeyi}, {Valdes}, {Valentini},
  {Valsan}, {van Bakel}, {van Beuzekom}, {van den Brand}, {van den Broeck},
  {Vander-Hyde}, {van der Schaaf}, {van Heijningen}, {Vanosky}, {van Putten},
  {van Remortel}, {Vardaro}, {Vargas}, {Varma}, {Vas{\'u}th}, {Vecchio},
  {Vedovato}, {Veitch}, {Veitch}, {Venkateswara}, {Venneberg}, {Venugopalan},
  {Verkindt}, {Verma}, {Veske}, {Vetrano}, {Vicer{\'e}}, {Viets},
  {Villa-Ortega}, {Vinet}, {Vitale}, {Vo}, {Vocca}, {von Reis}, {von Wrangel},
  {Vorvick}, {Vyatchanin}, {Wade}, {Wade}, {Wagner}, {Walet}, {Walker},
  {Wallace}, {Wallace}, {Walsh}, {Wang}, {Wang}, {Wang}, {Ward}, {Warner},
  {Was}, {Washimi}, {Washington}, {Watchi}, {Weaver}, {Wei}, {Weinert},
  {Weinstein}, {Weiss}, {Weller}, {Wellmann}, {Wen}, {We{\ss}els}, {Westhouse},
  {Wette}, {Whelan}, {White}, {Whiting}, {Whittle}, {Wilken}, {Williams},
  {Williams}, {Williamson}, {Willis}, {Willke}, {Wilson}, {Winkler}, {Wipf},
  {Wlodarczyk}, {Woan}, {Woehler}, {Wofford}, {Wong}, {Wu}, {Wu}, {Wu}, {Wu},
  {Wysocki}, {Xiao}, {Xu}, {Yamada}, {Yamamoto}, {Yamamoto}, {Yamamoto},
  {Yamamoto}, {Yamashita}, {Yamazaki}, {Yang}, {Yang}, {Yang}, {Yang}, {Yang},
  {Yap}, {Yeeles}, {Yelikar}, {Ying}, {Yokogawa}, {Yokoyama}, {Yokozawa},
  {Yoon}, {Yoshioka}, {Yu}, {Yu}, {Yuzurihara}, {Zadro{\.z}ny}, {Zanolin},
  {Zeidler}, {Zelenova}, {Zendri}, {Zevin}, {Zhan}, {Zhang}, {Zhang}, {Zhang},
  {Zhang}, {Zhang}, {Zhao}, {Zhao}, {Zhao}, {Zhao}, {Zhou}, {Zhu}, {Zhu},
  {Zimmerman}, {Zucker}, {Zweizig}, {Ligo Scientific Collaboration}, {VIRGO
  Collaboration}, \& {Kagra Collaboration}}]{Abbott21-PointSources}
{Abbott}, R., {Abbott}, T.~D., {Abraham}, S., {et~al.} 2021, \prd, 104, 022005,
  \dodoi{10.1103/PhysRevD.104.022005}

\bibitem[{{Abdollahi} {et~al.}(2020){Abdollahi}, {Acero}, {Ackermann},
  {Ajello}, {Atwood}, {Axelsson}, {Baldini}, {Ballet}, {Barbiellini},
  {Bastieri}, {Becerra Gonzalez}, {Bellazzini}, {Berretta}, {Bissaldi},
  {Blandford}, {Bloom}, {Bonino}, {Bottacini}, {Brandt}, {Bregeon}, {Bruel},
  {Buehler}, {Burnett}, {Buson}, {Cameron}, {Caputo}, {Caraveo}, {Casandjian},
  {Castro}, {Cavazzuti}, {Charles}, {Chaty}, {Chen}, {Cheung}, {Chiaro},
  {Ciprini}, {Cohen-Tanugi}, {Cominsky}, {Coronado-Bl{\'a}zquez}, {Costantin},
  {Cuoco}, {Cutini}, {D'Ammando}, {DeKlotz}, {de la Torre Luque}, {de Palma},
  {Desai}, {Digel}, {Di Lalla}, {Di Mauro}, {Di Venere}, {Dom{\'\i}nguez},
  {Dumora}, {Fana Dirirsa}, {Fegan}, {Ferrara}, {Franckowiak}, {Fukazawa},
  {Funk}, {Fusco}, {Gargano}, {Gasparrini}, {Giglietto}, {Giommi}, {Giordano},
  {Giroletti}, {Glanzman}, {Green}, {Grenier}, {Griffin}, {Grondin}, {Grove},
  {Guiriec}, {Harding}, {Hayashi}, {Hays}, {Hewitt}, {Horan},
  {J{\'o}hannesson}, {Johnson}, {Kamae}, {Kerr}, {Kocevski}, {Kovac'evic'},
  {Kuss}, {Landriu}, {Larsson}, {Latronico}, {Lemoine-Goumard}, {Li},
  {Liodakis}, {Longo}, {Loparco}, {Lott}, {Lovellette}, {Lubrano}, {Madejski},
  {Maldera}, {Malyshev}, {Manfreda}, {Marchesini}, {Marcotulli},
  {Mart{\'\i}-Devesa}, {Martin}, {Massaro}, {Mazziotta}, {McEnery}, {Mereu},
  {Meyer}, {Michelson}, {Mirabal}, {Mizuno}, {Monzani}, {Morselli},
  {Moskalenko}, {Negro}, {Nuss}, {Ojha}, {Omodei}, {Orienti}, {Orlando},
  {Ormes}, {Palatiello}, {Paliya}, {Paneque}, {Pei}, {Pe{\~n}a-Herazo},
  {Perkins}, {Persic}, {Pesce-Rollins}, {Petrosian}, {Petrov}, {Piron}, {Poon},
  {Porter}, {Principe}, {Rain{\`o}}, {Rando}, {Razzano}, {Razzaque}, {Reimer},
  {Reimer}, {Remy}, {Reposeur}, {Romani}, {Saz Parkinson}, {Schinzel},
  {Serini}, {Sgr{\`o}}, {Siskind}, {Smith}, {Spandre}, {Spinelli}, {Strong},
  {Suson}, {Tajima}, {Takahashi}, {Tak}, {Thayer}, {Thompson}, {Tibaldo},
  {Torres}, {Torresi}, {Valverde}, {Van Klaveren}, {van Zyl}, {Wood},
  {Yassine}, \& {Zaharijas}}]{Abdollahi20}
{Abdollahi}, S., {Acero}, F., {Ackermann}, M., {et~al.} 2020, \apjs, 247, 33,
  \dodoi{10.3847/1538-4365/ab6bcb}

\bibitem[{{Abeysekara} {et~al.}(2016){Abeysekara}, {Archambault}, {Archer},
  {Benbow}, {Bird}, {Buchovecky}, {Buckley}, {Byrum}, {Cardenzana}, {Cerruti},
  {Chen}, {Christiansen}, {Ciupik}, {Cui}, {Dickinson}, {Eisch}, {Errando},
  {Falcone}, {Fegan}, {Feng}, {Finley}, {Fleischhack}, {Fortin}, {Fortson},
  {Furniss}, {Gillanders}, {Griffin}, {Grube}, {Gyuk}, {H{\"u}tten},
  {H{\r{a}}kansson}, {Hanna}, {Holder}, {Humensky}, {Johnson}, {Kaaret}, {Kar},
  {Kelley-Hoskins}, {Kertzman}, {Kieda}, {Krause}, {Krennrich}, {Kumar},
  {Lang}, {Lin}, {Maier}, {McArthur}, {McCann}, {Meagher}, {Moriarty},
  {Mukherjee}, {Nieto}, {O'Brien}, {O'Faol{\'a}in de Bhr{\'o}ithe}, {Ong},
  {Otte}, {Park}, {Perkins}, {Petrashyk}, {Pohl}, {Popkow}, {Pueschel},
  {Quinn}, {Ragan}, {Ratliff}, {Reynolds}, {Richards}, {Roache}, {Santander},
  {Sembroski}, {Shahinyan}, {Staszak}, {Telezhinsky}, {Tucci}, {Tyler},
  {Vincent}, {Wakely}, {Weiner}, {Weinstein}, {Williams}, \&
  {Zitzer}}]{Abeysekara16}
{Abeysekara}, A.~U., {Archambault}, S., {Archer}, A., {et~al.} 2016, \apjl,
  818, L33, \dodoi{10.3847/2041-8205/818/2/L33}

\bibitem[{{Ackermann} {et~al.}(2012){Ackermann}, {Ajello}, {Allafort},
  {Baldini}, {Ballet}, {Bastieri}, {Bechtol}, {Bellazzini}, {Berenji}, {Bloom},
  {Bonamente}, {Borgland}, {Bouvier}, {Bregeon}, {Brigida}, {Bruel}, {Buehler},
  {Buson}, {Caliandro}, {Cameron}, {Caraveo}, {Casandjian}, {Cecchi},
  {Charles}, {Chekhtman}, {Cheung}, {Chiang}, {Cillis}, {Ciprini}, {Claus},
  {Cohen-Tanugi}, {Conrad}, {Cutini}, {de Palma}, {Dermer}, {Digel}, {Silva},
  {Drell}, {Drlica-Wagner}, {Favuzzi}, {Fegan}, {Fortin}, {Fukazawa}, {Funk},
  {Fusco}, {Gargano}, {Gasparrini}, {Germani}, {Giglietto}, {Giordano},
  {Glanzman}, {Godfrey}, {Grenier}, {Guiriec}, {Gustafsson}, {Hadasch},
  {Hayashida}, {Hays}, {Hughes}, {J{\'o}hannesson}, {Johnson}, {Kamae},
  {Katagiri}, {Kataoka}, {Kn{\"o}dlseder}, {Kuss}, {Lande}, {Longo}, {Loparco},
  {Lott}, {Lovellette}, {Lubrano}, {Madejski}, {Martin}, {Mazziotta},
  {McEnery}, {Michelson}, {Mizuno}, {Monte}, {Monzani}, {Morselli},
  {Moskalenko}, {Murgia}, {Nishino}, {Norris}, {Nuss}, {Ohno}, {Ohsugi},
  {Okumura}, {Omodei}, {Orlando}, {Ozaki}, {Parent}, {Persic}, {Pesce-Rollins},
  {Petrosian}, {Pierbattista}, {Piron}, {Pivato}, {Porter}, {Rain{\`o}},
  {Rando}, {Razzano}, {Reimer}, {Reimer}, {Ritz}, {Roth}, {Sbarra}, {Sgr{\`o}},
  {Siskind}, {Spandre}, {Spinelli}, {Stawarz}, {Strong}, {Takahashi}, {Tanaka},
  {Thayer}, {Tibaldo}, {Tinivella}, {Torres}, {Tosti}, {Troja}, {Uchiyama},
  {Vandenbroucke}, {Vianello}, {Vitale}, {Waite}, {Wood}, \&
  {Yang}}]{Ackermann12-SFGs}
{Ackermann}, M., {Ajello}, M., {Allafort}, A., {et~al.} 2012, \apj, 755, 164,
  \dodoi{10.1088/0004-637X/755/2/164}

\bibitem[{{Adelson}(1966)}]{Adelson66}
{Adelson}, R.~M. 1966, Journal of the Operational Research Society, 17, 73

\bibitem[{{Ajello} {et~al.}(2020){Ajello}, {Di Mauro}, {Paliya}, \&
  {Garrappa}}]{Ajello20}
{Ajello}, M., {Di Mauro}, M., {Paliya}, V.~S., \& {Garrappa}, S. 2020, \apj,
  894, 88, \dodoi{10.3847/1538-4357/ab86a6}

\bibitem[{{Annis}(1999)}]{Annis99}
{Annis}, J. 1999, Journal of the British Interplanetary Society, 52, 19.
\newblock \doarXiv{astro-ph/9901322}

\bibitem[{Baddeley(2007)}]{Baddeley07}
Baddeley, A. 2007, in Stochastic Geometry, ed. W.~{Weil} (Berlin: Springer),
  1--75, \dodoi{10.1007/978-3-540-38175-4_1}

\bibitem[{{Barbour} \& {Chryssaphinou}(2001)}]{Barbour01}
{Barbour}, A.~D., \& {Chryssaphinou}, O. 2001, Annals of Applied Probability,
  964, \dodoi{10.1214/aoap/1015345355}

\bibitem[{{Bas}(2019)}]{Bas19}
{Bas}, E. 2019, Basics of Probability and Stochastic Processes (Berlin:
  Springer), \dodoi{10.1007/978-3-030-32323-3}

\bibitem[{{Beck} {et~al.}(1998){Beck}, {Berkhuijsen}, \& {Hoernes}}]{Beck98}
{Beck}, R., {Berkhuijsen}, E.~M., \& {Hoernes}, P. 1998, \aaps, 129, 329,
  \dodoi{10.1051/aas:1998187}

\bibitem[{{Bell}(2003)}]{Bell03}
{Bell}, E.~F. 2003, \apj, 586, 794, \dodoi{10.1086/367829}

\bibitem[{{Brin}(1983)}]{Brin83}
{Brin}, G.~D. 1983, \qjras, 24, 283

\bibitem[{{Brown} {et~al.}(2011){Brown}, {Jannuzi}, {Floyd}, \&
  {Mould}}]{Brown11}
{Brown}, M. J.~I., {Jannuzi}, B.~T., {Floyd}, D. J.~E., \& {Mould}, J.~R. 2011,
  \apjl, 731, L41, \dodoi{10.1088/2041-8205/731/2/L41}

\bibitem[{{Brzycki} {et~al.}(2023){Brzycki}, {Siemion}, {de Pater}, {Cordes},
  {Gajjar}, {Lacki}, \& {Sheikh}}]{Brzycki23}
{Brzycki}, B., {Siemion}, A. P.~V., {de Pater}, I., {et~al.} 2023, \apj, 952,
  46, \dodoi{10.3847/1538-4357/acdee0}

\bibitem[{{Calabretta} {et~al.}(2014){Calabretta}, {Staveley-Smith}, \&
  {Barnes}}]{Calabretta14}
{Calabretta}, M.~R., {Staveley-Smith}, L., \& {Barnes}, D.~G. 2014, \pasa, 31,
  e007, \dodoi{10.1017/pasa.2013.36}

\bibitem[{{Carrigan}(2009)}]{Carrigan09}
{Carrigan}, Richard~A., J. 2009, \apj, 698, 2075,
  \dodoi{10.1088/0004-637X/698/2/2075}

\bibitem[{{Carroll-Nellenback} {et~al.}(2019){Carroll-Nellenback}, {Frank},
  {Wright}, \& {Scharf}}]{CarrollNellenback19}
{Carroll-Nellenback}, J., {Frank}, A., {Wright}, J., \& {Scharf}, C. 2019, \aj,
  158, 117, \dodoi{10.3847/1538-3881/ab31a3}

\bibitem[{{Castillo} {et~al.}(2005){Castillo}, {Hadi}, {Balakrishnan}, \&
  {Sarabia}}]{Castillo05}
{Castillo}, E., {Hadi}, A.~S., {Balakrishnan}, N., \& {Sarabia}, J.~M. 2005,
  {Extreme Value and Related Models with Applications in Engineering and
  Science} (Hoboken, NJ: Wiley-Interscience)

\bibitem[{{Caves} \& {Drummond}(1994)}]{Caves94}
{Caves}, C.~M., \& {Drummond}, P.~D. 1994, Reviews of Modern Physics, 66, 481,
  \dodoi{10.1103/RevModPhys.66.481}

\bibitem[{{Chabrier}(2003)}]{Chabrier03}
{Chabrier}, G. 2003, \apjl, 586, L133, \dodoi{10.1086/374879}

\bibitem[{{Chiu} {et~al.}(2013){Chiu}, {Stoyan}, {Kendall}, \&
  {Mecke}}]{Chiu13}
{Chiu}, S.~N., {Stoyan}, D., {Kendall}, W.~S., \& {Mecke}, J. 2013, {Stochastic
  Geometry and its Applications: Third Edition} (New York: Wiley),
  \dodoi{10.1002/9781118658222}

\bibitem[{{{\'C}irkovi{\'c}}(2015)}]{Cirkovic15}
{{\'C}irkovi{\'c}}, M.~M. 2015, Serbian Astronomical Journal, 191, 1,
  \dodoi{10.2298/SAJ1591001C}

\bibitem[{{{\'C}irkovi{\'c}}(2018)}]{Cirkovic18-Book}
---. 2018, {The Great Silence: Science and Philosophy of Fermi's Paradox} (New
  York: Oxford University Press)

\bibitem[{{{\'C}irkovi{\'c}} \& {Vukoti{\'c}}(2008)}]{Cirkovic08}
{{\'C}irkovi{\'c}}, M.~M., \& {Vukoti{\'c}}, B. 2008, Origins of Life and
  Evolution of the Biosphere, 38, 535, \dodoi{10.1007/s11084-008-9149-y}

\bibitem[{{Cocconi} \& {Morrison}(1959)}]{Cocconi59}
{Cocconi}, G., \& {Morrison}, P. 1959, \nat, 184, 844, \dodoi{10.1038/184844a0}

\bibitem[{{Condon}(1974)}]{Condon74}
{Condon}, J.~J. 1974, \apj, 188, 279, \dodoi{10.1086/152714}

\bibitem[{{Condon}(1992)}]{Condon92}
---. 1992, \araa, 30, 575, \dodoi{10.1146/annurev.aa.30.090192.003043}

\bibitem[{{Condon} {et~al.}(1998){Condon}, {Cotton}, {Greisen}, {Yin},
  {Perley}, {Taylor}, \& {Broderick}}]{Condon98}
{Condon}, J.~J., {Cotton}, W.~D., {Greisen}, E.~W., {et~al.} 1998, \aj, 115,
  1693, \dodoi{10.1086/300337}

\bibitem[{{Condon} {et~al.}(2012){Condon}, {Cotton}, {Fomalont}, {Kellermann},
  {Miller}, {Perley}, {Scott}, {Vernstrom}, \& {Wall}}]{Condon12}
{Condon}, J.~J., {Cotton}, W.~D., {Fomalont}, E.~B., {et~al.} 2012, \apj, 758,
  23, \dodoi{10.1088/0004-637X/758/1/23}

\bibitem[{{Corbet}(1997)}]{Corbet97}
{Corbet}, R.~H.~D. 1997, Journal of the British Interplanetary Society, 50, 253

\bibitem[{{Cordes} {et~al.}(1997){Cordes}, {Lazio}, \& {Sagan}}]{Cordes97}
{Cordes}, J.~M., {Lazio}, J.~W., \& {Sagan}, C. 1997, \apj, 487, 782,
  \dodoi{10.1086/304620}

\bibitem[{{Czech} {et~al.}(2021){Czech}, {Isaacson}, {Pearce}, {Cox}, {Sheikh},
  {Brzycki}, {Buchner}, {Croft}, {DeBoer}, {DeMarines}, {Drew}, {Gajjar},
  {Lacki}, {Lebofsky}, {MacMahon}, {Ng}, {de Pater}, {Price}, {Siemion}, {Van
  Rooyen}, \& {Pete Worden}}]{Czech21}
{Czech}, D., {Isaacson}, H., {Pearce}, L., {et~al.} 2021, \pasp, 133, 064502,
  \dodoi{10.1088/1538-3873/abf329}

\bibitem[{{Dale} {et~al.}(2001){Dale}, {Helou}, {Contursi}, {Silbermann}, \&
  {Kolhatkar}}]{Dale01}
{Dale}, D.~A., {Helou}, G., {Contursi}, A., {Silbermann}, N.~A., \&
  {Kolhatkar}, S. 2001, \apj, 549, 215, \dodoi{10.1086/319077}

\bibitem[{{Daley} \& {Vere-Jones}(2003)}]{Daley03}
{Daley}, D.~J., \& {Vere-Jones}, D. 2003, {An Introduction to the Theory of
  Point Processes. Volume I: Elementary Theory and Methods} (New York:
  Springer), \dodoi{10.1007/b97277}

\bibitem[{{Delhaize} {et~al.}(2017){Delhaize}, {Smol{\v{c}}i{\'c}},
  {Delvecchio}, {Novak}, {Sargent}, {Baran}, {Magnelli}, {Zamorani},
  {Schinnerer}, {Murphy}, {Aravena}, {Berta}, {Bondi}, {Capak}, {Carilli},
  {Ciliegi}, {Civano}, {Ilbert}, {Karim}, {Laigle}, {Le F{\`e}vre}, {Marchesi},
  {McCracken}, {Salvato}, {Seymour}, \& {Tasca}}]{Delhaize17}
{Delhaize}, J., {Smol{\v{c}}i{\'c}}, V., {Delvecchio}, I., {et~al.} 2017, \aap,
  602, A4, \dodoi{10.1051/0004-6361/201629430}

\bibitem[{{Dixon}(1985)}]{Dixon85}
{Dixon}, R.~S. 1985, in IAU Symposium, Vol. 112, The Search for
  Extraterrestrial Life: Recent Developments, ed. M.~D. {Papagiannis}
  (Dordrecht: D. Reidel Publishing Co.), 305--314,
  \dodoi{10.1007/978-94-009-5462-5_39}

\bibitem[{{Djorgovski} {et~al.}(2013){Djorgovski}, {Mahabal}, {Drake},
  {Graham}, \& {Donalek}}]{Djorgovski13}
{Djorgovski}, S.~G., {Mahabal}, A., {Drake}, A., {Graham}, M., \& {Donalek}, C.
  2013, in Planets, Stars and Stellar Systems. Volume 2: Astronomical
  Techniques, Software and Data, ed. T.~D. {Oswalt} \& H.~E. {Bond} (Springer),
  223, \dodoi{10.1007/978-94-007-5618-2_5}

\bibitem[{{Draine} \& {Lazarian}(1998)}]{Draine98}
{Draine}, B.~T., \& {Lazarian}, A. 1998, \apjl, 494, L19,
  \dodoi{10.1086/311167}

\bibitem[{{Dyson}(1960)}]{Dyson60}
{Dyson}, F.~J. 1960, Science, 131, 1667, \dodoi{10.1126/science.131.3414.1667}

\bibitem[{{Embrechts} {et~al.}(2013){Embrechts}, {Kl{\"u}ppelberg}, \&
  {Mikosch}}]{Embrechts13}
{Embrechts}, P., {Kl{\"u}ppelberg}, C., \& {Mikosch}, T. 2013, {Modelling
  Extremal Events: for Insurance and Finance}, Vol.~33 (Berlin: Springer),
  \dodoi{10.1007/978-3-642-33483-2}

\bibitem[{{Enriquez} {et~al.}(2017){Enriquez}, {Siemion}, {Foster}, {Gajjar},
  {Hellbourg}, {Hickish}, {Isaacson}, {Price}, {Croft}, {DeBoer}, {Lebofsky},
  {MacMahon}, \& {Werthimer}}]{Enriquez17}
{Enriquez}, J.~E., {Siemion}, A., {Foster}, G., {et~al.} 2017, \apj, 849, 104,
  \dodoi{10.3847/1538-4357/aa8d1b}

\bibitem[{{Evans} {et~al.}(1972){Evans}, {Hills}, {Rydbeck}, \&
  {Kollberg}}]{Evans72}
{Evans}, N.~J., I., {Hills}, R.~E., {Rydbeck}, O.~E., \& {Kollberg}, E. 1972,
  \pra, 6, 1643, \dodoi{10.1103/PhysRevA.6.1643}

\bibitem[{{Foellmi}(2009)}]{Foellmi09}
{Foellmi}, C. 2009, \aap, 507, 1719, \dodoi{10.1051/0004-6361/200911739}

\bibitem[{{Forgan}(2019)}]{Forgan19}
{Forgan}, D.~H. 2019, {Solving Fermi's Paradox} (Cambridge: Cambridge
  University Press), \dodoi{10.1017/9781316681510}

\bibitem[{{Forward}(1984)}]{Forward84}
{Forward}, R.~L. 1984, Journal of Spacecraft and Rockets, 21, 187,
  \dodoi{10.2514/3.8632}

\bibitem[{{Gajjar} {et~al.}(2021){Gajjar}, {Perez}, {Siemion}, {Foster},
  {Brzycki}, {Chatterjee}, {Chen}, {Cordes}, {Croft}, {Czech}, {DeBoer},
  {DeMarines}, {Drew}, {Gowanlock}, {Isaacson}, {Lacki}, {Lebofsky},
  {MacMahon}, {Morrison}, {Ng}, {de Pater}, {Price}, {Sheikh}, {Suresh},
  {Webb}, \& {Pete Worden}}]{Gajjar21}
{Gajjar}, V., {Perez}, K.~I., {Siemion}, A. P.~V., {et~al.} 2021, \aj, 162, 33,
  \dodoi{10.3847/1538-3881/abfd36}

\bibitem[{{Garrett}(2015)}]{Garrett15}
{Garrett}, M.~A. 2015, \aap, 581, L5, \dodoi{10.1051/0004-6361/201526687}

\bibitem[{{Geringer-Sameth} {et~al.}(2015){Geringer-Sameth}, {Koushiappas}, \&
  {Walker}}]{GeringerSameth15}
{Geringer-Sameth}, A., {Koushiappas}, S.~M., \& {Walker}, M.~G. 2015, \prd, 91,
  083535, \dodoi{10.1103/PhysRevD.91.083535}

\bibitem[{{Gray} \& {Mooley}(2017)}]{Gray17}
{Gray}, R.~H., \& {Mooley}, K. 2017, \aj, 153, 110,
  \dodoi{10.3847/1538-3881/153/3/110}

\bibitem[{{Greggio} \& {Renzini}(1990)}]{Greggio90}
{Greggio}, L., \& {Renzini}, A. 1990, \apj, 364, 35, \dodoi{10.1086/169384}

\bibitem[{{Griffith} {et~al.}(2015){Griffith}, {Wright}, {Maldonado}, {Povich},
  {Sigur{\dj}sson}, \& {Mullan}}]{Griffith15}
{Griffith}, R.~L., {Wright}, J.~T., {Maldonado}, J., {et~al.} 2015, \apjs, 217,
  25, \dodoi{10.1088/0067-0049/217/2/25}

\bibitem[{{Guerin} {et~al.}(2017){Guerin}, {Dussaux}, {Fouch{\'e}}, {Labeyrie},
  {Rivet}, {Vernet}, {Vakili}, \& {Kaiser}}]{Guerin17}
{Guerin}, W., {Dussaux}, A., {Fouch{\'e}}, M., {et~al.} 2017, \mnras, 472,
  4126, \dodoi{10.1093/mnras/stx2143}

\bibitem[{{Gumbel}(1958)}]{Gumbel58}
{Gumbel}, E.~J. 1958, Statistics of Extremes (New York: Columbia University
  Press), \dodoi{10.7312/gumb92958}

\bibitem[{{Haenggi}(2013)}]{Haenggi13}
{Haenggi}, M. 2013, Stochastic Geometry for Wireless Networks (Cambridge:
  Cambridge University Press), \dodoi{10.1017/cbo9781139043816}

\bibitem[{{Harp} {et~al.}(2016){Harp}, {Richards}, {Tarter}, {Dreher},
  {Jordan}, {Shostak}, {Smolek}, {Kilsdonk}, {Wilcox}, {Wimberly}, {Ross},
  {Barott}, {Ackermann}, \& {Blair}}]{Harp16}
{Harp}, G.~R., {Richards}, J., {Tarter}, J.~C., {et~al.} 2016, \aj, 152, 181,
  \dodoi{10.3847/0004-6256/152/6/181}

\bibitem[{{Harris}(1986)}]{Harris86}
{Harris}, M.~J. 1986, \apss, 123, 297, \dodoi{10.1007/BF00653949}

\bibitem[{{Harris}(2002)}]{Harris02}
---. 2002, Journal of the British Interplanetary Society, 55, 383,
  \dodoi{10.48550/arXiv.astro-ph/0112490}

\bibitem[{{Harwit}(1981)}]{Harwit81}
{Harwit}, M. 1981, {Cosmic discovery: The search, scope, and heritage of
  astronomy} (New York: Basic Books, Inc.)

\bibitem[{{Hecht}(1998)}]{Hecht98}
{Hecht}, E. 1998, {Optics: Fourth Edition} (San Francisco: Addison Wesley)

\bibitem[{{Hippke}(2018)}]{Hippke18-Messenger}
{Hippke}, M. 2018, Acta Astronautica, 151, 53,
  \dodoi{10.1016/j.actaastro.2018.05.038}

\bibitem[{{Hippke}(2021)}]{Hippke21}
---. 2021, \aj, 162, 1, \dodoi{10.3847/1538-3881/abf7b7}

\bibitem[{{Hippke} \& {Forgan}(2017)}]{Hippke17-XRay}
{Hippke}, M., \& {Forgan}, D.~H. 2017, arXiv e-prints, arXiv:1711.05761,
  \dodoi{10.48550/arXiv.1711.05761}

\bibitem[{{H{\"o}gbom}(1974)}]{Hogbom74}
{H{\"o}gbom}, J.~A. 1974, \aaps, 15, 417

\bibitem[{{Horowitz} \& {Sagan}(1993)}]{Horowitz93}
{Horowitz}, P., \& {Sagan}, C. 1993, \apj, 415, 218, \dodoi{10.1086/173157}

\bibitem[{{Howard} {et~al.}(2004){Howard}, {Horowitz}, {Wilkinson}, {Coldwell},
  {Groth}, {Jarosik}, {Latham}, {Stefanik}, {Willman}, {Wolff}, \&
  {Zajac}}]{Howard04}
{Howard}, A.~W., {Horowitz}, P., {Wilkinson}, D.~T., {et~al.} 2004, \apj, 613,
  1270, \dodoi{10.1086/423300}

\bibitem[{{Isaacson} {et~al.}(2017){Isaacson}, {Siemion}, {Marcy}, {Lebofsky},
  {Price}, {MacMahon}, {Croft}, {DeBoer}, {Hickish}, {Werthimer}, {Sheikh},
  {Hellbourg}, \& {Enriquez}}]{Isaacson17}
{Isaacson}, H., {Siemion}, A. P.~V., {Marcy}, G.~W., {et~al.} 2017, \pasp, 129,
  054501, \dodoi{10.1088/1538-3873/aa5800}

\bibitem[{{Jarrett} {et~al.}(2019){Jarrett}, {Cluver}, {Brown}, {Dale}, {Tsai},
  \& {Masci}}]{Jarrett19}
{Jarrett}, T.~H., {Cluver}, M.~E., {Brown}, M.~J.~I., {et~al.} 2019, \apjs,
  245, 25, \dodoi{10.3847/1538-4365/ab521a}

\bibitem[{{Jones}(1981)}]{Jones81}
{Jones}, E.~M. 1981, \icarus, 46, 328, \dodoi{10.1016/0019-1035(81)90136-6}

\bibitem[{{Jugaku} \& {Nishimura}(2004)}]{Jugaku04}
{Jugaku}, J., \& {Nishimura}, S. 2004, in Bioastronomy 2002: Life Among the
  Stars, ed. R.~{Norris} \& F.~{Stootman}, Vol. 213 (San Francisco:
  Astronomical Society of the Pacific), 437, \dodoi{10.1017/s0074180900193672}

\bibitem[{{Kardashev}(1964)}]{Kardashev64}
{Kardashev}, N.~S. 1964, \sovast, 8, 217

\bibitem[{{Kashibadze} {et~al.}(2020){Kashibadze}, {Karachentsev}, \&
  {Karachentseva}}]{Kashibadze20}
{Kashibadze}, O.~G., {Karachentsev}, I.~D., \& {Karachentseva}, V.~E. 2020,
  \aap, 635, A135, \dodoi{10.1051/0004-6361/201936172}

\bibitem[{{Kingman}(1993)}]{Kingman93}
{Kingman}, J. F.~C. 1993, {Poisson Processes} (Oxford: Clarendon Press),
  \dodoi{10.1093/oso/9780198536932.001.0001}

\bibitem[{{Kogan}(1999)}]{Kogan99}
{Kogan}, L. 1999, \pasp, 111, 510, \dodoi{10.1086/316345}

\bibitem[{{Kudale} \& {Chengalur}(2017)}]{Kudale17}
{Kudale}, S., \& {Chengalur}, J.~N. 2017, Experimental Astronomy, 44, 97,
  \dodoi{10.1007/s10686-017-9547-0}

\bibitem[{{Kuiper} \& {Morris}(1977)}]{Kuiper77}
{Kuiper}, T.~B.~H., \& {Morris}, M. 1977, Science, 196, 616,
  \dodoi{10.1126/science.196.4290.616}

\bibitem[{{Kulkarni}(1989)}]{Kulkarni89}
{Kulkarni}, S.~R. 1989, \aj, 98, 1112, \dodoi{10.1086/115202}

\bibitem[{{Lacki}(2015{\natexlab{a}})}]{Lacki15-ATTs}
{Lacki}, B.~C. 2015{\natexlab{a}}, arXiv e-prints, arXiv:1501.07309.
\newblock \doarXiv{1501.07309}

\bibitem[{{Lacki}(2015{\natexlab{b}})}]{Lacki15-yEv}
---. 2015{\natexlab{b}}, arXiv e-prints, arXiv:1503.01509.
\newblock \doarXiv{1503.01509}

\bibitem[{{Lacki}(2020)}]{Lacki20-LensFlare}
---. 2020, \apj, 905, 18, \dodoi{10.3847/1538-4357/abc1e3}

\bibitem[{{Learned} {et~al.}(2012){Learned}, {Kudritzki}, {Pakvasa}, \&
  {Zee}}]{Learned08}
{Learned}, J.~G., {Kudritzki}, R.~P., {Pakvasa}, S., \& {Zee}, A. 2012,
  Contemporary Physics, 53, 113, \dodoi{10.1080/00107514.2011.640142}

\bibitem[{{Learned} {et~al.}(1994){Learned}, {Pakvasa}, {Simmons}, \&
  {Tata}}]{Learned94}
{Learned}, J.~G., {Pakvasa}, S., {Simmons}, W.~A., \& {Tata}, X. 1994, \qjras,
  35, 321

\bibitem[{{Lebofsky} {et~al.}(2019){Lebofsky}, {Croft}, {Siemion}, {Price},
  {Enriquez}, {Isaacson}, {MacMahon}, {Anderson}, {Brzycki}, {Cobb}, {Czech},
  {DeBoer}, {DeMarines}, {Drew}, {Foster}, {Gajjar}, {Gizani}, {Hellbourg},
  {Korpela}, {Lacki}, {Sheikh}, {Werthimer}, {Worden}, {Yu}, \&
  {Zhang}}]{Lebofsky19}
{Lebofsky}, M., {Croft}, S., {Siemion}, A. P.~V., {et~al.} 2019, \pasp, 131,
  124505, \dodoi{10.1088/1538-3873/ab3e82}

\bibitem[{{Lehmer} {et~al.}(2010){Lehmer}, {Alexander}, {Bauer}, {Brandt},
  {Goulding}, {Jenkins}, {Ptak}, \& {Roberts}}]{Lehmer10}
{Lehmer}, B.~D., {Alexander}, D.~M., {Bauer}, F.~E., {et~al.} 2010, \apj, 724,
  559, \dodoi{10.1088/0004-637X/724/1/559}

\bibitem[{{Lehmer} {et~al.}(2012){Lehmer}, {Xue}, {Brandt}, {Alexander},
  {Bauer}, {Brusa}, {Comastri}, {Gilli}, {Hornschemeier}, {Luo}, {Paolillo},
  {Ptak}, {Shemmer}, {Schneider}, {Tozzi}, \& {Vignali}}]{Lehmer12}
{Lehmer}, B.~D., {Xue}, Y.~Q., {Brandt}, W.~N., {et~al.} 2012, \apj, 752, 46,
  \dodoi{10.1088/0004-637X/752/1/46}

\bibitem[{{Lehmer} {et~al.}(2016){Lehmer}, {Basu-Zych}, {Mineo}, {Brandt},
  {Eufrasio}, {Fragos}, {Hornschemeier}, {Luo}, {Xue}, {Bauer}, {Gilfanov},
  {Ranalli}, {Schneider}, {Shemmer}, {Tozzi}, {Trump}, {Vignali}, {Wang},
  {Yukita}, \& {Zezas}}]{Lehmer16}
{Lehmer}, B.~D., {Basu-Zych}, A.~R., {Mineo}, S., {et~al.} 2016, \apj, 825, 7,
  \dodoi{10.3847/0004-637X/825/1/7}

\bibitem[{{Li} {et~al.}(2020){Li}, {Gajjar}, {Wang}, {Siemion}, {Zhang},
  {Zhang}, {Yue}, {Zhu}, {Jin}, {Li}, {Berger}, {Brzycki}, {Cobb}, {Croft},
  {Czech}, {DeBoer}, {DeMarines}, {Drew}, {Emilio Enriquez}, {Gizani},
  {Korpela}, {Isaacson}, {Lebofsky}, {Lacki}, {MacMahon}, {Nanez}, {Niu},
  {Pei}, {Price}, {Werthimer}, {Worden}, {Gerry Zhang}, {Zhang}, \& {FAST
  Collaboration}}]{Li20}
{Li}, D., {Gajjar}, V., {Wang}, P., {et~al.} 2020, Research in Astronomy and
  Astrophysics, 20, 078, \dodoi{10.1088/1674-4527/20/5/78}

\bibitem[{{Licquia} \& {Newman}(2015)}]{Licquia15}
{Licquia}, T.~C., \& {Newman}, J.~A. 2015, \apj, 806, 96,
  \dodoi{10.1088/0004-637X/806/1/96}

\bibitem[{{Lien} \& {Fields}(2012)}]{Lien12}
{Lien}, A., \& {Fields}, B.~D. 2012, \apj, 747, 120,
  \dodoi{10.1088/0004-637X/747/2/120}

\bibitem[{{Lingam} \& {Loeb}(2021)}]{Lingam21}
{Lingam}, M., \& {Loeb}, A. 2021, {Life in the Cosmos: From Biosignatures to
  Technosignatures} (Cambridge, MA: Harvard University Press),
  \dodoi{10.4159/9780674259959}

\bibitem[{{Malyshev} \& {Hogg}(2011)}]{Malyshev11}
{Malyshev}, D., \& {Hogg}, D.~W. 2011, \apj, 738, 181,
  \dodoi{10.1088/0004-637X/738/2/181}

\bibitem[{{Margot} {et~al.}(2021){Margot}, {Pinchuk}, {Geil}, {Alexander},
  {Arora}, {Biswas}, {Cebreros}, {Desai}, {Duclos}, {Dunne}, {Lin Fu}, {Goel},
  {Gonzales}, {Gonzalez}, {Jain}, {Lam}, {Lewis}, {Lewis}, {Li}, {MacDougall},
  {Makarem}, {Manan}, {Molina}, {Nagib}, {Neville}, {O'Toole}, {Rockwell},
  {Rokushima}, {Romanek}, {Schmidgall}, {Seth}, {Shah}, {Shimane}, {Singhal},
  {Tokadjian}, {Villafana}, {Wang}, {Yun}, {Zhu}, \& {Lynch}}]{Margot21}
{Margot}, J.-L., {Pinchuk}, P., {Geil}, R., {et~al.} 2021, \aj, 161, 55,
  \dodoi{10.3847/1538-3881/abcc77}

\bibitem[{{McConnachie}(2012)}]{McConnachie12}
{McConnachie}, A.~W. 2012, \aj, 144, 4, \dodoi{10.1088/0004-6256/144/1/4}

\bibitem[{{Messerschmitt}(2015)}]{Messerschmitt15}
{Messerschmitt}, D.~G. 2015, Acta Astronautica, 107, 20,
  \dodoi{10.1016/j.actaastro.2014.11.007}

\bibitem[{{Meyer} {et~al.}(2004){Meyer}, {Zwaan}, {Webster}, {Staveley-Smith},
  {Ryan-Weber}, {Drinkwater}, {Barnes}, {Howlett}, {Kilborn}, {Stevens},
  {Waugh}, {Pierce}, {Bhathal}, {de Blok}, {Disney}, {Ekers}, {Freeman},
  {Garcia}, {Gibson}, {Harnett}, {Henning}, {Jerjen}, {Kesteven}, {Knezek},
  {Koribalski}, {Mader}, {Marquarding}, {Minchin}, {O'Brien}, {Oosterloo},
  {Price}, {Putman}, {Ryder}, {Sadler}, {Stewart}, {Stootman}, \&
  {Wright}}]{Meyer04}
{Meyer}, M.~J., {Zwaan}, M.~A., {Webster}, R.~L., {et~al.} 2004, \mnras, 350,
  1195, \dodoi{10.1111/j.1365-2966.2004.07710.x}

\bibitem[{{Mineo} {et~al.}(2012){Mineo}, {Gilfanov}, \&
  {Sunyaev}}]{Mineo12-ISM}
{Mineo}, S., {Gilfanov}, M., \& {Sunyaev}, R. 2012, \mnras, 426, 1870,
  \dodoi{10.1111/j.1365-2966.2012.21831.x}

\bibitem[{{Moffett} {et~al.}(2016){Moffett}, {Ingarfield}, {Driver},
  {Robotham}, {Kelvin}, {Lange}, {Me{\v{s}}tri{\'c}}, {Alpaslan}, {Baldry},
  {Bland-Hawthorn}, {Brough}, {Cluver}, {Davies}, {Holwerda}, {Hopkins},
  {Kafle}, {Kennedy}, {Norberg}, \& {Taylor}}]{Moffett16}
{Moffett}, A.~J., {Ingarfield}, S.~A., {Driver}, S.~P., {et~al.} 2016, \mnras,
  457, 1308, \dodoi{10.1093/mnras/stv2883}

\bibitem[{{Murphy} {et~al.}(2011){Murphy}, {Condon}, {Schinnerer}, {Kennicutt},
  {Calzetti}, {Armus}, {Helou}, {Turner}, {Aniano}, {Beir{\~a}o}, {Bolatto},
  {Brandl}, {Croxall}, {Dale}, {Donovan Meyer}, {Draine}, {Engelbracht},
  {Hunt}, {Hao}, {Koda}, {Roussel}, {Skibba}, \& {Smith}}]{Murphy11}
{Murphy}, E.~J., {Condon}, J.~J., {Schinnerer}, E., {et~al.} 2011, \apj, 737,
  67, \dodoi{10.1088/0004-637X/737/2/67}

\bibitem[{{Nityananda}(1994)}]{Nityananda94}
{Nityananda}, R. 1994, in IAU Symposium, Vol. 158, Very High Angular Resolution
  Imaging, ed. J.~G. {Robertson} \& W.~J. {Tango} (Dordrecht: Kluwer), 11,
  \dodoi{10.1007/978-94-011-0880-5_2}

\bibitem[{{Norris} {et~al.}(2011){Norris}, {Hopkins}, {Afonso}, {Brown},
  {Condon}, {Dunne}, {Feain}, {Hollow}, {Jarvis}, {Johnston-Hollitt}, {Lenc},
  {Middelberg}, {Padovani}, {Prandoni}, {Rudnick}, {Seymour}, {Umana},
  {Andernach}, {Alexander}, {Appleton}, {Bacon}, {Banfield}, {Becker}, {Brown},
  {Ciliegi}, {Jackson}, {Eales}, {Edge}, {Gaensler}, {Giovannini}, {Hales},
  {Hancock}, {Huynh}, {Ibar}, {Ivison}, {Kennicutt}, {Kimball}, {Koekemoer},
  {Koribalski}, {L{\'o}pez-S{\'a}nchez}, {Mao}, {Murphy}, {Messias},
  {Pimbblet}, {Raccanelli}, {Randall}, {Reiprich}, {Roseboom},
  {R{\"o}ttgering}, {Saikia}, {Sharp}, {Slee}, {Smail}, {Thompson}, {Urquhart},
  {Wall}, \& {Zhao}}]{Norris11}
{Norris}, R.~P., {Hopkins}, A.~M., {Afonso}, J., {et~al.} 2011, \pasa, 28, 215,
  \dodoi{10.1071/AS11021}

\bibitem[{{Nyland} {et~al.}(2017){Nyland}, {Young}, {Wrobel}, {Davis},
  {Bureau}, {Alatalo}, {Morganti}, {Duc}, {de Zeeuw}, {McDermid}, {Crocker}, \&
  {Oosterloo}}]{Nyland17}
{Nyland}, K., {Young}, L.~M., {Wrobel}, J.~M., {et~al.} 2017, \mnras, 464,
  1029, \dodoi{10.1093/mnras/stw2385}

\bibitem[{{Oliver} \& {Billingham}(1971)}]{Oliver71}
{Oliver}, B.~M., \& {Billingham}, J. 1971, {Project Cyclops: A Design Study of
  a System for Detecting Extraterrestrial Intelligent Life}, Vol.
  NASA-CR-114445 (Mountain View, CA: NASA Ames Research Center)

\bibitem[{{Paris}(2010)}]{Paris10}
{Paris}, R.~B. 2010, in NIST Handbook of Mathematical Functions, ed. F.~W.~J.
  {Olver}, D.~W. {Lozier}, R.~F. {Boisvert}, \& C.~W. {Clark} (Cambridge
  University Press), 173--192

\bibitem[{{Price}(2021)}]{Price16}
{Price}, D.~C. 2021, in The WSPC Handbook of Astronomical Instrumentation,
  Volume 1: Radio Astronomic al Instrumentation, ed. A.~{Wolszczan} (Singapore:
  World Scientific), 159--179, \dodoi{10.1142/9789811203770_0007}

\bibitem[{{Price} {et~al.}(2020){Price}, {Enriquez}, {Brzycki}, {Croft},
  {Czech}, {DeBoer}, {DeMarines}, {Foster}, {Gajjar}, {Gizani}, {Hellbourg},
  {Isaacson}, {Lacki}, {Lebofsky}, {MacMahon}, {Pater}, {Siemion}, {Werthimer},
  {Green}, {Kaczmarek}, {Maddalena}, {Mader}, {Drew}, \& {Worden}}]{Price20}
{Price}, D.~C., {Enriquez}, J.~E., {Brzycki}, B., {et~al.} 2020, \aj, 159, 86,
  \dodoi{10.3847/1538-3881/ab65f1}

\bibitem[{{Radhakrishnan}(1999)}]{Radhakrishnan99}
{Radhakrishnan}, V. 1999, in Astronomical Society of the Pacific Conference
  Series, Vol. 180, Synthesis Imaging in Radio Astronomy II, ed. G.~B.
  {Taylor}, C.~L. {Carilli}, \& R.~A. {Perley} (San Francisco: Astronomical
  Society of the Pacific), 671

\bibitem[{{Raimondo} {et~al.}(2005){Raimondo}, {Brocato}, {Cantiello}, \&
  {Capaccioli}}]{Raimondo05}
{Raimondo}, G., {Brocato}, E., {Cantiello}, M., \& {Capaccioli}, M. 2005, \aj,
  130, 2625, \dodoi{10.1086/497591}

\bibitem[{{Rampadarath} {et~al.}(2012){Rampadarath}, {Morgan}, {Tingay}, \&
  {Trott}}]{Rampadarath12}
{Rampadarath}, H., {Morgan}, J.~S., {Tingay}, S.~J., \& {Trott}, C.~M. 2012,
  \aj, 144, 38, \dodoi{10.1088/0004-6256/144/2/38}

\bibitem[{{Roy} \& {Olver}(2010)}]{Roy10}
{Roy}, R., \& {Olver}, F. W.~J. 2010, in NIST Handbook of Mathematical
  Functions, ed. F.~W.~J. {Olver}, D.~W. {Lozier}, R.~F. {Boisvert}, \& C.~W.
  {Clark} (Cambridge University Press), 103--134

\bibitem[{{Sabater} {et~al.}(2019){Sabater}, {Best}, {Hardcastle}, {Shimwell},
  {Tasse}, {Williams}, {Br{\"u}ggen}, {Cochrane}, {Croston}, {de Gasperin},
  {Duncan}, {G{\"u}rkan}, {Mechev}, {Morabito}, {Prandoni}, {R{\"o}ttgering},
  {Smith}, {Harwood}, {Mingo}, {Mooney}, \& {Saxena}}]{Sabater19}
{Sabater}, J., {Best}, P.~N., {Hardcastle}, M.~J., {et~al.} 2019, \aap, 622,
  A17, \dodoi{10.1051/0004-6361/201833883}

\bibitem[{{Scheuer}(1957)}]{Scheuer57}
{Scheuer}, P.~A.~G. 1957, Proceedings of the Cambridge Philosophical Society,
  53, 764, \dodoi{10.1017/S0305004100032825}

\bibitem[{{Scheuer}(1974)}]{Scheuer74}
---. 1974, \mnras, 166, 329, \dodoi{10.1093/mnras/166.2.329}

\bibitem[{{Schwartz} \& {Townes}(1961)}]{Schwartz61}
{Schwartz}, R.~N., \& {Townes}, C.~H. 1961, \nat, 190, 205,
  \dodoi{10.1038/190205a0}

\bibitem[{{Schwarz}(1978)}]{Schwarz78}
{Schwarz}, U.~J. 1978, \aap, 65, 345

\bibitem[{{Sheikh} {et~al.}(2019){Sheikh}, {Wright}, {Siemion}, \&
  {Enriquez}}]{Sheikh19}
{Sheikh}, S.~Z., {Wright}, J.~T., {Siemion}, A., \& {Enriquez}, J.~E. 2019,
  \apj, 884, 14, \dodoi{10.3847/1538-4357/ab3fa8}

\bibitem[{{Shostak} {et~al.}(1996){Shostak}, {Ekers}, \& {Vaile}}]{Shostak96}
{Shostak}, S., {Ekers}, R., \& {Vaile}, R. 1996, \aj, 112, 164,
  \dodoi{10.1086/117996}

\bibitem[{{Siemion} {et~al.}(2013){Siemion}, {Demorest}, {Korpela},
  {Maddalena}, {Werthimer}, {Cobb}, {Howard}, {Langston}, {Lebofsky}, {Marcy},
  \& {Tarter}}]{Siemion13}
{Siemion}, A. P.~V., {Demorest}, P., {Korpela}, E., {et~al.} 2013, \apj, 767,
  94, \dodoi{10.1088/0004-637X/767/1/94}

\bibitem[{{Silva} {et~al.}(1998){Silva}, {Granato}, {Bressan}, \&
  {Danese}}]{Silva98}
{Silva}, L., {Granato}, G.~L., {Bressan}, A., \& {Danese}, L. 1998, \apj, 509,
  103, \dodoi{10.1086/306476}

\bibitem[{{Smith} {et~al.}(2012){Smith}, {Gomez}, {Eales}, {Ciesla}, {Boselli},
  {Cortese}, {Bendo}, {Baes}, {Bianchi}, {Clemens}, {Clements}, {Cooray},
  {Davies}, {De Looze}, {di Serego Alighieri}, {Fritz}, {Gavazzi}, {Gear},
  {Madden}, {Mentuch}, {Panuzzo}, {Pohlen}, {Spinoglio}, {Verstappen},
  {Vlahakis}, {Wilson}, \& {Xilouris}}]{Smith12}
{Smith}, M.~W.~L., {Gomez}, H.~L., {Eales}, S.~A., {et~al.} 2012, \apj, 748,
  123, \dodoi{10.1088/0004-637X/748/2/123}

\bibitem[{{Speagle} {et~al.}(2014){Speagle}, {Steinhardt}, {Capak}, \&
  {Silverman}}]{Speagle14}
{Speagle}, J.~S., {Steinhardt}, C.~L., {Capak}, P.~L., \& {Silverman}, J.~D.
  2014, \apjs, 214, 15, \dodoi{10.1088/0067-0049/214/2/15}

\bibitem[{{Spekkens} {et~al.}(2013){Spekkens}, {Mason}, {Aguirre}, \&
  {Nhan}}]{Spekkens13}
{Spekkens}, K., {Mason}, B.~S., {Aguirre}, J.~E., \& {Nhan}, B. 2013, \apj,
  773, 61, \dodoi{10.1088/0004-637X/773/1/61}

\bibitem[{{Stappers} {et~al.}(2011){Stappers}, {Hessels}, {Alexov}, {Anderson},
  {Coenen}, {Hassall}, {Karastergiou}, {Kondratiev}, {Kramer}, {van Leeuwen},
  {Mol}, {Noutsos}, {Romein}, {Weltevrede}, {Fender}, {Wijers}, {B{\"a}hren},
  {Bell}, {Broderick}, {Daw}, {Dhillon}, {Eisl{\"o}ffel}, {Falcke},
  {Griessmeier}, {Law}, {Markoff}, {Miller-Jones}, {Scheers}, {Spreeuw},
  {Swinbank}, {Ter Veen}, {Wise}, {Wucknitz}, {Zarka}, {Anderson}, {Asgekar},
  {Avruch}, {Beck}, {Bennema}, {Bentum}, {Best}, {Bregman}, {Brentjens}, {van
  de Brink}, {Broekema}, {Brouw}, {Br{\"u}ggen}, {de Bruyn}, {Butcher},
  {Ciardi}, {Conway}, {Dettmar}, {van Duin}, {van Enst}, {Garrett}, {Gerbers},
  {Grit}, {Gunst}, {van Haarlem}, {Hamaker}, {Heald}, {Hoeft}, {Holties},
  {Horneffer}, {Koopmans}, {Kuper}, {Loose}, {Maat}, {McKay-Bukowski},
  {McKean}, {Miley}, {Morganti}, {Nijboer}, {Noordam}, {Norden}, {Olofsson},
  {Pandey-Pommier}, {Polatidis}, {Reich}, {R{\"o}ttgering}, {Schoenmakers},
  {Sluman}, {Smirnov}, {Steinmetz}, {Sterks}, {Tagger}, {Tang}, {Vermeulen},
  {Vermaas}, {Vogt}, {de Vos}, {Wijnholds}, {Yatawatta}, \&
  {Zensus}}]{Stappers11}
{Stappers}, B.~W., {Hessels}, J.~W.~T., {Alexov}, A., {et~al.} 2011, \aap, 530,
  A80, \dodoi{10.1051/0004-6361/201116681}

\bibitem[{{Stigler}(1973)}]{Stigler73}
{Stigler}, S.~M. 1973, The Annals of Statistics, 472,
  \dodoi{10.1214/aos/1176342412}

\bibitem[{{Strong} {et~al.}(2010){Strong}, {Porter}, {Digel},
  {J{\'o}hannesson}, {Martin}, {Moskalenko}, {Murphy}, \& {Orlando}}]{Strong10}
{Strong}, A.~W., {Porter}, T.~A., {Digel}, S.~W., {et~al.} 2010, \apjl, 722,
  L58, \dodoi{10.1088/2041-8205/722/1/L58}

\bibitem[{{Suazo} {et~al.}(2022){Suazo}, {Zackrisson}, {Wright}, {Korn}, \&
  {Huston}}]{Suazo22}
{Suazo}, M., {Zackrisson}, E., {Wright}, J.~T., {Korn}, A.~J., \& {Huston}, M.
  2022, \mnras, 512, 2988, \dodoi{10.1093/mnras/stac280}

\bibitem[{{Subotowicz}(1979)}]{Subotowicz79}
{Subotowicz}, M. 1979, Acta Astronautica, 6, 213,
  \dodoi{10.1016/0094-5765(79)90157-7}

\bibitem[{{Sullivan} {et~al.}(1978){Sullivan}, {Brown}, \&
  {Wetherill}}]{Sullivan78}
{Sullivan}, W.~T., I., {Brown}, S., \& {Wetherill}, C. 1978, Science, 199, 377,
  \dodoi{10.1126/science.199.4327.377}

\bibitem[{{Tamm} {et~al.}(2012){Tamm}, {Tempel}, {Tenjes}, {Tihhonova}, \&
  {Tuvikene}}]{Tamm12}
{Tamm}, A., {Tempel}, E., {Tenjes}, P., {Tihhonova}, O., \& {Tuvikene}, T.
  2012, \aap, 546, A4, \dodoi{10.1051/0004-6361/201220065}

\bibitem[{{Tan} \& {Kurtsiefer}(2017)}]{Tan17}
{Tan}, P.~K., \& {Kurtsiefer}, C. 2017, \mnras, 469, 1617,
  \dodoi{10.1093/mnras/stx968}

\bibitem[{{Tan} {et~al.}(2014){Tan}, {Yeo}, {Poh}, {Chan}, \&
  {Kurtsiefer}}]{Tan14}
{Tan}, P.~K., {Yeo}, G.~H., {Poh}, H.~S., {Chan}, A.~H., \& {Kurtsiefer}, C.
  2014, \apjl, 789, L10, \dodoi{10.1088/2041-8205/789/1/L10}

\bibitem[{{Tarter}(1985)}]{Tarter85}
{Tarter}, J. 1985, in IAU Symposium, Vol. 112, The Search for Extraterrestrial
  Life: Recent Developments, ed. M.~D. {Papagiannis} (Dordrecht: D. Reidel
  Publishing Co.), 271--290, \dodoi{10.1007/978-94-009-5462-5_37}

\bibitem[{{Tarter}(2001)}]{Tarter01}
{Tarter}, J. 2001, \araa, 39, 511, \dodoi{10.1146/annurev.astro.39.1.511}

\bibitem[{{Temi} {et~al.}(2007){Temi}, {Brighenti}, \& {Mathews}}]{Temi07}
{Temi}, P., {Brighenti}, F., \& {Mathews}, W.~G. 2007, \apj, 660, 1215,
  \dodoi{10.1086/513690}

\bibitem[{{Tonry} \& {Schneider}(1988)}]{Tonry88}
{Tonry}, J., \& {Schneider}, D.~P. 1988, \aj, 96, 807, \dodoi{10.1086/114847}

\bibitem[{{Tremblay} \& {Tingay}(2020)}]{Tremblay20}
{Tremblay}, C.~D., \& {Tingay}, S.~J. 2020, \pasa, 37, e035,
  \dodoi{10.1017/pasa.2020.27}

\bibitem[{{Vernstrom} {et~al.}(2014){Vernstrom}, {Scott}, {Wall}, {Condon},
  {Cotton}, {Fomalont}, {Kellermann}, {Miller}, \& {Perley}}]{Vernstrom14}
{Vernstrom}, T., {Scott}, D., {Wall}, J.~V., {et~al.} 2014, \mnras, 440, 2791,
  \dodoi{10.1093/mnras/stu470}

\bibitem[{{Voros}(2013)}]{Voros14}
{Voros}, J. 2013, arXiv e-prints, arXiv:1412.4011.
\newblock \doarXiv{1412.4011}

\bibitem[{{Wasserman}(2004)}]{Wasserman04}
{Wasserman}, L. 2004, All of Statistics: A Concise Course in Statistical
  Inference (New York: Springer New York), \dodoi{10.1007/978-0-387-21736-9}

\bibitem[{{Webb}(2015)}]{Webb15}
{Webb}, S. 2015, {If the Universe Is Teeming with Aliens... Where is
  Everybody?} (Cham, Switzerland: Springer Cham),
  \dodoi{10.1007/978-3-319-13236-5}

\bibitem[{{Wilson} {et~al.}(2009){Wilson}, {Rohlfs}, \&
  {H{\"u}ttemeister}}]{Wilson09}
{Wilson}, T.~L., {Rohlfs}, K., \& {H{\"u}ttemeister}, S. 2009, {Tools of Radio
  Astronomy} (Berlin: Springer-Verlag), \dodoi{10.1007/978-3-540-85122-6}

\bibitem[{{Wlodarczyk-Sroka} {et~al.}(2020){Wlodarczyk-Sroka}, {Garrett}, \&
  {Siemion}}]{WlodarczykSroka20}
{Wlodarczyk-Sroka}, B.~S., {Garrett}, M.~A., \& {Siemion}, A.~P.~V. 2020,
  \mnras, 498, 5720, \dodoi{10.1093/mnras/staa2672}

\bibitem[{{Worden} {et~al.}(2017){Worden}, {Drew}, {Siemion}, {Werthimer},
  {DeBoer}, {Croft}, {MacMahon}, {Lebofsky}, {Isaacson}, {Hickish}, {Price},
  {Gajjar}, \& {Wright}}]{Worden17}
{Worden}, S.~P., {Drew}, J., {Siemion}, A., {et~al.} 2017, Acta Astronautica,
  139, 98, \dodoi{10.1016/j.actaastro.2017.06.008}

\bibitem[{{Wright} {et~al.}(2014{\natexlab{a}}){Wright}, {Griffith},
  {Sigurdsson}, {Povich}, \& {Mullan}}]{Wright14-Search}
{Wright}, J.~T., {Griffith}, R.~L., {Sigurdsson}, S., {Povich}, M.~S., \&
  {Mullan}, B. 2014{\natexlab{a}}, \apj, 792, 27,
  \dodoi{10.1088/0004-637X/792/1/27}

\bibitem[{{Wright} {et~al.}(2018){Wright}, {Kanodia}, \& {Lubar}}]{Wright18}
{Wright}, J.~T., {Kanodia}, S., \& {Lubar}, E. 2018, \aj, 156, 260,
  \dodoi{10.3847/1538-3881/aae099}

\bibitem[{{Wright} {et~al.}(2014{\natexlab{b}}){Wright}, {Mullan},
  {Sigurdsson}, \& {Povich}}]{Wright14-Paradox}
{Wright}, J.~T., {Mullan}, B., {Sigurdsson}, S., \& {Povich}, M.~S.
  2014{\natexlab{b}}, \apj, 792, 26, \dodoi{10.1088/0004-637X/792/1/26}

\bibitem[{{Yamamoto} \& {Haus}(1986)}]{Yamamoto86}
{Yamamoto}, Y., \& {Haus}, H.~A. 1986, Reviews of Modern Physics, 58, 1001,
  \dodoi{10.1103/RevModPhys.58.1001}

\bibitem[{{Yun} {et~al.}(2001){Yun}, {Reddy}, \& {Condon}}]{Yun01}
{Yun}, M.~S., {Reddy}, N.~A., \& {Condon}, J.~J. 2001, \apj, 554, 803,
  \dodoi{10.1086/323145}

\bibitem[{{Zackrisson} {et~al.}(2015){Zackrisson}, {Calissendorff}, {Asadi}, \&
  {Nyholm}}]{Zackrisson15}
{Zackrisson}, E., {Calissendorff}, P., {Asadi}, S., \& {Nyholm}, A. 2015, \apj,
  810, 23, \dodoi{10.1088/0004-637X/810/1/23}

\bibitem[{{Zheng} {et~al.}(2017){Zheng}, {Tegmark}, {Dillon}, {Kim}, {Liu},
  {Neben}, {Jonas}, {Reich}, \& {Reich}}]{Zheng17}
{Zheng}, H., {Tegmark}, M., {Dillon}, J.~S., {et~al.} 2017, \mnras, 464, 3486,
  \dodoi{10.1093/mnras/stw2525}

\bibitem[{{Zmuidzinas}(2003)}]{Zmuidzinas03}
{Zmuidzinas}, J. 2003, \ao, 42, 4989, \dodoi{10.1364/AO.42.004989}

\bibitem[{{Zmuidzinas}(2015)}]{Zmuidzinas15}
---. 2015, \apj, 813, 17, \dodoi{10.1088/0004-637X/813/1/17}

\end{thebibliography}

\end{document}